\documentclass[10pt]{article}
\usepackage{amsmath, amsthm, amssymb,amsfonts}
\usepackage{geometry}
\usepackage{verbatim, mathtools,enumerate, physics,framed, mathrsfs}
\usepackage{color}
\usepackage{todonotes}
\usepackage{graphicx, caption}

\definecolor{shadecolor}{rgb}{0.98, 0.98, 0.9}

\usepackage{cancel,ulem}

\def\nn{\nonumber}
\usepackage{tikz}
\def\pa{\partial}
\usetikzlibrary{calc}
\usetikzlibrary{decorations.markings}

\usepackage[ colorlinks=true, pdfstartview=FitV, linkcolor=blue, citecolor=blue, urlcolor=blue]{hyperref}

\def\scr{\mathscr}

\def\Ai{ {\mathrm {Ai}}}

\def\z{\zeta}
\theoremstyle{plain}
\def\d{{\mathrm d}}
\def \le{\left}
\def \ri{\right}
\newtheorem{theorem}{Theorem}[section]
\newtheorem{corollary}[theorem]{Corollary}
\newtheorem{lemma}[theorem]{Lemma}
\newtheorem{rhp}[theorem]{Riemann-Hilbert Problem}
\newtheorem{proposition}[theorem]{Proposition}
\newtheorem{conjecture}[theorem]{Conjecture}

 \def\bea#1\eea{\begin{align}#1\end{align}}
 \def\be#1\ee{\begin{align}#1\end{align}}

\theoremstyle{definition}
\newtheorem{definition}{Definition}[section]

\newtheorem{remark}[theorem]{Remark}

\newtheorem{assumption}[theorem]{Assumption}

\theoremstyle{remark}

\def\N{\mathbb{N}}
\def\Z{\mathbb{Z}}
\def \wt{\widetilde}

\def\R{\mathbb{R}}
\def\C{\mathbb{C}}

\def\ds{\displaystyle}
\def\Res{\text{Res}}

\def\P2{\mathrm{P}_{{\mathrm{II}}}}
\def\bigO{\mathcal{O}}

\newcommand{\So}{S_{\rm{odd}}}

\numberwithin{equation}{section}

\renewcommand{\S}{\mathbb{S}}
\def\St{\mathcal{S}}

\newcommand{\wkb}[2]{\psi_{#2}^{(#1)} }

\newcommand{\VJM}{V_{\rm JM}}
\newcommand{\VST}{V_{\rm ST}}

\def\QED{\hfill $\blacksquare$\par\vskip 10pt}

\newcommand{\Tau}{\mathrm{T}}

\title{Painlev\'e equation, anharmonic oscillators and orthogonality}

\author{M. Bertola, E. Chavez, T. Grava}

\begin{document}
\centerline
{\bf \Large Exactly solvable anharmonic oscillator, }
\centerline
{\bf \Large degenerate orthogonal polynomials and Painlev\'e\ II}

\begin{center}
M. Bertola$^{\dagger\ddagger\star}$ \footnote{Marco.Bertola@\{concordia.ca, sissa.it\}} 
E. Chavez-Heredia$^{\ddagger\diamondsuit}$ \footnote{Eduardo.ChavezHeredia@bristol.ac.uk}
T. Grava $^{\ddagger\diamondsuit}$ \footnote{Tamara.Grava@sissa.it}.
\\
\bigskip
\begin{minipage}{0.7\textwidth}
\begin{small}
\begin{enumerate}
\item [${\dagger}$] {\it  Department of Mathematics and
Statistics, Concordia University\\ 1455 de Maisonneuve W., Montr\'eal, Qu\'ebec,
Canada H3G 1M8} 
\item[${\ddagger}$] {\it SISSA, International School for Advanced Studies, via Bonomea 265, Trieste, Italy  and INFN sezione di Trieste }
\item[${\star}$] {\it Centre de recherches math\'ematiques,
Universit\'e de Montr\'eal\\ C.~P.~6128, succ. centre ville, Montr\'eal,
Qu\'ebec, Canada H3C 3J7}
\item[${\diamondsuit}$] {\it School of Mathematics, University of Bristol, Fry Building, Bristol,
BS8 1UG, UK}
\end{enumerate}
\end{small}
\end{minipage}
\end{center}

\centerline{Abstract.}

The paper addresses  a conjecture of Shapiro and Tater on the similarity between two sets of points in the complex plane; on one side is the values
of $t\in \C$ for which the spectrum of the quartic anharmonic oscillator in the complex plane
\begin{equation*}
\dv[2]{y}{x} - \left( x^4 + tx^2 + 2Jx \right)y = \Lambda y,
\end{equation*}
with certain boundary conditions, has repeated eigenvalues.
On the other side is the set of zeroes of the Vorob'ev-Yablonskii polynomials, i.e. the poles of rational solutions of the second Painlev\'e\ equation. 
Along the way, we indicate a surprising and deep connection between the anharmonic oscillator problem and  certain degenerate orthogonal polynomials. 
\tableofcontents

\newpage

\section{Introduction and results}
The second Painlev\'e\ equation is an ODE in the complex domain given by
\begin{equation} \label{eq:painleve2} \tag{PII}
    \dv[2]{u}{t} = 2u^3 + t u + \alpha, 
\end{equation}
with $\alpha \in \C$ and $t \in \C$.
It admits rational solutions when $\alpha = n \in \Z$ and this was recognized  by Vorob'ev and Yablonskii in two separate papers  \cite{V,Y}. For $\alpha = n$ there are $n^2$ poles of the rational solution, of which $n(n-1)/2$ correspond to poles {with} residue $+1$ and the remaining $n(n+1)/2$ correspond to poles with residue $-1$. Both sets of poles are zeros of certain polynomials defined recursively (see below), and  that are  referred to as Vorob'ev-Yablonsky \eqref{eq:VY} polynomials.

\begin{figure}
\centering
\includegraphics[width=0.6\textwidth]{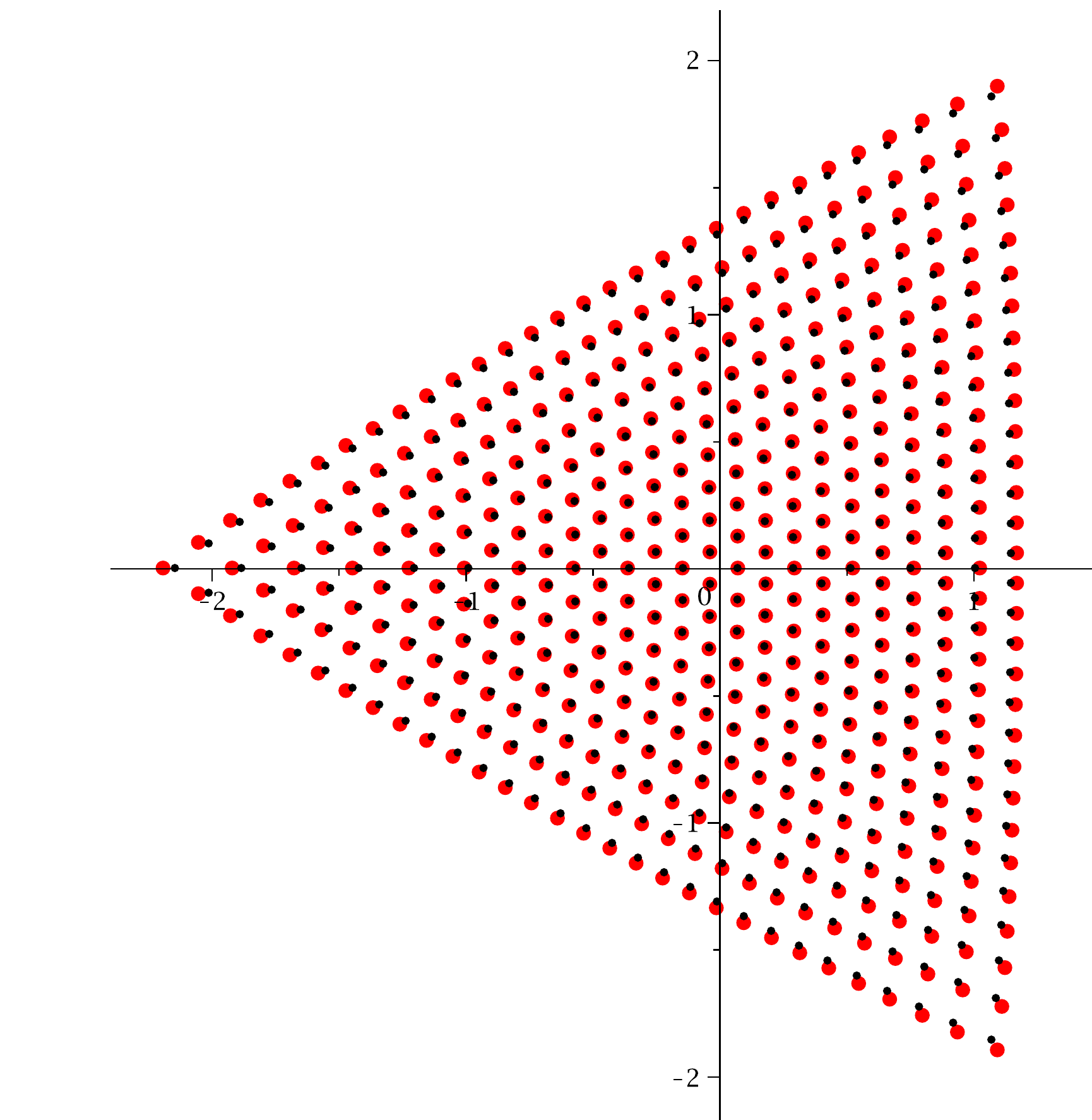}
\caption{Scaled roots of the Vorob'ev-Yablonsky polynomials  $Y_{n}(\hbar^{-2/3}s)$ with $\hbar^{-1}=n+1/2$ in red, and roots of $D_n(\hbar^{-2/3}s)$ with $\hbar^{-1}=n+1$ in black,  for $n=40$. This is the natural scaling with regards to the ``exact'' WKB analysis, as it will be demonstrated below.}
\label{fig:matching_dots}
\end{figure}

The study of the asymptotic behaviour of the poles of  rational solutions  of Painlev\'e equations  has received significant attention both in the community of researchers interested in  Painlev\'e theory \cite{CM,Mas18} and also due to their occurrence in the description of asymptotic behaviour in the semiclassical limit of the Sine-Gordon equation \cite{buckingham_miller, BM14,BM15, BB15}.
The poles form a very regular ``triangular'' pattern as it can be seen in Fig. \ref{fig:matching_dots}.

A seemingly disconnected problem consists in the study of the spectrum of the following boundary value problem
for  the anharmonic oscillator
\begin{align}
    \label{eq:shapiro-ODE}
    &\dv[2]{y(x)}{x} - \left( x^4 + tx^2 + 2Jx \right)y(x) = \Lambda y(x)\\
    \label{boundary}
    & y(x)\to 0 \mbox{    as $x\to\infty$ and $\arg(x)=\pi,  \pm \pi /3$,}
\end{align}
where $t$ and $J$ are in general complex parameters,  and $x$,  $y(x)$ are the complex independent and dependent variable, respectively. In other words \eqref{eq:shapiro-ODE} is an ODE in the complex domain.

The equation is equivalent to a tri-confluent Heun equation by a gauge transformation $y(x)=p(x) {\rm e}^{x^3/3+t/2 x} $. 
  Following the definition in \cite{BB98} a quantum mechanical potential   is {\it Quasi Exactly  Solvable}  (QES)
if  a finite portion of the energy spectrum and associated eigenfunctions can be found exactly and in closed form, while the 
 potential  is {\it  Exactly  Solvable}  (ES) if all  the spectrum and associated eigenfunctions can be found exactly and in closed form.  In particular  
it was shown in \cite{ST22} that the eigenvalue problem \eqref{eq:shapiro-ODE} with  only {\it two out of the three} boundary conditions  at infinity is {\it Quasi Exactly Solvable}.
Here we show that the quartic anharmonic oscillator in \eqref{eq:shapiro-ODE},  with the boundary conditions \eqref{boundary} is Exactly Solvable and admits a solution if and only if  $J=n+1\in \N$ is  a positive integer. The corresponding eigenfunctions   are quasi--polynomials (i.e. a polynomial times a fixed exponential factor).  

These polynomials have been studied in the works of Eremenko and Gabrielov \cite{Eremenko,Eremenko2} and also Mukhin and Tarasov \cite{MT2013}. We actually find several additional structures relating these polynomials to a special type of orthogonality (see below).
For every given $t\in \C$, the corresponding eigenvalues of these quasi--polynomial eigenfunctions are the zeros of a secular equation of degree $n+1$ in the shifted variable $\lambda=\Lambda-\frac{t^2}{4}$.
There are particular discrete values of $t\in \C$ for which the spectrum of the boundary problem \eqref{eq:shapiro-ODE}-\eqref{boundary} becomes {\it degenerate}, namely, one of the eigenvalues has higher algebraic multiplicity.  

For a fixed $J\in\N$, these particular values of the parameter $t\in \C$ form a pattern shown in Fig.~\ref{fig:matching_dots} that is ostensibly very similar to the pattern of the poles of the rational solutions of the Painlev\'e II equation (i.e. the zeroes of the Vorob'ev--Yablonskii polynomials). 
\begin{figure} [ht]
    \centering
    \includegraphics[width=0.6\textwidth]{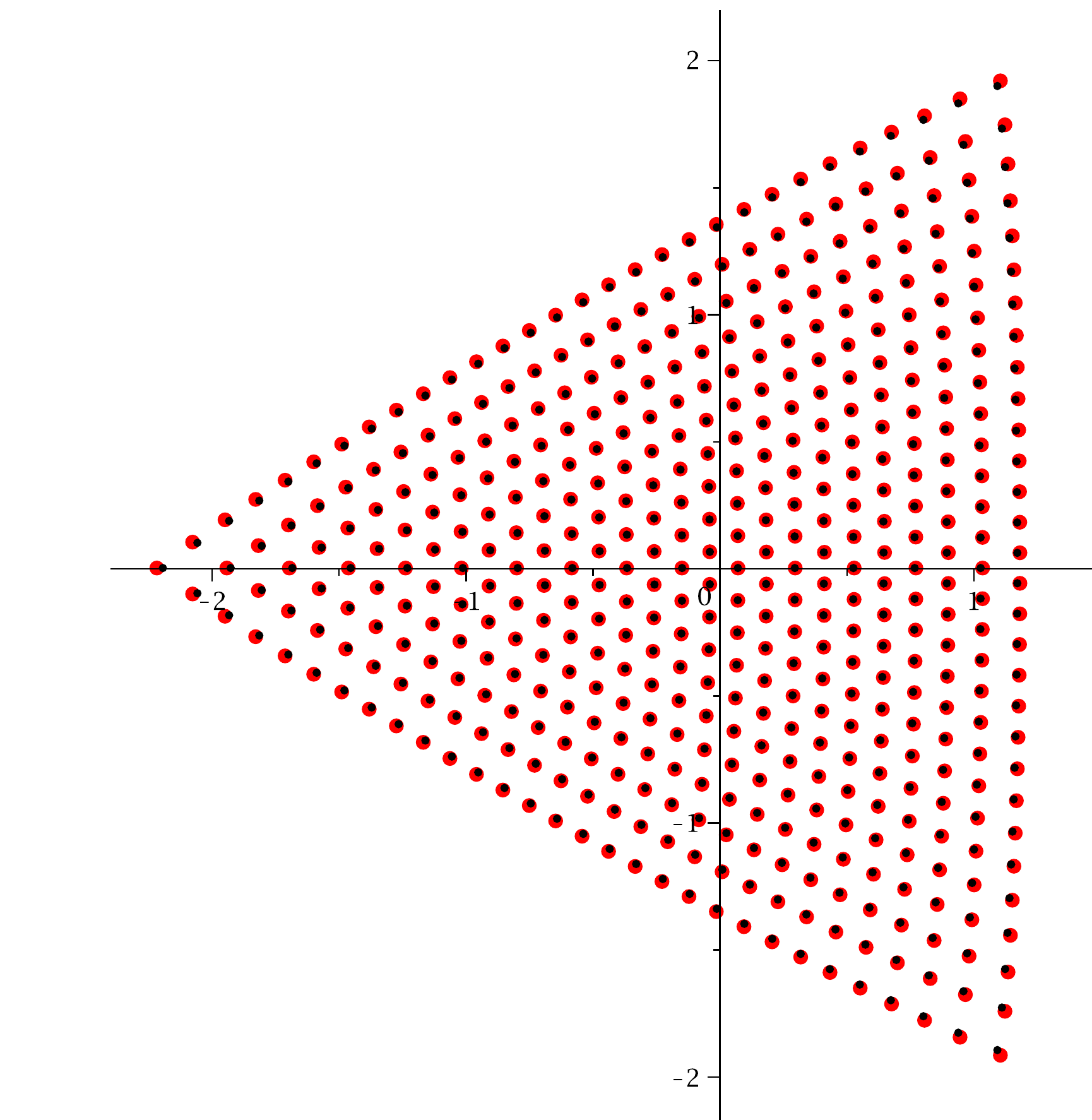}
    \caption{Scaled roots of the Vorob'ev-Yablonsky polynomials  $Y_{n}(n^{2/3}s)$ in red, and roots of the discriminant $D_n(n^{2/3}s)$ in black, for $n=30$.
    This particular scaling was conjectured by Shapiro and Tater in \cite{ST22}.}
    \label{fig:conjecture-scaling}
\end{figure}
This similarity is the object of a conjecture  that  B. Shapiro and M. Tater floated  several years ago, but only recently formalized in \cite{ST22} (Conjecture 2 ibidem). 

The present paper addresses this Shapiro-Tater (ST) conjecture. Further, along the way,  we  investigate the rather surprisingly deep connections of the boundary value problem \eqref{eq:shapiro-ODE}-\eqref{boundary}  with ``degenerate'' orthogonal polynomials: in fact we show that the polynomials arising from the solutions of \eqref{eq:shapiro-ODE}-\eqref{boundary}    satisfy an  excess of  orthogonality conditions. 
We also quantify the conjecture by estimating the discrepancy of the patterns of the two lattices coming from the VY polynomials on one side and the ST boundary problem on the other, explaining also why they have the regular pattern that appears numerically.

For the zeros of the VY polynomials, the analysis of these patterns was extensively already explained in \cite{BM14, BM15, BB15}; the rather surprising result is the explanation of the connection between the two seemingly distant problems that we have briefly outlined above.

\paragraph{Detailed results.} We now explain the conjecture in more detail.
Suppose we are looking for {\it quasi-polynomials} solutions of \eqref{eq:shapiro-ODE},
that is, solutions of the form
\begin{equation}
\label{quasi-polynomials}
    y(x)= p(x)e^{\theta(x)}, \quad \text{ where } \theta(x) = \frac{x^3}{3}+\frac{tx}{2}
\end{equation}
and $p(x)$ is a polynomial. 
We may sometimes use the notation $\theta(z; t)$ to emphasize the dependence of $\theta$ on the parameter t.  A substitution into \eqref{eq:shapiro-ODE} leads to
\[
\dv[2]{p}{x}  + (2x^2+t)\dv{p}{x} - 2(J-1)xp=\lambda\,p,\;\; 
    \text{with } \lambda = \Lambda  - \frac{t^2}{4}.
\]
Upon setting  $J=n+1$, one notices that the ODE
\begin{equation}
\label{eq:ST-polynomial-ODE}
    \dv[2]{p}{x}  + (2x^2+t)\dv{p}{x} - 2nx p = \lambda p, 
    \end{equation}
preserves the finite dimensional linear space of polynomials of degree at most $n$.
 The left hand side of \eqref{eq:ST-polynomial-ODE} is  a linear operator acting on the space of polynomials  of degree at most $n$  that can be written as  a $(n+1)\times(n+1)$ matrix $M_n(t)$ with respect to the usual basis
$(1,x, x^2, \dots , x^n)^t$, namely
\begin{equation}\label{STMn}
 M_n(t) :=
\begin{bmatrix}
0 & -2n &         &        &        &    &    \\ 
t & 0   & -2(n-1) &        &        &    &    \\
2 & 2t  & 0       &        &        &    &    \\
  & 6   & 3t      & 0      & \ddots &    &    \\
  &     & 12      & \ddots & \ddots & -4 &    \\
  &     &         & \ddots &        & 0  & -2 \\
  &     &         &        & n(n-1) & nt & 0  
\end{bmatrix}\,.
\end{equation}
We observe that the matrix  $M_n(t)$ is  a  $4$-diagonal matrix.
Eigenvalues of  this matrix correspond to the eigenvalues $\lambda$ in \eqref{eq:ST-polynomial-ODE},
and the eigenvectors correspond to quasi-polynomial eigenfunctions of 
the differential operator \eqref{eq:shapiro-ODE} via the identification \eqref{quasi-polynomials}.
Therefore we can compute the spectrum of \eqref{eq:shapiro-ODE} from   the characteristic polynomial
\begin{equation}
 \label{eq:spectral-polynomial}
    C_n(t,\lambda) := \det(M_n(t) - \lambda I).
\end{equation}
Now consider the discriminant of $C_n(t,\lambda)$ with respect to $\lambda$, i.e. 
\begin{equation}
    D_n(t) := \text{Disc}_\lambda(C_n)(t).\label{STDn}
\end{equation}
The roots of $D_n(t)=0$ are precisely the values of $t$ such that 
the matrix $M_n(t)$ has repeated eigenvalues (algebraic multiplicity greater than 1).

The Shapiro-Tater conjecture \cite{ST22} relates the zeros of $D_n(t)$ to
the poles of  rational solutions of \ref{eq:painleve2}.
The Painlev\'e II equation has rational solutions $u(t)$
if and only if $\alpha = n \in \Z$, and we denote them by $u_n(t)$.
Furthermore $u_n(t)$ is given explicitly in terms of the
Vorob'ev-Yablonsky polynomials $Y_n(t)$ in the form:
\begin{equation} 
\label{ratlsoln}
    u_n(t) = \dv{t} \log \frac{Y_{n-1}(t)}{Y_{n}(t)}
\end{equation}
The VY polynomials are constructed recursively \cite{V,Y} as follows:
 
\begin{equation}
\label{eq:VY}
\tag{VY}
{Y}_{n+1}(t){Y}_{n-1}(t) = t {Y}_n^2(t)-4\Big[{Y}_n''(t) {Y}_n(t)-\big({Y}_n'(t)\big)^2\Big],\hspace{0.5cm}n\geq 1,
  \ t\in\mathbb{C}\nonumber
\end{equation}
with ${Y}_0(t)=1, {Y}_1(t)=t$.
They can also be represented in terms of Schur function indexed by the ``staircase partition'' \cite{KajiwaraOhta} 
\begin{equation}
    Y_n (t)= \le(-\frac 4 3\ri) ^{n(n+1)/6} \le(\prod_{k=1}^n(2k-1)!!\ri) S_{(n,n-1,\dots, 1)} \le(\le(-\frac 3 4\ri)^\frac 13 t, 0,1,0,0,\dots\ri).
\end{equation}
The conjecture of Shapiro and Tater \cite{ST22} is the numerical observation that can be summarized in the conjecture {\it loosely} formulated below.
\begin{conjecture}
The roots of  the rescaled discriminant  $D_n(s) = \text{Disc}_\lambda(C_n)(n^\frac 23 s) $  in \eqref{STDn}  and the roots of  the rescaled Vorob'ev-Yablonsky  polynomials  $Y_n( n^\frac 2 3 s)$ 
form two coinciding lattices as $n \to \infty$.
\end{conjecture}
The evidence leading Shapiro and Tater to make their conjecture was purely numerical, as seen in Fig.  \ref{fig:matching_dots}.
The numerical picture seems so precise that one may at first be tempted to compare the polynomials $Y_n$ with $D_n$; however a simple inspection shows that their
coefficients are not close to each other, as seen from  Table \ref{table:polynomial-comparison}.
\begin{table} 
    \centering
    \begin{tabular}{ c|p{0.9\linewidth}  } 
     \hline
     $n$ & $ D_n(t)$ \\ 
     \hline
     1 & $t$  \\ 
     2 & $\ds t^3 + \frac {27}8$  \\[12pt]
     3 & $\ds t^6 + \frac{35}2 t^3 - \frac{243}4$\\ [12pt]
     4 & $\ds t^{10} + \frac{215}4t^7 +  \frac{89}8 t^4 + \frac{4084101}{512} t$ \\[12pt]
     5 & $\ds t^{15} + \frac{255}2 t^{12} +  \frac{76211}{32}  t^9 +  \frac{3730405}{64} t^6 -  \frac{8700637815}{4096} t^3 -  \frac{125005275}{32}$ \\ [12pt]
     \hline
    \end{tabular}

    \begin{tabular}{ c|p{0.9\linewidth}  } 
     \hline
     $n$ & $ Y_n(t)$ \\ 
     \hline
     1 & $t$  \\ 
     2 & $t^3 + 4$  \\
     3 & $t^6 + 20t^3 - 80$\\ 
     4 & $t^{10} + 60t^7 + 11200t$ \\
     5 & $t^{15} + 140t^{12} + 2800t^9 + 78400t^6 - 3136000t^3 - 6272000$ \\ 
     \hline
    \end{tabular}
    \caption{The first five monic Vorob'ev--Yablonskii polynomials $Y_n(t)$ and discriminant polynomials  $D_n(t)$ .}
    \label{table:polynomial-comparison}
\end{table}
\begin{remark}
When $\alpha= n$ the rational solutions $u_n(t)$ of \eqref{eq:painleve2} have two types of poles: those of residue $+1$ and those of residue $-1$. 
This follows from the fact that all the zeroes of $Y_n(t)$ are simple,
$Y_n(t), Y_{n+1}(t)$ do not share any roots \cite{taneda2000}and that rational solutions have the shape
\begin{equation} 
\label{rational}
    u_n(t) = \dv{t} \log \frac{Y_{n-1}(t)}{Y_{n}(t)} 
           =\frac{Y'_{n-1}(t)}{Y_{n-1}(t)} - \frac{Y'_n(t)}{Y_n(t)}.
\end{equation}
Therefore for fixed $n \in \N$ the poles of residue $+1$ correspond to zeroes of 
$Y_{n-1}(t)$ and poles of residue $-1$ correspond to zeroes of $Y_n(t)$
\end{remark}

The results of the paper can be grouped into two categories: the first consists in ``structural'' analysis of the boundary value problem \eqref{eq:shapiro-ODE}-\eqref{boundary}, in Section \ref{study}. The second category involves the asymptotic study for large $n$ and the description of the ST lattice, Sections \ref{section:exact-wkb-analysis},  \ref{quantcondJM}, \ref{quantcondST}. The first set of results can be summarized in the following points:
\begin{enumerate}
\item The boundary problem \eqref{eq:shapiro-ODE}-\eqref{boundary}  has solutions if and only if $J= n+1\in \N$; in this case $\Lambda$ takes at most $n+1$  values for each $t\in \C$. See Proposition \ref{propQES}. 
For fixed $J \in \N$ these values $(t,\Lambda)$ are called the {\it Exactly-Solvable} (ES) spectrum.
\item For given $J=n+1\in \N$ and $(t,\Lambda)$ in the ES spectrum, the corresponding solution of \eqref{eq:shapiro-ODE} is of the form $y(x) = p_n(x) {\rm e}^{2 \theta(x ;t)}$ with $p_n(x)$ a polynomial of degree $n$ and  $\theta(x;t)=\frac{x^3}{3}+\frac{tx}{2}$.  This polynomial is {\it degenerate} orthogonal in the sense that
\be   
\label{degortho}
\le(\varkappa   \int_{\infty_1}^{\infty_3} + \wt \varkappa  \int_{\infty_3}^{\infty_5} \ri)p_n(x) x^j {\rm e}^{2 \theta(x ;t)} \d x = 0, \ \ j=0,\dots, n,
\ee
where  $\varkappa$ and $\tilde{\varkappa}$ are some constants. Here   $\infty_k$ denotes a contour that tends to infinity with asymptotic direction of argument $\frac {i\pi  k}3$ with $k=1,3,5$. 
Observe that the orthogonality involves all powers including the $n$-th power, and this means that the $(n+1)$-st Hankel determinant of the moments of the above pairing is zero.  This result is contained in Section \ref{sec32} and in particular Theorem \ref{thm:quaspol_is_OP}.
These results extend the several properties established for these polynomials in \cite{Eremenko, Eremenko2, MT2013}.

\item In fact more is true: given $t\in \C$ suppose that  $\varkappa ,\wt \varkappa$  are such that \eqref{degortho} admits a nontrivial polynomial $p_n$ for solution. Then $y(x) = p_n(x){\rm e}^{\frac {x^3}3+tx}$ is a solution of the boundary problem \eqref{eq:shapiro-ODE}--\eqref{boundary}. This is proved in Theorem \ref{thm:OP_is_quaspolys}. 
\item The two above results, together, yield a complete characterization of the ES spectrum of the quartic anharmonic oscillator  with boundary conditions \eqref{boundary}  in terms of degenerate orthogonality. See Corollary \ref{thm:quasi-polys_iff_orthogonal}  
\item We then investigate the consequences of the requirement that an eigenvalue is repeated; we  prove that this is equivalent to the additional condition that {\it both} integrals  in \eqref{degortho} for the degenerate orthogonal polynomials vanish independently,  that is equivalent to:
\be
\int_{\infty_{2j+1}}^{\infty_{2j+3}} p_n(x)^2{ \rm e}^{2\theta(x;t)} \d x = 0 , \quad j=0,1.
\ee
Note that their linear combination in \eqref{degortho} vanishes due to the degeneracy of the orthogonality pairing, so that this condition implies only one additional constraint on the value of $t$. This is proven in Theorem \ref{thm:repeated_eigenval_condition}.
Another amusing consequence is that in these cases the antiderivative of $p_n(x)^2{ \rm e}^{\frac {2x^3}3+tx}$ can be shown to be also a quasipolynomial. See Corollary \ref{corquaspol}.
\end{enumerate}
{\color{black} The second set of results involves the asymptotic analysis for large $n$ and contains the actual proof of the  Shapiro-Tate conjecture. This begins in Section \ref{section:exact-wkb-analysis}  where we introduce the following rescaled variables for the Shapiro-Tater and Jimbo-Miwa  cases

\begin{align*} 
\mbox{Shapiro-Tater},\quad    z &= \hbar^{\frac{1}{3}}x,\qquad
    {s} = \hbar^{\frac{2}{3}}t,\qquad
    E = \hbar^{\frac{4}{3}}\Lambda,\quad \hbar^{-1}=n+1\\
    \mbox{Jimbo-Miwa},\quad    z &= \hbar^{\frac{1}{3}}x,\qquad
    {s} = \hbar^{\frac{2}{3}}a,\qquad
    \hat{b} = \hbar^{\frac{4}{3}}b, \;\;E=\frac{7s^2}{36}+10\hat{b}, \quad \hbar^{-1}=n+\frac{1}{2}\,.
\end{align*}
The reason for this scaling is that it yelds the same  WKB-type equation   with a small parameter $\hbar$ and a $n$-independent quartic potential
\begin{align*}
    &\hbar^2\dv[2]{y}{z} - Q(z;{s},E)y =0,\\
      & Q(z;{s},E) = z^4 + {s} z^2 +2z +E\,.
\end{align*}
    This puts both the ST and JM anharmonic oscillators on the same footing and allows us to use the exact WKB method to compute asymptotic expressions for the Stokes phenomenon of both systems simultaneously. }
\begin{enumerate}
\item In Section \ref{section:exact-wkb-analysis} we recall the ``exact WKB analysis'' following \cite{kawai_takei_algebraic_analysis_sing_perturbation} and set up notation used in later sections.
This allows us to express the Stokes' parameters for the quartic anharmonic oscillator in terms of the so--called exact Vor\"os symbols, namely, integrals of formal series in the small parameter $\hbar$. This section does not contain new results and is mostly a preparation for the two subsequent ones.

\item In Section \ref{quantcondJM} we use exact WKB analysis to re-derive the asymptotic conditions  for the zeros of the Vorob'ev Yablonskii polynomials already appeared in \cite{BM14,BM15, BB15} in a different way, based on the matching of the Stokes' phenomenon of the associated linear problem for the second Painlev\'e\ equation \eqref{eq:painleve2}, transformed into a problem for the anharmonic oscillator using an idea originally used by Masoero in \cite{Masoero_Non, Masoero_2010}.
This analysis produces the asymptotic conditions in Theorem \ref{thm:quantisation-rational-PII} that implicitly describe the location of the zeros  of the \ref{eq:VY} polynomials in terms of certain contour integrals. 

\item In Section \ref{quantcondST} we derive a similar asymptotic description of the points of the ES spectrum that correspond to multiple eigenvalues by combining our previous results in Section \ref{study} and the exact WKB analysis. The key result in this section is Theorem \ref{integralA1} which leads to the quantization conditions \eqref{QuantST}.

\item Finally in Section \ref{comparing-quantization} we proceed with the comparison  of the two sets of quantization conditions that describe the two lattices of points.
These are the equations \eqref{QuantST} and \eqref{QuantJM}, { that we report hereafter (to leading order):
\color{black}
\begin{align*}
&2(n+1)\int_{\tau_1}^{\tau_0} \sqrt{Q(z_+;s,E)} \d z = \ln \le( \frac {-1}{1+\boldsymbol \tau(s,E)}
\ri) - 2i\pi (m_1+1)
\nn
\\
&2(n+1)\int_{\tau_2}^{\tau_0} \sqrt{Q(z_+;s,E)} \d z = \ln \le( -1 -\frac 1{\boldsymbol \tau(s,E)}
\ri) - 2i\pi (m_2+1)
\nn
\\
&2(n+1)\int_{\tau_3}^{\tau_0} \sqrt{Q(z_+;s,E)} \d z = \ln \le( {\boldsymbol \tau(s,E)}
\ri)- 2i\pi (m_3+1)
\nn\\
&\boldsymbol \tau(s,E) = \frac
{\ds \int_{\tau_1}^{\tau_0}\frac {\d z}{\sqrt{Q(z_+;s,E)} }}{\ds \int_{\tau_2}^{\tau_0}\frac {\d z}{\sqrt{Q(z_+;s,E)} }},\quad \Im(\boldsymbol \tau(s,E))>0.\nn
\end{align*}
where $\tau_j$, $j=0,1,2,3$ are the zeros of $Q(z_+;s,E)$. The three integers satisfy  $m_1+m_2+m_3 = n-1$ due to the fact that the sum of the three integrals on the left is $-2i\pi(n+1)$ while the sum of the three logarithms is $2i\pi$ (principal determination) due to the definition of $\boldsymbol \tau(s,E) $. 
On the other hand, the quantization conditions for the Vorob'ev-Yablonskii zeroes, to the same order of approximation, read
\begin{align*}
&\le(2n+1\ri)\int_{\tau_j}^{\tau_0} \sqrt{Q(z_+;s,E)} \d z =-{i\pi} - 2i\pi k_j
\nn\\
&k_1+k_2+k_3=n-1.
\end{align*}
}

In Proposition \ref{locallattice}
we show that both lattices form a regular pattern where the local lattice generators are slowly modulated vectors. 
In the scaled $s = t\hbar^\frac 23$ plane (where $\hbar^{-1} = n+\frac 1 2$ for the VY polynomials and $\hbar^{-1} = n+1$  for the ST case) the two lattices fill a region of triangular shape that was analyzed in \cite{BM14,BM15, BB15, ST22} with corners at $s_j = {\rm e}^{\frac {i\pi}3 j} \frac 3{2^{1/3}}$.
Within this region the number of points is $\mathcal O(n^2)$ so that the relative spacing is $\mathcal O(n^{-1})$. 

Our results indicate that near the origin the two lattices have a discrepancy (in the re-scaled $s$--plane) of order $\mathcal O(n^{-2})$, see Theorem \ref{thmorigin}; as we move away from the origin, this discrepancy progressively accumulates so that, in a fixed small disk around $s_0\neq 0$, their discrepancy would be only $\mathcal O(n^{-1})$.
This would seem at first inconclusive because the lattices have a separation already of order $\mathcal O(n^{-1})$.
However, as shown in Prop. \ref{locallattice}, the lattices are very regular and have the same local geometry. This contributes to the {\it impression} 
of them being almost identical. This drift of the lattices is rather visible in Figure \ref{fig:matching_dots}.
In particular, this settles the Conjecture of \cite{ST22}.

\item In Appendix \ref{Numerics} we report on the numerical verifications of the quantization conditions we have derived; moreover we observe the curious and mostly accidental fact that if we scale the ST lattice by $n^{-\frac 2 3}$ instead of the more natural $(n+1)^{-\frac 23}$ the numerical coincidence become even more surprising. However, as we document numerically, the order of discrepancies of the two lattices of the ST problem and VY zeros  is the same with either scaling.
\end{enumerate}

\section{Link to Painlev\'e}
In this section we explain the link between the anharmonic oscillator  
\eqref{eq:shapiro-ODE} 
and the second Painlev\'e transcendent, following the ideas first pioneered by  Its and Novkshenov  \cite{its-novok} and  later by Masoero \cite{Masoero_Non,Masoero_2010}.
The \ref{eq:painleve2} equation can be expressed as  the compatibility condition of two linear systems called Lax pairs.
There are several such Lax pairs but for our purposes we will only need the Jimbo-Miwa Lax pair \cite{Jimbo_Miwa_Ueno_2,FIKN}, namely:
\begin{equation} 
\label{eq:JM_Lax_pair}
    \begin{dcases}
    \pdv{\Phi}{x} =A(x,t) \Phi,\\ 
    \pdv{\Phi}{t} = B(x,t) \Phi,
    \end{dcases}
\end{equation}
with the coefficient functions
\begin{gather}
    A(x,t; \alpha) = \left(x^2 + w + \frac{t}{2}\right) \sigma_3 + (x-u)v \sigma_+  - \frac{2}{v}\left( x w + uw + -\alpha + \frac{1}{2}\right) \sigma_{-}  \\ 
    B(x,t) = \frac{x}{2}  \sigma_3 + \frac{v}{2} \sigma_+ -\frac{w}{v} \sigma_-
\end{gather}
where $\alpha \in \C$ is a parameter, $u=u(t), v=v(t), w=w(t)$ are meromorphic
functions of $t$ and 
\begin{equation}
    \sigma_3   = \begin{bmatrix} 1 & 0 \\ 0 & -1 \end{bmatrix},\quad 
    \sigma_{+} = \begin{bmatrix} 0 & 1 \\ 0 & 0 \end{bmatrix}, \quad 
    \sigma_{-} = \begin{bmatrix} 0 & 0 \\ 1 & 0 \end{bmatrix}.
\end{equation}
The compatibility condition $\pa_x \pa_t \Phi = \pa_t \pa_x\Phi$ of the system 
\eqref{eq:JM_Lax_pair} is equivalent to 
\begin{equation}
    \pdv{A}{t} - \pdv{B}{x} +[A,B] = 0,
\end{equation}
which in turn yields the following system of ODEs for $u,v,w$:
\begin{equation} \label{eq:OJM_compatibility_system}
    \dv{u}{t} = u^2 + w + \frac{t}{2}, \quad \dv{v}{t} = -u v, \quad \dv{w}{t} = -2 u w + \alpha
    -\frac{1}{2}
    .
\end{equation}
Eliminating $w$ from the system $u(t)$ satisfies
\eqref{eq:painleve2} with parameter $\alpha$; 
similarly eliminating $u(t)$ 
gives that $w(t)$ satisfies PXXXIV \cite{ince_differential_eq,FIKN}.


Now we find local solutions of the system \eqref{eq:OJM_compatibility_system} near a pole
of the Painlev\'e transcent $u(t)$. This will be useful in the upcoming computations.
\begin{proposition}
Let $t=a$ be a pole of residue $+1$ of the Painlev\'e II function $u(t)$ with parameter $\alpha$. 
Then near $t=a$ we have the following Laurent series expansion of \eqref{eq:OJM_compatibility_system}:
\begin{align} \label{eq:P2-laurent}
    u(t) &= \frac{1}{t-a} - \frac{a}{6}(t-a)-\frac{\alpha + 1}{4}(t-a)^2 + b (t-a)^3 + \bigO{((t-a)^4)}, \nonumber\\
    w(t) &= -\frac{2}{(t-a)^2} - \frac{a}{3}-\frac{1}{2} (t-a) 
    + \left( -\frac{a^2}{36}+b \right)(t-a)^2 + \bigO{((t-a)^3)}, \nonumber \\
    v(t) &= c \left( \frac{1}{t-a}+\frac{a}{12}(t-a) + \frac{\alpha+1}{12}(t-a)^2 + \bigO{((t-a)^3)}\right),
\end{align}
where $b$ is arbitrary and $c\neq 0$ is a constant of integration.

Similarly, if $t=a$ is a pole of residue $-1$, then near $t=a$ we have the following Laurent expansions:
\begin{align}
\label{expu}
    u(t) &= \frac{-1}{t-a} + \frac{a}{6}(t-a)-\frac{\alpha - 1}{4}(t-a)^2 + b (t-a)^3 + \bigO{((t-a)^3)}, \\
    w(t) &= \left( \frac{1}{2}-\alpha \right)(t-a)+ 
    \left( 5b - \frac{a^2}{36}\right)(t-a)^2 +\frac{a(2\alpha-1)}{6}(t-a)^3 + \bigO((t-a)^4),\\
    v(t) &= c \left( (t-a) -\frac{a}{12} (t-a)^3 + \bigO((t-a)^4),\right)
\end{align}
where, again, $b$ is arbitrary and $c\neq 0$ is a constant of integration.
\end{proposition}
\noindent{\bf Proof.}
It is well known that all poles of PII are simple and with residue $\pm 1$  (see e.g. \cite{its-novok}). For every such pole $t=a$,
and arbitrary $b\in\C$ there exists a solution of PII with the prescribed Laurent expansions, where all
the coefficients of order $4$ or higher are polynomials in $a,b$ (see \cite{gromak_painleve_diff_eq_in_C}).
Furthermore, the Laurent expansion for $w(t),v(t)$ can be obtained by substituting $u(t)$ in the 
system \eqref{eq:OJM_compatibility_system} and comparing the coefficients. 
\QED

\subsection{Gauge transformation}
With a particular gauge transformation we can convert the  matrix ODE 
$\pdv{x}\Phi= A(x,t;\alpha) \Phi$ to a scalar ODE; the procedure results in adding an {\it apparent singularity} in the equation at the position $x = u(t)$. Thus, when the independent variable $t$ tends to one of the poles of the solution $u(t)$ to the Painlev\'e\ equation, the 
 singularity ``escapes'' to infinity and in the limit we obtain a {\it polynomial} ODE.  This  gives us an eigenvalue problem similar to the one of 
Shapiro and Tater \eqref{eq:shapiro-ODE}.

To implement this idea in detail, first we outline the general gauge transformation as  used by 
\cite{Masoero_2010}. 
Take a traceless $2\times 2$ matrix ODE system $\Phi_x = M(x) \Phi$, where
\begin{equation}
M(x) = 
\begin{bmatrix}
m_{11}(x) & m_{12}(x) \\
m_{21}(x) & -m_{11}(x) \\ 
\end{bmatrix}.
\end{equation}
We wish to turn this into a scalar ODE of the form $y''-V(x)y=0$, which we can do by 
using the gauge transformation 
\begin{equation}
\label{Gauge}
W(x) =  G(x)\Phi(x),\qquad
G(x) :=  
\begin{bmatrix}
{m_{12}(x)}^{-\frac{1}{2}} & & 0 \\
 & & \\
-\frac{m_{12}'(x)}{2 m_{12}(x)^{\frac{3}{2}}} + \frac{m_{11}(x)}{m_{12}(x)^{\frac{1}{2}}} & &    m_{12}(x)^{\frac{1}{2}}
\end{bmatrix}
\end{equation} 
We find that $W(x)$ solves the matrix ODE
$\dv{x}W = \widehat{M}(x) W(x)$
where 
\begin{equation}
\widehat{M}(x) = G_x G^{-1} + G M G^{-1} = 
\begin{bmatrix}
0 & 1 \\ 
V(x) & 0 \\
\end{bmatrix},
\end{equation}
where the function $V(x)$ is:
\begin{equation}
    V(x) = m_{12} m_{21} + m_{11}^2 + m_{11}' - m_{11} \frac{m_{12}'}{m_{12}} - \frac{m_{12}''}{2m_{12}} + \frac{3}{4}\left(\frac{m_{12}'}{m_{12}}\right)^2.
\end{equation}
Thus the matrix ODE system $W_x = \widehat{M}(x)W(x)$
is then  equivalent to the scalar ODE $$y''(x) =V(x) y(x)$$ and $W$ is the Wronskian matrix of a pair of independent solutions (whence the choice of symbol).\\[4pt]

Let us now apply this transformation to the matrix ODE from the Jimbo-Miwa Lax pair 
\eqref{eq:JM_Lax_pair}, i.e. we set $M(x) = A(x,t)$ and perform the aforementioned gauge transformation.
Since the entries of $A(x,t)$ depend on $t$, our potential $V$ will be a function of both $x$ and $t$,
namely: 
\begin{gather}
\begin{aligned}
    V(x,t) =&\frac{1}{\left( x-u(t)\right)^2} \Bigg[
 2\,w \left( t \right)  u(t) ^{4} + \left( -4\,xw \left( t \right) -2\,\alpha+1 \right)u(t)^{3} \\
+& \left(  w( t )^2+ \left( 2\,{x}^{2}+\,t \right) w(t) +x^{4}+tx^{2}+
   ( 6\alpha-1)x+\frac{{t}^{2}}{4} \right) u(t)^{2}\\ 
+& \left( -2\,x w(t)^{2} + ( -2\,tx+1 ) w (t)    
  -2\,x^{5} -2\,tx^{3}-\frac{1}{2}\,t^{2}x-6\,\alpha\,x^{2}+\frac{t}{2}
 \right) u(t) \\
+& \, x^{2} w (t)^{2}+\left( \,t{x}^{2}-x \right) w(t) 
+x^6+tx^4+\frac{t^2x^2}{4}+2\,\alpha\,x^{3}-\frac{tx}{2}+\frac{3}{4} 
\Bigg]\,.
\label{eq:VMaso}
\end{aligned}
\end{gather}

We are interested in the expression of this potential at  a pole $t=a$ of PII equation. 

\begin{proposition}
Let $t=a$ be a pole  with residue $\pm 1$  of the \ref{eq:painleve2} solution $u(t)$ with parameter $\alpha$.
Then the  $ \lim_{t\to a}V(x,t)$  exists and we have for a pole of residue $+1$
\begin{equation}
 \VJM(x;a,b,\alpha):=\lim_{t\to a}V(x,t)=x^4+ax^2+ (2\alpha-1)x + \left( \frac{7a^2}{36}- 10b\right)
 \end{equation}
and for a pole of residue $-1$
 \begin{equation}
 \label{eq:JM_potential}
\lim_{t\to a}V(x,t)= x^4+ax^2+ (2\alpha+1)x + \left( \frac{7a^2}{36}+ 10b\right)= \VJM(x;a,-b,\alpha+1) 
 \end{equation}
\end{proposition}
\noindent{\bf Proof.}
The result is obtained from a straightforward computation where we substitute the Laurent series expansions \eqref{eq:P2-laurent} into \eqref{eq:VMaso} and take the limit as $t\to a$.
The details are left as exercise (it is helpful to use a computer algebra program for this purpose).
\QED 
The importance of this transformation for us is the following: the gauge transformation \eqref{Gauge}
introduces only a singularity at the zeros of $m_{1,2}(x)$, which in our case is only at $x= u(t)$, namely, the {\it value} of the Painlev\'e\ transcendent solution $u(t)$. The singularity is clearly only a square-root type singularity with local monodromy $-1$. 
Other than this, the Stokes phenomenon of the ODE $y''(x) = V(x;t)y(x)$ is unchanged and  independent of $t$ by construction. { This is evident from the fact that the transformation of the system effected by \eqref{Gauge} is a {\it left} multiplication by a simple algebraic matrix $G(x)$ with at most square--root singularities  around which the whole matrix has a scalar monodromy of multiplication by $-1$.  However the Stokes' phenomenon involves {\it right} multiplications of the solution by constant matrices, which are thus the same as for the original system.}

This important observation has the consequence that as $t$ approaches one of the poles of  the given solution $u(t)$ of the Painlev\'e\ equation, the additional singularity moves off to infinity. 
Thus we have the following simple but essential statement that we formalize in the proposition below.
\begin{proposition}
\label{propMasoero}
Let $u(t)$ be a solution of the Painlev\'e\ II equation corresponding to a particular set of Stokes data for the ODE in \eqref{eq:JM_Lax_pair}. Let $t=a$ be a pole of $u(t)$ with residue $-1$  and $b$ the coefficient as in \eqref{expu}.  
Then the Stokes phenomenon of the ODE 
\bea
\label{217}
y''(x) = \le(x^4 + ax^2 +(2\alpha+1) x + \Lambda\ri) y(x),\ \ \ \Lambda:= \frac {7a^2}{36} + 10 b
\eea
is the same as the original Stokes phenomenon of the ODE \eqref{eq:JM_Lax_pair}. 
\end{proposition}
A similar statement holds for the poles with residue $+1$ but we make the choice of considering only   those with negative residue because 
 as formula \eqref{rational}  shows the poles with positive residue are the zero of $Y_{n-1}(t)$ and the poles  with negative residue are the zeros of $Y_n(t)$.  {\color{black} The  complete  proof of the above proposition is equivalent to the one presented in  \cite{Masoero_2010} Appendix B.}
%
\section{A study of quasi-polynomials}

\label{study}
In this section we find a characterization of the (quasi)-polynomials corresponding to a repeated eigenvalue for the operator \eqref{eq:ST-polynomial-ODE}, namely,  the Shapiro-Tater eigenvalue problem:
\begin{align}
       \label{eq:ST_eval_problem}
   & \dv[2]{y}{x}- (x^4 +tx^2 +2Jx +\Lambda)y =0\\
    \label{boundary1}
    &y(s {\rm e}^{{k}\pi i/3}) \to 0, \ \ \ s \to +\infty,\  {k=1,3,5}.
    \end{align}

We will call the set  $(t, J, \lambda)$ for which there is a solution of the problem \eqref{eq:ST_eval_problem}-\eqref{boundary1},  the {\it Exactly Solvable} (ES) spectrum. 
 Our  setting is different from  \cite{ST22}, where the authors considered a modified eigenvalue problem \eqref{eq:ST_eval_problem} with  only {\it two} boundary conditions.  In this case 
only a  finite portion of the spectrum can be  computed explicitly, and for this reason it is a  {\it Quasi Exactly Solvable} (QES) spectrum, the naming comes from \cite{BB98}.

We will see below that with three  boundary conditions at infinity as in \eqref{boundary1}, the whole  spectrum can be characterised by the vanishing of a finite determinant, and it is therefore {\it Exactly Solvable}.
The first result in Proposition \ref{propQES} shows  that the boundary conditions in \eqref{boundary1}  is compatible only with $J$ being a positive integer.
We  then compute the Stokes phenomenon of the quasi-polynomial solutions explicitly, as shown in \eqref{thm:stokes_quasi_polys}.
Finally we relate the problem to (degenerate) orthogonality for a suitable class of non-hermitean orthogonal polynomials, see Theorem \ref{thm:quaspol_is_OP} and Theorem \ref{thm:OP_is_quaspolys} which are the main results of this section.

\begin{lemma}
\label{lemmasimpleeval}
The equation \eqref{eq:ST_eval_problem} admits quasi--polynomial solutions of the form
\begin{equation}
    y(x) = p(x)e^{\theta(x;t)} \quad \mbox{where  $\;\;\theta(x;t) =\frac{x^3}{3}+\frac{tx}{2}$},
\end{equation}
with $p(x)$ a polynomial of degree $n$ if and only if $J=n+1$ and $\lambda=\Lambda-\frac{t^2}{4}$ is an eigenvalue of the operator 
\begin{equation}
\label{Lhat}
\mathcal{L}_{J}:=
    \dv[2]{}{x}+ 2\left(x^2+ \frac{t}{2} \right)  \dv{}{x} -  2(J-1) x  
    \end{equation}
 acting on the space of  polynomials of degree up to n.
\end{lemma}
\noindent {\bf Proof.}
Substituting $y(x)=p(x) {\rm e}^{\theta(x)} $ in the ODE    \eqref{eq:ST_eval_problem}  gives an equivalent  differential equation for  the function $p(x)$:
\begin{equation} \label{eq:polynomial_ODE}
  \mathcal{L}_{J}(p(x)) = \lambda p(x) \quad \text{ where } \quad\lambda = \Lambda - \frac{t^2}{4}\,,
\end{equation}
and $  \mathcal{L}_{J}$ as in \eqref{Lhat}.
One can readily see that if $J=n+1$ then the operator $\mathcal{L}_{n+1}$ in \eqref{eq:polynomial_ODE} preserves the space of  polynomials of  degree at most $n$ and then $\Lambda$ is, by definition, an eigenvalue of \eqref{eq:ST_eval_problem}.

Viceversa, if $p(x)$ is a polynomial of degree $n$ and solves \eqref{eq:polynomial_ODE} then one finds by inspection that the l.h.s   is a polynomial of degree $n+1$ whose  leading coefficient is $ 2 (n-J+1)$ while the r.h.s. is a polynomial of degree $n$. Thus $J=n+1$. Then $  \mathcal{L}_{n+1}$   preserves the space of polynomials of degree $n$ and $\Lambda$ (and the corresponding $\lambda$ as per \eqref{eq:polynomial_ODE}) is an eigenvalue of the corresponding finite dimensional operator. \QED

\subsection{Stokes phenomenon}
\begin{figure}
    \centering
    \includegraphics[width=.5\textwidth]{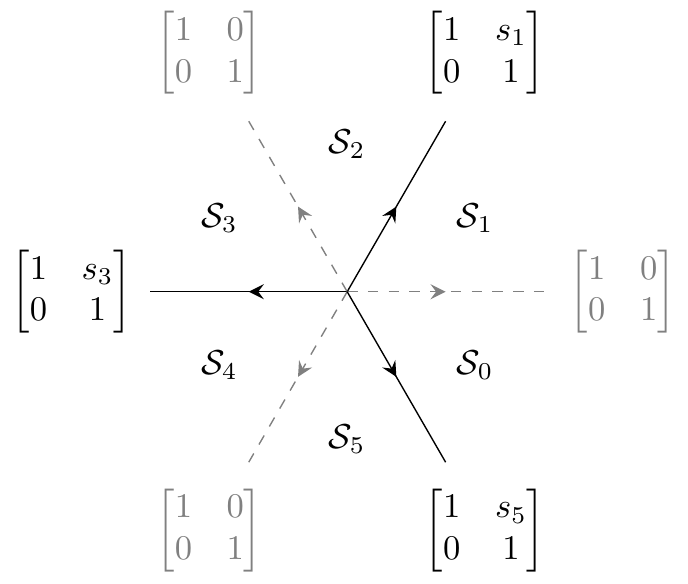}\\
    \caption{Stokes matrices and Stokes sectors for the Shapiro-Tater eigenvalue problem \eqref{eq:ST_eval_problem} with quasi-polynomial solutions.
     The Stokes matrices $\S_0, \S_2, \S_4$ are all the identity.}  
    \label{fig:ST_stokes_data}
\end{figure}

\begin{proposition}
\label{propQES}
    The eigenvalue problem \eqref{eq:ST_eval_problem}  with the boundary conditions \eqref{boundary1}  require that $J=n+1, \ n=0,1,\dots$ and that $\lambda$ in \eqref{eq:polynomial_ODE} is an eigenvalue of the operator $  \mathcal{L}_J$.
     In particular the solutions are quasipolynomials as in Lemma \ref{lemmasimpleeval}.
\end{proposition}
\noindent{\bf Proof.}
The equation \eqref{eq:ST_eval_problem} can be written as a first order system for the Wronkstian matrix
 $\mathcal{W}(x)$:
\bea
\label{eq_W}
 \dv{\mathcal{W}}{x}  = \begin{bmatrix}
0&1\\
x^4+tx^2 + 2Jx +\Lambda & 0
\end{bmatrix}\mathcal{W}(x),\ \  \Lambda = \lambda +\frac {t^2}4.
\eea
This equation has a singularity at infinity with Poincar\'e\ rank  $3$.
Following the ordinary asymptotic analysis \cite{Wasov} we see that we have formal-series solutions  of the form 
\begin{align}
    \mathcal{W}_{\text{form}}(x)
    \sim & 
    x^{-\sigma_3}( \mathbb{I}-  \sigma_++\sigma_- ) \left( \mathbb{I} + \bigO(x^{-1}) \right) x^{J\sigma_3} e^{\theta(x;a) \sigma_3},\quad \theta(x;t)=\frac{x^3}{3}+\frac{tx}{2} \\ 
    = & 
    \begin{bmatrix}
    x^{J-1} +\mathcal O(x^{J-2})& -x^{-J-1}+\mathcal O(x^{-J-2}) \\
    x^{J+1}+\mathcal O(x^J) & x^{-J+1}+\mathcal O(x^{-J})
    \end{bmatrix}  {\rm e}^{\theta(x;t)\sigma_3}, \quad x \to \infty.
    \label{eq:formal}
\end{align}
We define the Stokes sectors as shown in Figure \ref{fig:ST_stokes_data}, i.e. $\St_k$ is the sector of opening $\pi/3$ centered
around the rays of argument $\pi/6+ (k-1) \pi/3$: 
\begin{equation} \label{eq:stokes_sector}
    \St_k = \left\{ z \in \C : \abs{\arg(z)-  \frac{\pi}{6}{ -(k-1)}\frac{\pi}{3}} < \frac{\pi}{6}\right\}, \quad k = 0,1,2,3,4,5.
\end{equation}
Each open sector $\Omega_k =\left\{ z \in \C : \abs{\arg(z)- \frac{\pi}{6}{-(k-1)}\frac{\pi}{3}} < \frac{\pi}{6}+\delta\right\}$  with $\delta>0$, strictly contains a Stokes' sector $S_k$.
Then there exists a unique solution $W^{(k)}(x)$  of \eqref{eq_W} in $\Omega_k$ asymptotic to $   W_{\text{form}}(x)$, namely 
\[
\mathcal{W}^{(k)}(x)\simeq  \mathcal{W}_{\text{form}}(x),\quad x\in\Omega_k,\;\;k=0,\dots,5.
\]
Since $\mathcal{W}^{(k)}(x)$ and $\mathcal{W}^{(k+1)}(x)$ have the same asymptotic expansion in $\Omega_{k}\cap \Omega_{k+1}$, there is a constant matrix $\S_k$ (called {\it Stokes' { matrix}}), such that
\[
\mathcal{W}^{(k+1)}(x)=\mathcal{W}^{(k)}(x)\S_k, \quad  x\in \Omega_{k}\cap \Omega_{k+1}, \quad k=0,1,\dots,5,
\]
where  $\S_{2k}=\begin{bmatrix}
1 & 0\\
s_{2k} &1
\end{bmatrix}$ and  $\S_{2k+1}=\begin{bmatrix}
1 & s_{2k+1}\\
0 &1
\end{bmatrix}$ for $k=0,1,2$ { (the parameters $s_k$ are called {\it Stokes' multipliers})}.
By standard methods one can see that the Stokes phenomenon consists of the following relation
\[
\mathcal{W}^{(6)}(x)=\mathcal{W}^{(0)}(x)\S_0\S_1\dots\S_5,\quad \mathcal{W}^{(6)}(x)=\mathcal{W}^{(0)}(x)e^{2\pi i J\sigma_3}
\]
which gives the relation
\begin{equation}
\begin{bmatrix}
1 & 0\\
s_0 &1
\end{bmatrix}\begin{bmatrix}
1 & s_1\\
0 &1
\end{bmatrix}
\begin{bmatrix}
1 & 0\\
s_2 &1
\end{bmatrix}
\begin{bmatrix}
1 & s_3\\
0 &1
\end{bmatrix}
\begin{bmatrix}
1 & 0\\
s_4 &1
\end{bmatrix}
\begin{bmatrix}
1 & s_5\\
0 &1
\end{bmatrix}
\begin{bmatrix}
{\rm e}^{-2i\pi J} & 0\\
0 &{\rm e}^{2i\pi J}
\end{bmatrix}= \mathbb{I}
\label{nomono}
\end{equation}

The first six matrices are  the Stokes matrices associated with the directions $\arg(x) = k\frac {\pi}3$, $k=0,1,..,5$ and the last matrix is the formal monodromy matrix.
The boundary conditions \eqref{boundary1} imply that $s_0=s_2=s_4=0$ because it means that the recessive solution along the direction $\arg(x) = \frac {\pi}3$ is also recessive along the directions $\arg(x) = k\frac {\pi}3$, $k=3,5$. 
But then the matrix equation \eqref{nomono} implies that
\be
\label{Stokes}
s_1+s_3+s_5=0
\ee
and  ${\rm e}^{2i\pi J}=1$ and hence $J$ must be an integer. 

To show that $J=n+1$ is a {\it positive} integer and that $\lambda$ is an eigenvalue of $\mathcal{L_J}$ we proceed as follows. 
{
 Given that now the Stokes matrices are all upper triangular, the first column of the solution is an entire function which is asymptotic to the first column of  the formal-series solution \eqref{eq:formal} along all directions. }
But then  the asymptotic \eqref{eq:formal} implies that the $(1,1)$ entry is of the form $p(x) {\rm e}^{\theta(x;t)}$ with $p(x)$ entire and bounded at infinity by $x^{J-1}$. Then Liouville's theorem implies that if $J\geq 1$ then $p(x)$ is a polynomial, and for $J=0,-1,-2,\dots$ $p(x)$ should vanish at infinity and hence it should be identically zero, leading to a contradiction.

We have now established that the only solutions of the eigenvalue problem \eqref{eq:ST_eval_problem} are quasipolynomials and therefore the hypothesis of Lemma \ref{lemmasimpleeval} prevail, thus showing that  $\lambda$ is the claimed eigenvalue.
\QED

\begin{figure}
    \centering
    \includegraphics[width=.4\textwidth]{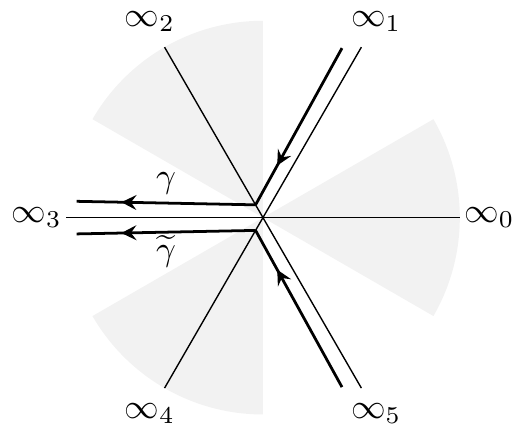}\\
    \caption{Directions at infinity $\infty_k$ of argument $k \frac{i \pi}{3}$ and { the oriented contours $\gamma$ and $\tilde{\gamma}$ from  $\infty_1 $ to $\infty_3$ and $\infty_3 $ to  $ \infty_5$ respectively.}
    The shaded regions denote the sectors of dominance of ${\rm e}^{\theta(x;t)}$ where $\theta(x,t) = \frac{x^3}{3}+\frac{tx}{2}$.
    The unshaded regions denote the sectors of recessiveness.}
    \label{fig:dominant-recessive}
\end{figure}

Next we consider the properties of the operator \eqref{eq:ST_eval_problem} (or equivalently of \eqref{eq:polynomial_ODE}) under the assumption that $J=n+1\in \mathbb N$ and $\lambda$ (respectively $\Lambda$) is an eigenvalue. We start from further  analysis of the corresponding Stokes phenomenon. 
\begin{theorem} \label{thm:stokes_quasi_polys}
Let $J=n+1$, $n\in \N$ and $\Lambda$ be an eigenvalue of the boundary value problem  \eqref{eq:ST_eval_problem}-\eqref{boundary1}
with eigenfunction the quasi-polynomial  $F(x) = p_n(x){\rm e}^{\theta(x)}$.
 Let $G_k$,   be  solutions  of the ODE   \eqref{eq:ST_eval_problem}   linearly independent from $F(x)$  and 
  which can be expressed as 
\begin{equation} \label{eq:def_G_in_sectors}
    G_k(x) = F(x) \int_{\infty_k}^x\frac {\dd \zeta}{F(\zeta)^2} ,   \quad k= 0,2,4.
\end{equation}
Here $\infty_k$ indicates that the contour of integration extends to infinity along the direction  $\arg(z)= {k \frac {\pi} {3}}$.

Then the  Stokes phenomenon for the solutions $[F,G_k]$  is given by the following equations:
\begin{equation}
 \label{eq:exact_RHP_jumps}
\begin{split}
    [F,G_2] = & [F,G_0] 
    \begin{bmatrix}
    1 & s_1 \\ 0 & 1
    \end{bmatrix}, \quad  \quad  s_1 := \int_{\infty_2}^{\infty_0}\frac {\dd \zeta}{F(\zeta)^2} \\
    [F,G_4] = & [F,G_2] 
    \begin{bmatrix}
    1 & s_3 \\ 0 & 1
    \end{bmatrix}, \quad \quad   s_3 := \int_{\infty_4}^{\infty_2} \frac {\dd \zeta}{F(\zeta)^2} \\
    [F,G_0] = & [F,G_4] 
    \begin{bmatrix}
    1 & s_5 \\ 0 & 1
    \end{bmatrix}, \quad \quad   s_5 := \int_{\infty_0}^{\infty_4}\frac {\dd \zeta}{F(\zeta)^2} .
\end{split}
\end{equation}
Furthermore the Stokes  parameters $s_j$ satisfy \eqref{Stokes}.
\end{theorem}
\noindent{\bf Proof.}
Let $p_n(x)$ be a polynomial solution of degree $n$ of \eqref{eq:polynomial_ODE} with $J=n+1$.
We can obtain a second linearly independent solution $q$  of \eqref{eq:polynomial_ODE} 
using the Wronskian identity:
\begin{equation}
    \dv{p_n}{x}q - p_n \dv{q}{x} = e^{-2 \theta}.
\end{equation}
The solution is written as: 
\begin{equation}
    q(x) :=p_n(x) \int_{x_0}^x \left( p_n(\zeta) e^{\theta(\zeta;t)} \right)^{-2} \dd \z
\end{equation} 
with $x_0$ is arbitrary.
We have thus found two linearly independent solutions of \eqref{eq:polynomial_ODE}, and in turn
we found two particular solutions of the eigenvalue problem \eqref{eq:ST_eval_problem}, namely:
\begin{align}
\nonumber 
    F(x) :=& p_n(x) e^{\theta(x;t)}, \\
    \label{G22k}
    G_{2k}(x) :=& F(x) \int_{\infty_{2k}}^x \frac{ \dd \z}{F(\z)^2}, \quad k =0,1,2.
\end{align}
Note that since $\exp(\theta(x;t)) \to 0$ along rays  $\arg(x) = \pi/3, \pi, 5\pi/3$ (see Fig. \ref{fig:dominant-recessive}), the function $F(x)$ satisfies the  boundary conditions \eqref{boundary1}, { while $G_{2k}(x)$ is unbounded as $x\to\infty$ along the same directions.\color{black}}

%

We can split the integral representation of $G_{2k}$ in \eqref{G22k} as follows: for $k=0,1,2$ we have
\begin{align}
    G_{2k+2}(x)
    = & F(x) \int_{\infty_{2k+2}}^x \frac{ \dd \z}{F(\z)^2} \\ 
    = &F(x) \left( \int_{\infty_{2k+2}}^{\infty_{2k}} \frac{ \dd \z}{F(\z)^2} + \int_{\infty_{2k}}^x  \frac{ \dd \z}{F(\z)^2} \right)\\
    = & s_{2k+1} F(x) + G_{2k}(x)
\end{align} 
were the indices are taken $\mod{6}$ and we have defined $s_{2k+1}, k=0,1,2 $ to be the integral of
$F(\z)^{-2} \dd \z$ between $\infty_{2k+2}$ and $\infty_{2k}$, owing to the observation that 
 its contour of integration  crosses the Stokes line of argument $(2k+1)\pi/3$ with $k=0,1,2$.

The specific contour of integration of the Stokes parameters $s_{2k+1}$ defined in \eqref{eq:exact_RHP_jumps} does not matter  as long as it avoids the poles of   the integrand $F(\z)^{-2} \dd \z$.  
Indeed the integrand $F(\z)^{-2} \dd \z$ has zero residue 
in the finite complex plane  because  if there were a non-zero residue at a pole, 
the function  $G_{2k}(x)$ would have non-trivial monodromy around that pole, but this is not the case since $G_{2k}$ is a solution to the linear ODE system   
\eqref{eq:ST_eval_problem} which has analytic coefficients in the finite complex plane $\C$. One 
may also  verify directly that the ODE \eqref{eq:ST_eval_problem} implies the vanishing of these residues.
{ To prove the condition  \eqref{Stokes}  for the  Stokes  parameters  $s_j$ we observe that  sum of the Stokes parameters  corresponds to an integral of the function $F^{-2}(\z)$  on a closed contour in the complex plane and by the residue theorem it vanishes since  the integrand has second order poles with zero residue.}
\QED 

\subsection{Connection to degenerate orthogonal polynomials}
\label{sec32}
In this section we characterise the quasi-polynomial solutions to \eqref{eq:ST_eval_problem} as
degenerate orthogonal polynomials in the following sense.

\begin{definition}
Let $\theta(x)$ be a fixed polynomial of degree $d+1$ and positive leading coefficient,  and let $\Gamma = \sum_{j=1}^d s_{2j-1} \gamma_{2j-1}$, where $\gamma_{2j-1}$ are the ``wedge'' contours  extending from $\infty_{2j-1}$ to $\infty_{2j+1}$  
where $\infty_\ell$ denotes the point at infinity in the directions $\arg(x) = \frac {\ell \pi}{d+1}$. 
Consider the non-hermitian orthogonal polynomials $p_n(x)$ satisfying
\begin{equation}
    \langle p_n,x^k \rangle := \sum_{j=1}^d s_{2j-1} \int_{\gamma_{2j-1}} p_n(x) x^k e^{2 \theta(x)} \dd x =  \int_\Gamma p_n(x) x^k e^{2 \theta(x)} \dd x = 0, \quad k=0,\dots, n-1.
\end{equation}
We say the polynomial $p_n$ is $l$\textbf{-degenerate orthogonal} if we additionally have
\begin{equation}
    \langle p_n, x^k\rangle = 0 \quad k = n, n+1, \dots, n + l -1 .
\end{equation}
\end{definition}

To prove that the quasi-polynomials are degenerate orthogonal polynomials we will interpret the jumps in
Theorem \ref{thm:stokes_quasi_polys} as a Riemann-Hilbert problem for orthogonal polynomials
\cite{deift_orthogonal_polys_and_random_matrices_book}.

\begin{theorem} \label{thm:quaspol_is_OP}
Supppose that $F(x) = p_n(x)e^{\theta(x;t)}$ is a quasi-polynomial solution of the ODE \eqref{eq:ST_eval_problem}.
Then $p_n(x)$ is a $1$-degenerate non-hermitian orthogonal polynomial with respect to the 
weight $w(x) = e^{2 \theta(x,t) } \dd x$ on the contour
\begin{equation}
    \Gamma = \varkappa  \gamma + \widetilde \varkappa  \widetilde{\gamma}
\end{equation}
where $\varkappa  = s_1$ and $\widetilde \varkappa  = s_5$   are defined  in \eqref{eq:exact_RHP_jumps}. The contour  $\gamma$ is the wedge contour  from 
$\infty_1$ to  $ \infty_3$ and $\widetilde{\gamma}$ is the wedge contour  from  $\infty_5$ to  $ \infty_3$  (see Figure~\ref{fig:dominant-recessive}).
\end{theorem}

\noindent{\bf Proof.}
Consider the quasi-polynomial $F(x) := p_n(x) {\rm e}^{\theta(x;t)}$ which solves \eqref{eq:ST_eval_problem} with $J=n+1$.
Define $G_{k}(x), k=0,2,4$ as in \eqref{eq:def_G_in_sectors}, which also solves the same ODE, and let us set 
\begin{equation}
\Psi_{k}(x) := [F(x),G_{k}(x)], \quad k=0,2,4,
\end{equation} to be a fundamental solution
of \eqref{eq:ST_eval_problem}.
We interpret Theorem \ref{thm:stokes_quasi_polys} as an $2\times2$ Riemann-Hilbert problem solved by $\Psi_k(x)$ in the sectors of opening $2\pi/3$ :
\begin{equation}
    \St_{k}\cup \St_{k+1}, \quad k=0,2,4.
\end{equation}
The jump conditions \eqref{eq:exact_RHP_jumps} imply that $G_k$ satisfies:
\begin{equation}
    {\rm e}^{\theta(x;t)}G_{k}(x) = {\rm e}^{\theta(x;t)} G_{k+2}(x) + s_{k+1} p_n(x) {\rm e}^{2\theta(x;t)}, \quad k = 0,2,4.
\end{equation}
The three functions $G_{k}$, $k=0,2,4$ define a piecewise--analytic function $G(x)$ with discontinuities along the three rays $[0,\infty_{j}]$ with $j=1,3,5$.  { The function $G(x)$   satisfies  the  jump conditions}
\begin{equation}
 {\rm e}^{\theta(x;t)}G_+(x)=  {\rm e}^{\theta(x;t)}G_-(x)  + s_{j} p_n(x) {\rm e}^{2\theta(x;t)}, \quad x \in [0,\infty_j], \quad j=1,3,5,
\label{RHPG}
\end{equation}
{ where $G_{\pm}(x)$ are the boundary values of the function $G(x)$ on the right and left boundaries  of the oriented rays $ [0,\infty_j]$, $j=1,3,5$.}
By the Sokhotski--Plemelj formula we can express $G$ as the following Cauchy transform:
\begin{equation}
    G(x) := \frac{{\rm e}^{-\theta(x;t)}}{2 \pi i} 
     \int_\Gamma 
    \frac{p_n(\z){\rm e}^{2 \theta(\z;t)}}{\z-x} \dd \z,\quad \Gamma=s_1\gamma+s_5\tilde{\gamma},
    \label{eq:G(x)}
\end{equation}
with $\gamma$ and $ \widetilde{\gamma}$ as in Fig.\ref{fig:dominant-recessive}.  Note that the two contours overlap with the same orientation on the ray $[0, \infty_3]$.
{
Note that the function $G$ defined by \eqref{eq:G(x)} coincides with $G_k$ in each appropriate sector, since they satisfy the same RHP and $G(x){\rm e}^{\theta(x)}$ and $G_k(x) {\rm e}^{\theta(x)}$ behave like $\bigO(x^{-1})$ at infinity. 
To prove this claim  it is sufficient to use the formula \eqref{eq:G(x)} for $G$ and the asymptotics
\eqref{eq:formal} for $G_k$.}

This  considerations  imply the degenerate orthogonality of the polynomials $p_n(x)$ as follows.
From \eqref{eq:formal} with $J=n+1$ we see the Wronskian matrix of the ODE \eqref{eq_W} for the pair of solutions $(F,G_k)$ has the following asymptotic expansion (using $' = \dv{x}$):
\begin{align}
    \begin{bmatrix} F(x) & G_k(x) \\ F'(x) & G_k'(x) \end{bmatrix} 
    \sim & 
    \begin{bmatrix}
    x^n     {\rm e}^{\theta(x)} & -x^{-n-2}{\rm e}^{-\theta(x)} \\
    x^{n+2} {\rm e}^{\theta(x)} &  x^{-n}{\rm e}^{-\theta(x)}
    \end{bmatrix} \left( \mathbb{I} + \bigO(x^{-1}) \right),
    \quad x \to \infty.
    \label{eq:formal2}
\end{align}
In fact, with $F$ and $G$ given as above, it follows that  \eqref{eq:formal2} is the asymptotic expansion of $F$ and $G$ in each of the six sectors.
Thus we find that the asymptotic
\begin{align}
    G(x) =& \frac{{\rm e}^{-\theta(x)}}{2\pi i} \int_{\Gamma} \frac{p_n(\z) {\rm e}^{2 \theta(\z;t)}}{\z-x} \dd \z \\ 
    = & 
    -\frac{1}{x} \frac{{\rm e}^{-\theta(x)}}{2\pi i} \int_\Gamma p_n(\z) {\rm e}^{2 \theta(\z;t)}
    \left( 1 + \frac{\z}{x} + \dots + \frac{\z^{n+1}}{x^{n+1}} + \dots \right) \dd \z \sim -x^{-n-2} {\rm e}^{-\theta(x;a)}
\end{align}
implies the vanishing of the integrals:
\begin{equation}
    \int_{\Gamma} p_n(\z) \z^k {\rm e}^{2\theta(\z;t)} \dd \z, \quad k =0,1, \dots , n .
\end{equation}
Therefore the polynomials $p_n(x)$ are \textit{degenerate} orthogonal as claimed.
\QED 

One can also prove the converse of this result, but to do so we first need the following lemma.

\begin{lemma} \label{thm:apparent_singularity_lemma}
Consider the second order ODE  in the complex plane 
\begin{equation} \label{eq:ODE-order2}
   \dv[2]{y}{x} -V(x)y =0.
\end{equation}
 {Suppose that $x=x_*$  is a (possible) singularity of the potential $V(x)$ where  it has a    pole of order
at most $2$ . Additionally, assume that $x=x^*$ is an apparent singularity  ( namely  two  linearly independent 
solutions of the ODE \eqref{eq:ODE-order2} are analytic at $x=x^*$).  }
Then, $V(x)$  { is   analytic at $x=x^*$}.
\end{lemma}
\noindent{\bf Proof.}
We argue by contradiction and consider separately the cases when $V(x)$ has a double pole and when $V(x)$ has a simple pole at $x=x^*$. In both cases we make use of the indicial equation and its basic properties, which can be found in more detail in \cite{olver_asymptotics_1974}. 
\begin{itemize}
    \item \textit{Case 1: Simple pole.}
    Suppose that near the singular point $x=x_*$ the potential is of the form
    \begin{equation}
        V(x) = \frac{a}{\zeta} + b + \bigO(\zeta) 
    \end{equation}
    with $a\neq 0$ and $\zeta = x-x_*$.
    The indicial equation of the ODE  becomes
    \begin{equation}
        d(d-1)=0\,.
    \end{equation}
   It  has two solutions $d_1=1$ and $d_2=0$ differing by a non-zero integer $d_1-d_2=1$, meaning that  that there are two linearly independent solutions $y_1(x), y_2(x)$ such that
    \begin{align}
    y_1(x) = \bigO(\zeta^{d_1}), \\
    y_2(x) = \bigO(\zeta^{d_2})   
    \end{align}
    as $x \to x_*$ \cite[p.151]{olver_asymptotics_1974}.
    But a simple computation shows that 
    $y_2(x) = 1 + \bigO(\zeta)$
    cannot solve the differential equation \eqref{eq:ODE-order2} if $V(x)$ has a simple pole at $x=x_*$. This is a contradiction.
    
    \item \textit{Case 2: Double pole}.
    Suppose that near the singular point $x=x_*$ the potential has the shape
    \begin{equation}
        V(x) = \frac{a}{\zeta^2} + \frac{b}{\zeta} + \bigO(1) 
    \end{equation}
    with $a\neq 0$ and again we denote $\zeta = x-x_*$.
    The indicial equation now gives
    \begin{align}
        a =& d(d-1) \ .\label{indicial}
    \end{align}
    Let $d_1,d_2$ be the two solutions of the {indicial equation} \eqref{indicial}.
    By the assumption that the solutions to the ODE are all analytic, we must have that 
    $d_1$ and $d_2$ must be non-negative integer and not equal to each other (if $d_1=d_2$ then
    one of the solutions has a logarithmic singularity, see \cite{olver_asymptotics_1974}).
    Note that $d_1 = d_2$ if and only if $a=-1/4$, in which case $d_1= d_2 = 1/2$, so we may simply
    assume that $d_1,d_2$ are non-negative integers.
    Rewriting the equation $d_1(d_1 - 1) = d_2 (d_2-1)$ as 
    \begin{equation}
        (d_1 - d_2)(d_1+d_2 -1) = 0
    \end{equation}
    we see that the only non-negative integer solutions are $d_1 = 0$ and $d_2 =1$ and vice versa.
    In either case from the indicial equation we find that $a=0$, which is a contradiction.
    Therefore $V(x)$ cannot have a double pole. \QED
\end{itemize}

We are ready to prove the converse of Theorem \ref{thm:quaspol_is_OP}, namely
that these degenerate orthogonal polynomials satisfy the Shapiro-Tater ODE \eqref{eq:ST_eval_problem}.

\begin{theorem} \label{thm:OP_is_quaspolys}
Suppose that $p_n(x)$ is a $1$-degenerate orthogonal polynomial with respect to the weight 
$w(x) = {\rm e}^{2\theta(x;t)}$ on the { weighted} contour $\Gamma := \varkappa  \gamma + \tilde \varkappa  \tilde{\gamma}$  { for $(\varkappa, \tilde \varkappa)\neq (0,0)$, in the notations}  of  Theorem \ref{thm:quaspol_is_OP}.
Then  $F(x) = p_n(x) {\rm e}^{\theta(x;t)}$ is a quasi-polynomial solution of 
the boundary problem \eqref{eq:ST_eval_problem}-\eqref{boundary1} with $J=n+1$.
\end{theorem}
\noindent{\bf Proof.}
Suppose that $p_n(x)$ is a degenerate orthogonal polynomial with respect to the weight
$w(x)= {\rm e}^{\theta(x;t)}$ on the (weighted) contour $\Gamma$, i.e.
\begin{equation}
    \int_\Gamma p_n(\z) z^k {\rm e}^{2 \theta(\z;t)} \dd \z = 0 \quad k=0,1, \cdots , n.
\end{equation}
Define the functions 
\begin{align}
F(x)& :=   p_n(x) {\rm e}^{\theta(x;t)} \\ 
G(x)& := \frac{{\rm e}^{-\theta(x;t)}}{2 \pi i} \int_\Gamma \frac{p_n(\z) {\rm e}^{2 \theta(\z;t)}\dd \z}{\z-x}.
\end{align}
We claim that the Wronskian $W = FG' - F'G$ is constant. 
The degenerate orthogonality condition implies that
\begin{equation}
 G(x) \simeq \left( h_n x^{-n-2} + \bigO(x^{-n-3}) \right) {\rm e}^{-\theta(x;t)},
 \quad  x \to \infty
 \end{equation} 
where the leading factor is 
\begin{equation}
h_n := -\frac{1}{2 \pi i} \int_\Gamma p_n(\z) {\rm e}^{2\theta(\z;t)} \z^{n+1} \dd \z.
\end{equation}
Furthermore, by differentiating (denoting $' = \dv{x}$) we find
\begin{align}
    G'(x) =& -\theta'(x) G(x) + \frac{{\rm e}^{-\theta(x;t)}}{2 \pi i} \int_\Gamma \frac{p_n(\z) {\rm e}^{2\theta(\z;t)}}{(\z-x)^2} \dd \z \\
    = & -\theta'(x) G(x) + \frac{1}{x^2} \frac{{\rm e}^{-\theta(x;t)}}{2 \pi i}  \int_\Gamma p_n(\z) {\rm e}^{2\theta(\z;t)} \left( 1 + 2 \frac{\z}{x} + 3  \frac{\z^2}{x^2} + \dots \right)  \dd \z \\
    =&  {\rm e}^{-\theta(x;t)} \left( - h_n x^{-n} + \bigO(x^{-n-1}) \right)+ \frac{1}{x^2} \frac{{\rm e}^{-\theta(x;t)}}{2 \pi i} \bigO(x^{-n-1}) \\
    = &  {\rm e}^{-\theta(x;t)} \left( - h_n x^{-n} + \bigO(x^{-n-1}) \right).
\end{align} 
Additionally, since $p_n(x)$ is a monic polynomial,
we have that $F(x) \sim x^n {\rm e}^{\theta(x;t)}$ and $F'(x) \sim x^{n+2} {\rm e}^{\theta(x;t)}$, which means that the Wronskian is bounded at infinity, i.e. $W(x) = F'(x) G(x) - F(x) G'(x) = -2 h_n + \bigO(x^{-1})$ as $x \to \infty$.

The Wronskian  $W(x)$  has no jump-discontinuities  since {the boundary values $W_{\pm}(x)$ } for  $x\in \gamma, \widetilde{\gamma}$  satisfied
\begin{equation}
    W_+ = F' G_+ -F G_+' = F' (s_k F +G_-) -F(s_kF' +G_-') = F'G_- -FG_-' = W_-.
\end{equation}
Furthermore, since $W$ is built from locally analytic functions, it also follows that $W(x)$ has no poles.
Thus $W$ is an entire function; furthermore it is bounded at infinity since, in each sector, $F' G, F G'$ are bounded. By the Liouville theorem we conclude that  $W(x)$ must be a constant, i.e.  
\begin{equation}
    W(x) = F'(x) G(x) - F(x) G'(x) \equiv -2h_n.
\end{equation}
Differentiating this equation gives that $F''/F \equiv G''/G$.  Let us denote by $V(x)$ this ratio; then both $F$ and $G$ satisfy a 2nd order linear ODE of the form:
\begin{equation}
    y'' - V(x)y=0 \quad \text{ with potential }  V(x) := \frac{F''(x)}{F(x)}.
\end{equation}
We can rewrite the potential using the defining expression $F = p_n {\rm e}^\theta$ in terms of the polynomial $p_n$,  which gives us:
\begin{equation} \label{eq:shape_of_V}
    V(x) = \theta''(x) + (\theta'(x))^2 + 2\theta'(x) \frac{p_n'(x)}{p_n(x)} + \frac{p''_n(x)}{p_n(x)}.
\end{equation}

Let $c$ be one of the zeros of $p_n$ of multiplicity $d$ then we can write  $p_n(x) = (x-c)^d h(x)$ and then we  expand \eqref{eq:shape_of_V}
near $x=c$:
\begin{equation}
    V(x)= \frac{d(d-1)}{(x-c)^2} + \frac{2d}{x-c} \left( \frac{h'(c)}{h(c)} + \theta'(c) \right) + \bigO(1), \ \ \ x\to c.
\end{equation}
This shows that $V(x)$ may have at most a second-order pole: since now  $F(x)$ and $G(x)$ are both analytic near $x=c$ and
both satisfy the ODE $y''-V(x) y=0$ we deduce that all the singularities of the ODE are apparent. 
We can therefore, apply Lemma \ref{thm:apparent_singularity_lemma}, from which it follows that $V(x)= F''/F$ is entire.
We conclude that $d=1$, namely all the zeros 
$x_1, \dots, x_n$ of the polynomial $p_n(x)$ are simple and obtain
\begin{equation}
V(x)=\theta''(x) + (\theta'(x))^2+2nx+2\sum_{j=1}^nx_j+\sum_{j=1}^n\frac{1}{x-x_j}\left[\theta'(x_j)+\sum_{\substack{k=1 \\ k\neq j}}^n\frac{1}{x_j-x_k}\right]
\end{equation}
which implies that 
\begin{equation}
\label{eq:Fekete-equilibrium}
\theta'(x_j)=\sum_{\substack{k=1 \\ k\neq j}}^n\frac{1}{x_k-x_j},\quad j=1,\dots,n.
\end{equation}
This gives the potential
\begin{equation}\label{eq:quartic-fekete}
V(x)=x^4+tx^2+2x(n+1)+\frac{t^2}{4}+2\sum_{j=1}^nx_j
\end{equation}
which coincides with the potential in \eqref{eq:ST_eval_problem} with $\Lambda=\frac{t^2}{4}+2\sum_{j=1}^nx_j$
\QED 

\begin{remark}
The proof of the above theorem shows that   $V(x)$ is a quartic polynomial if and only if all zeros $x_1,\dots, x_n$ of $p_n(x)$ 
are simple and satisfy a Fekete type equilibrium property \eqref{eq:Fekete-equilibrium}.
In fact, this property also hold true for the zeros of non-hermitean semiclassical degenerate orthogonal polynomials; this interesting observation is generalized and expanded in \cite{BGH}.
\end{remark}

To summarise, we have shown the following:

\begin{corollary} \label{thm:quasi-polys_iff_orthogonal}
The following statements are equivalent:
\begin{enumerate}
    \item  the boundary problem in \eqref{eq:ST_eval_problem}-\eqref{boundary1}  admits a quasi-polynomial solution $F(x) =p_n(x) {\rm e}^{\theta(x,a)}$ with $p_n(x)$  a polynomial  of degree $n$;
    \item { The polynomial $p_n(x)$ (of degree $n$) is a $1$--degenerate orthogonal polynomial with respect to  the weight }$w(x) = {\rm e}^{2\theta(x;t)} $ on the contour $\Gamma := \varkappa  \gamma + \tilde \varkappa  \tilde{\gamma}$  {\color{black} for
some values of $(\varkappa , \tilde \varkappa) \neq  (0, 0)$} and  with $\gamma$ and $\tilde{\gamma}$ as in Fig.\ref{fig:dominant-recessive}. Furthermore, the coefficients $\varkappa  = s_1, \tilde \varkappa  = s_5$ are expressible in terms of $p_n(x)$ by the formulas \eqref{eq:exact_RHP_jumps}. 
\end{enumerate}
\end{corollary}

\subsection{Repeated eigenvalues}
Recall that the characteristic polynomial
$C_n(t,\lambda)$ in \eqref{eq:spectral-polynomial} gives us the eigenvalues
of \eqref{eq:ST-polynomial-ODE}, which in turn characterizes the  Exactly Solvable spectrum of the 
eigenvalue problem of \eqref{eq:ST_eval_problem} with boundary condtions \eqref{boundary1}.

In this section we prove the following crucial result.
\begin{theorem} \label{thm:repeated_eigenval_condition}
The following statements are equivalent:
\begin{enumerate}
\item The value of $t\in \C$ is such that the Exactly Solvable  spectrum of \eqref{eq:ST_eval_problem}-\eqref{boundary1}  has a repeated eigenvalue;
\item { The value $t\in \C$ is such that there exist $n\in \N$ and $\Lambda \in \C$ so that  for $J=n+1$ the problem \eqref{eq:ST_eval_problem}, \eqref{boundary1}  has a nontrivial solution and furthermore $D_n(t)=0$, with $D_n$  the discriminant in  \eqref{STDn}};
\item there is a quasi-polynomial solution $p_n(x) {\rm e}^{\theta(x;t)}$ of \eqref{eq:ST_eval_problem}-\eqref{boundary1}  that satisfies 
\begin{equation}
\label{458} 
    \int_{\gamma} \left( p_n(x) e^{\theta(x;{ t})} \right)^2 \dd x = 0,\;\;\;   \int_{ \widetilde{\gamma}} \left( p_n(x) e^{\theta(x;{ t})} \right)^2 \dd x = 0
\end{equation}
where $\gamma$ and $\widetilde{\gamma}$ are defined as in Theorem \ref{thm:OP_is_quaspolys} (contours from $\infty_1 $ to $\infty_3$ and $\infty_3 $ to  $ \infty_5$ respectively, see Fig.\ref{fig:dominant-recessive}).
\end{enumerate}

\end{theorem}
\noindent{\bf Proof.}
\noindent {\bf [$(1.)\Rightarrow (2.)$].} If $y(x)$ is a solution of the eigenvalue problem  \eqref{eq:ST_eval_problem}  then $y$ must be a quasipolynomial  according to Proposition \ref{propQES}, and we must have that $J=n+1$ and $\Lambda$ ($\lambda = \Lambda-\frac {t^2} 4$) is an eigenvalue.

If this eigenvalue is repeated then the derivative of the characteristic polynomial $C_n(t,\lambda)$  in \eqref{eq:spectral-polynomial} must also vanish. 

\noindent {\bf [$(2.)\Rightarrow (1.)$].} This is immediate consequence of Lemma \ref{lemmasimpleeval} together with the fact that the derivative of a polynomial vanish at each root of multiplicity higher than one.

\noindent {\bf [$(2.)\Rightarrow (3.)$].} 
The condition $(2.)$ means the  eigenvalue has algebraic multiplicity (at least) $2$ and so we can consider the generalized eigenvector. 
Let us remind the reader here that if $v$ is the eigenvector  of the matrix $M_n(t)$  in \eqref{STMn}  with   eigenvalue $\lambda$, then the generalized eigenvector $w$ satisfies the equation $(M-\lambda I)w=v$. It follows that  the generalized eigenvector  equation for a {\it polynomial} $r(x)$ (of degree $\leq n$)  takes the form of  the following  differential equation for  $r(x)$:
\begin{equation}
\label{genevect}
    \dv[2]{r(x)}{x} + 2\left(x^2 + \frac{t}{2}\right) \dv{r(x)}{x} - (2
    (J-1)x + \lambda) r(x)
    = p_n(x).
\end{equation}
{ The fact that the above non-homogeneous equation admits a {\it polynomial} solution $r(x)$ is essential to bear in mind: such solution is not unique since we can add to it an arbitrary multiple of  $p_n(x)$.}
{ The associated homogeneous differential equation has two solutions, one of which is  
 $p_n(x)$ (a polynomial of degree $n=J-1$) and the other  one is obtained from \eqref{eq:def_G_in_sectors} and can be written as follows:}
 $$
 q_k(x) = p_n(x) \int_{\infty_k}^x F(\z)^{-2} \dd \z, \ \ \ F(x):= p_n(x){\rm e}^{\theta(x;{  t})}, 
 $$
 where we can choose any $k=0,2,4$ { for the basepoint of integration}. 
 { Consequently, a  particular solution $r_0(z)$ of \eqref{genevect} is found by the standard } ``variation of parameters'' as follows:
\begin{equation} 
\label{r0}
    r_0(x) := p_n(x) 
    \underbrace{\int_{\infty_0}^x F(\z)^{-2} 
    \left(\int_{\infty_{1}}^\z F(s)^2 \dd s \right) \dd \z}_{H(x)}.
\end{equation}
{ Before proceeding with the proof we claim and prove that $H(x)$ is bounded by a constant as  $x\to \infty_{1}$. 
To see this it is sufficient to show that  the inner integral in the defintion of  $H(x)$ tends to zero as $\mathcal O(\z^{2n-2}) {\rm e}^{2\theta(\z;t)}$ (recall that the real part of $\theta$  goes to $-\infty$ along the direction $\infty_1$ and hence this is exponentially small).
To prove this claim we write $p_n^2(s) = 2Q_{2n-2}(s) \theta'(s) + R(s)$, with $Q_{2n-2}$ the quotient (of degree $2n-2$ and $R$ the remainder  (of degree $1$) polynomials of division by $2\theta'$. Then integrating by parts we obtain
\be
\int_{\infty_1} ^\z F(s)^2\d s = Q_{2n-2}(\z) {\rm e}^{2\theta(\z)} + \int_{\infty_1}^\z R(s){\rm e}^{2\theta(s)} \dd s.
\ee
To estimate the last integral we ``force'' again an integration by parts:
\bea
\int_{\infty_1}^\z R(s){\rm e}^{2\theta(s)} \dd s=& 
\frac {R(\z)}{2\theta'(\z)}{\rm e}^{2\theta(\z)} -\int_{\infty_1}^\z \le(\frac{R(s)}{2\theta'(s)}\ri)'{\rm e}^{2\theta(s)} \dd s.
\eea
In the remaining integral  the prefactor to the exponential is $\mathcal O(s^{-2})$ and thus the integral is easily estimated to be $\mathcal O(1) {\rm e}^{2\theta}$.
}
 With the claim proved, let us proceed: we know that \eqref{genevect} has by assumption a  polynomial solution and we are going to show now that $r_0(x)$ itself is such a polynomial.
 Indeed, the general  solution of \eqref{genevect} is obtained by adding an arbitrary  linear combination of $p_n(x), q_0(x)$ to $r_0(x)$ in \eqref{r0}.   We then must show that here are constants $A,B$ such that  $r_0(x) +A p_n(x) + B q_0(x)$ is a polynomial of degree at most $n$; clearly here only the value of $B$ is relevant (since $p_n$ is already a polynomial).  Thus the issue boils down to  whether  there is a value of $B$ for which $r_0(x) + B q_0(x)$ is a polynomial.  
We first observe that  the only possible value for $B$ must be zero. 
{Indeed consider the asymptotic behaviour  for  $x\to\infty_1$. The  integral $H(x)$  is bounded by a constant as $x\to\infty_1$.}
However, $q_0(x)$ has a dominant  exponential growth in this direction  and hence the expression $r_0(x) + Bq_0(x)$ has polynomial growth in this direction only if $B=0$. 

{ We thus conclude that since \eqref{genevect} has a (pencil of) polynomial solution(s), this must be given by $r_0(x)+ Ap_n(x)$ so that }  $r_0$ must itself be a polynomial. 

{ We now establish \eqref{458}. To this end} we consider the behaviour of $r_0(x)$ near $\infty_{3}$ and $ \infty_5$. We can write, for example for $\infty_3$, 
\begin{align}
\label{462}
    r_0(x) &= p_n(x) \int_{\infty_0}^x F(\z)^{-2} 
    \left(\int_{\infty_{1}}^{\infty_{3}}   F(s)^2 \dd s + \int_{\infty_{3}}^\z   F(s)^2 \dd s  \right) \dd \z\nonumber
    \\
    &= \left(\int_{\infty_{1}}^{\infty_{3}}   F(s)^2 \dd s\right) q_0(x) +
    p_n(x) \int_{\infty_0}^x F(\z)^{-2} 
    \left(\int_{\infty_{3}}^\z   F(s)^2 \dd s  \right) \dd \z
\end{align}
The second term above is polynomially bounded    near $\infty_3$,   by the same argument used to show that $H$ is bounded near $\infty_{1}$. But since $q_0$ is exponentially dominant also near $\infty_{1}$ we deduce  the  condition  $\int_{\infty_{1}}^{\infty_{3}}   F(s)^2 \dd s=0$. One similarly deduces $\int_{\infty_{5}}^{\infty_{3}}   F(s)^2 \dd s=0$, which establishes \eqref{458}.

\noindent {\bf [$(3.)\Rightarrow (2.)$].} 
Consider the expression \eqref{r0}. It is easy to see directly that it satifies the generalized eigenvector equation \eqref{genevect}; we must only verify that the conditions \eqref{458} guarantee that $r_0(z)$ is a polynomial. But this follows again from the Liouville theorem and using \eqref{462}.
\QED 
\begin{remark}
     The Theorem \ref{thm:repeated_eigenval_condition} seems at first sight  nothing short of a miracle; indeed once we fix $J=n+1\in \mathbb N$, then  the ODE \eqref{eq:ST_eval_problem} has only two continuous parameters $t, \lambda$.  
     
However the multiple eigenvalue condition apparently involves now three equations which are:\\
 (i) the existence of a quasi-polynomial solution which determines  $\lambda$ as a function of $t$;
\\
 (ii) the two equations \eqref{458} for the parameter $t$.

However, the system  is actually not overdetermined because of the following reasoning: if, for given $J=n+1\in \mathbb N$ the pair $(t,\Lambda)$ is in the ES spectrum, then we have shown in Theorem \ref{thm:quaspol_is_OP} that $p_n(x)$ is a {\it degenerate} orthogonal polynomial, namely, 
\bea
\varkappa  \int_\gamma p_n^2(x){\rm e}^{2\theta(x;t)}\d x +\widetilde \varkappa  \int_{\widetilde \gamma} p_n^2(x){\rm e}^{2\theta(x;t)}\d x =0,
\eea
with $\gamma, \widetilde \gamma, \varkappa  , \widetilde\varkappa  $ defined in the same theorem. 
The coefficients  $\varkappa  ,\widetilde \varkappa  $ cannot be both vanishing for otherwise the Stokes phenomenon of the ODE  would be trivial (which is not possible). Then the two equations \eqref{458} yield only one additional constraint.\par\vskip 5pt

  \end{remark}
 
 The following corollary is also a nice property, although it will not be used in the rest.
 \begin{corollary}
 \label{corquaspol}
 Suppose that $y(x) = p_n(x)^{\theta(x;t)} $ is a solution of the boundary problem \eqref{eq:ST_eval_problem} and hence $t$, $J=n+1$ and $\Lambda$  are in the ES spectrum. Suppose also that $\Lambda = \lambda + \frac {t^2}4$ is a repeated eigenvalue. 
 Then the antiderivative of $y^2(x)$ is also a quasipolynomial, 
 \bea
 \int^x p_n(\z)^2{\rm e}^{2\theta(\z;t)}\d \z = q_{2n+2}(x) {\rm e}^{2\theta(x;t)}.
 \eea
  \end{corollary}
  \noindent{\bf Proof.}
  Take the integration basepoint from $\infty_1$ and consider the function 
  \bea
  K(x) = \int_{\infty_1}^x p_n(z)^2{\rm e}^{2\theta(z;t)}\d z
  \eea
  Standard asymptotic analysis shows that $K(x)$ is of the form $\mathcal O(x^{2n+2}){\rm e}^{2\theta(x;t)}$ along the (dominant) direction towards $\infty_{0}, \infty_2,$ and $\infty_4$.

  Moreover towards $\infty_1$ we have 
 \bea
q(x):=   K(x){\rm e}^{-2\theta(x;t)} = \mathcal O(x^{2n+2}).
 \eea
Thus the function  $q(x)=  K(x){\rm e}^{-2\theta(x;t)}$  is polynomially bounded along the sectors containing $\infty_{j}, $ $j\in\{0,1,2,4\}$. If we can show that it is also polynomially bounded in the sectors containing $\infty_{3}$ and $\infty_{5}$ then we conclude, by the Liouville's theorem, that it is a polynomial. 

Consider $x$ tending to infinity along the sector that contains $\infty_3$; we have 
\bea
q(x)&= {\rm e }^{-2\theta(x;t)}\int_{\infty_1}^x p_n(\z)^2{\rm e}^{2\theta(\z;t)}\d \z =\nn \\
&=
{\rm e}^{-2\theta(x;t)}\int_{\infty_3}^x p_n(\z)^2{\rm e}^{2\theta(\z;t)}\d \z  +
{\rm e }^{-2\theta(x;t)}\int_{\infty_1}^{\infty_3} p_n(\z)^2{\rm e}^{2\theta(\z;t)}\d \z .
\eea

The second term above is just a multiple of ${\rm e}^{-2\theta(x;t)}$ which is not polynomially bounded near $\infty_{2j+1}$. However the coefficient is precisely one of the integrals in \eqref{458} which has been proved to vanish. Thus $q(x)$ is also polynomially bounded near $\infty_3$. Similarly we can prove that it is polynomially bounded at $\infty_5$ and the proof is thus complete.   \QED

\section{Exact WKB analysis} \label{section:exact-wkb-analysis}
In this section we give a brief overview of the exact WKB method and establish useful notations for later.
The material here is well-known 
and  we follow  the exposition in \cite{Iwaki_Nakanishi_2014} and \cite{kawai_takei_algebraic_analysis_sing_perturbation}.
In order to apply the exact WKB method to \eqref{eq:shapiro-ODE} and \eqref{eq:JM_potential}
we need to scale them appropriately to obtain an ODE of the form $y'' - \hbar^{-2} Q y =0$, where $\hbar$ is a small parameter.

We introduce the scaling for the Shapiro-Tater potential 
\begin{equation*}
\VST = x^4+tx^2 + 2(n+1)x + \Lambda
\end{equation*}
\begin{align} 
    z &= (n+1)^{-1/3}x,\qquad
    {s} = (n+1)^{-2/3}t,\qquad
    E = (n+1)^{-4/3}\Lambda\label{eq:wkb-scaling}
\end{align}
which implies $\dv{y}{x} =(n+1)^{-2/3} \dv{y}{z}$.

A similar scaling applies in the case of  the Jimbo--Miwa potential (for the poles with residue $-1$), 
\begin{equation*}
\VJM (x;a,b)= x^4+ a x^2 + (2n+1)x + \le(\frac {7a^2}{36}  + 10b\ri)
\end{equation*}

\bea
\label{scalingJMU}
    z = \le(n+\frac 12\ri)^{-1/3}x,\qquad 
    {s}  =\le(n+\frac 12\ri) ^{-2/3}a,\qquad
    \widehat{b} = \le(n+\frac 12\ri)^{-4/3}b.
\eea
The result of these scaling is the exact same potential with the identification of $E = \frac {7{s}^2}{36} + 10 \widehat{b}$, but with different scaling factors.
If  we  set
\bea
\label{defQ}
Q(z;{s},E) = z^4 + {s} z^2 +2z +E,
\eea
then  in either cases 
we obtain an $n$-independent potential, namely:
\bea
    &\dv[2]{y}{z} - \le(n+1 \ri)^2 Q(z;{s},E)y =0, \quad \text{for the Shapiro-Tater case;}
    \label{eqsST}
\\    &\dv[2]{y}{z} - \le(n+\tfrac 1 2 \ri)^2 Q(z;{s},E)y =0, \quad \text{for the Jimbo--Miwa case.}
\label{eqsJM}    
\eea
    
\subsection{Schr\"odinger, Riccati and WKB}
Keeping with the tradition we will denote by $\hbar^{-1}$ the large parameter in either  \eqref{eqsJM}, \eqref{eqsST}; namely $\hbar =(n+1)^{-1}$ in the Shapiro-Tater  (ST) case or $\hbar = (n+1/2)^{-1}$ in the Jimbo-Miwa (JM)  case. 

 Consider the following Schr\"odinger equation with small parameter $\hbar$ and polynomial  potential
$Q(x)$ \eqref{defQ}:
\begin{equation} \label{eq:schrodinger_wkb_equation}
    \dv[2]{y}{z} - \hbar^{-2} Q(z;{s},E) y =0. 
\end{equation}z
We now explain how to construct the \textit{WKB solutions} of \eqref{eq:schrodinger_wkb_equation}.
The formal series ansatz
\begin{equation} \label{eq:wkb_ansatz}
    y(z) = \exp\left( \int^z S(\z, \hbar) \dd \z \right),
\end{equation}
with  
\begin{equation}
    S(z,\hbar) := h^{-1} S_{-1}(z)   +  \sum_{k \ge 0} \hbar^{k} S_k(z) 
\end{equation}
implies that $S(z,\hbar^{-1})$ satisfies the following Riccati equation:
\begin{equation}  \label{eq:ricatti_S}
    S^2(z) + \dv{S(z)}{z} = \hbar^{-2} Q(z).
\end{equation}
Comparing each power of $\hbar$ we get a recursive relation for the coefficients 
$S_k(z)$ of the formal series:
\begin{align}
    S_{-1}(z)^2 = & \quad Q(z), \\
    2 S_{-1}S_{k+1} =&- \dv{S_k}{z}-\sum_{\substack{n + m = k \\ n,m \ge 0}}S_{n} S_{m} \quad (k\ge -1).
\end{align}
The first three coefficients $S_k(z)$ are:
 \begin{align}
     S_{-1}(z) =  \sqrt{Q(z)}, \qquad
     S_{0}(z)  = -\frac{1}{4} \frac{Q'(z)}{Q(z)},   \qquad
     S_{1}(z)  =  \frac{4 Q(z) Q(z)'' - 5 (Q(z)')^2}{32 Q(z)^\frac 5 2}. 
 \end{align}

The choice of sign for $S_{-1}(z) = \pm \sqrt{Q(z)}$ gives two 
solutions of the Riccati equation \eqref{eq:ricatti_S}, denoted $S_+,
S_-$ respectively; if we change sign in $S_{-1}$  then all the odd $S_{2k+1}$ change sign while the even $S_{2k}$ remain unchanged. 
These correspond to the two linearly independent solutions of \eqref{eq:schrodinger_wkb_equation}.
We define the odd and even parts of $S(z;\hbar)$:
\begin{align}
    \So(z;\hbar)  := & \frac{1}{2} \big(S_+(z,\hbar)- S_-(z,\hbar)\big) \\
    S_{\rm {even}}(z;\hbar) := & \frac{1}{2} \big(S_+(z,\hbar) + S_-(z,\hbar)\big)
\end{align}
Since our potential $Q(z)$ is independent of $\hbar$, it follows that $\So$ only
contains odd powers of $\hbar$, namely:
\begin{align} \label{eq:sodd}
    \So(z;\hbar)  = \frac 1\hbar \sqrt{Q(z)} + \hbar S_1(z) + \hbar^3 S_3(z) + \dots . 
\end{align}
We will only care about $\So$ since it can be shown \cite{kawai_takei_algebraic_analysis_sing_perturbation} 
that $S_{\text{even}}$ can be written in terms of $\So$  as:
\begin{equation}
    S_{\rm even}(z,\hbar) = -\frac{1}{2} \dv{x} \log \So(z,\hbar ).
\end{equation}

Another important fact is the following: 
the differential $\So(z) \d z$ has a pole at $z=\infty$ which comes solely from the term $S_{-1}(z) = \sqrt Q(z)$; namely, $S_{2k+1}(z) = \mathcal O(z^{-2})$ as $z\to\infty$.  This property can also be shown inductively.

These facts motivate the following definition. As is customary in exact WKB theory, we will refer to the zeroes $\Tau=\{\tau_0, \tau_1, \tau_2, \tau_3\}$ of $Q(z)$ as {\bf turning points}.

\begin{definition}[WKB solutions] \label{def:wkb_sols_turning_pt}
The \textit{ WKB solutions} to \eqref{eq:schrodinger_wkb_equation}
are formal power series in $\hbar$ given in terms of $\So$ in \eqref{eq:sodd}.
We give two different normalizations that will be used throughout this paper.

\begin{itemize}
    \item {\bf Near a turning point} $\tau$ of the potential $Q(z)$
    we define the {\it normalized WKB solutions} to be:
    \begin{equation}
    \label{inttau}
        \psi^{(\tau)}_{\pm}(z,\hbar) := \frac{1}{\sqrt{\So(z,\hbar)}}
        \exp \left( \pm \int^z_\tau \So(\z,\hbar) \dd \z \right)
    \end{equation}
    \item {\bf Near infinity} we define the {\it normalized  WKB solutions} to be:
    \begin{equation} 
        \psi_\pm^{(\infty)}(z,\hbar) := \frac{1}{\sqrt{\So(z,\hbar)}}
        \exp \Big( \pm  R(z;\hbar) \Big)
    \end{equation}
    where
    \begin{align}
    \label{defR}\nn
        R (z;\hbar) := &
        \frac{1}{\hbar}\lim_{p \to \infty} \left[ \int_{p}^{z} \sqrt{Q(\z)} \dd \z
        - \left(\frac{p^3}{3}+\frac{s}{2}p + \log p\right)\right] 
         + \sum_{j\geq 0} \hbar ^{2j+1} \int_{\infty}^z S_{2j+1}(\z)\d \z
        \\
         =& \frac 1 \hbar \le(\frac{z^3}{3} + \frac{s}{2}z + \log(z)\ri) + \bigO(z^{-1})\bigO(\hbar), \quad (z \to \infty),
    \end{align}
\end{itemize}
where all logarithms are principal.  Notice that $  R (z;\hbar) $ is the  anti-derivative  of $\So(z,\hbar)$ that does not have a  constant term in the expansion as $|z|\to\infty$.
\end{definition}
\begin{remark}
{
The integral in the exponent of $\wkb{\tau}{\pm}$ is to be understood term-wise in each coefficient of the powers of $\hbar$.
Additionally, $\So(z,\hbar)$ 
is multivalued on $\C$ with branch points at the zeros
 $\Tau := \{\tau_0, \tau_1,\tau_2,\tau_3\}$
of $Q(z)$. Therefore the integral should be considered 
on the elliptic compact Riemann surface  $\overline{\Sigma}$  obtained from the affine curve 
\begin{equation}
\label{Rsurf} 
\Sigma  = \Big\{ (w,z) \in \C^2 : w^2 = Q(z;s,E)  \Big\}
\end{equation}
by adding two points $P_{\infty}^{\pm}$ at infinity.
The projection $\pi:\overline{\Sigma} \mapsto \overline{\C}   $ from the Riemann surface $\overline{\Sigma}$ to the extended complex plane $\overline{\C}$, 
 maps $(w,z)\mapsto z $. The projection $\pi$ realizes $\overline{\Sigma} $
as a double cover of $\overline{\C} $ ramified at the zeros  of $Q$ (turning points).  The pre-image of any point $z\in \overline{\C}$  are  the two points
$\pi^{-1}(z)=(z,\pm w)$  on the two sheets of the Riemann surface where the numbering of the sheets is such that $(z,w=\sqrt{Q(z)})$ belongs to the first sheet.
Choosing branch cuts and the first  sheet of the    Riemann surface, we can talk about the integral from a turning point $\tau$   to $z$ by defining:
\begin{equation} \label{eq:def-integral-branch-pt}
    \int_\tau^z \So(\z,\hbar) \dd \z := \frac{1}{2} \int_{\gamma(z)} \So(\z,\hbar) \dd \z
\end{equation}
where the contour $\gamma(z)$ lives in the Riemann surface $\overline{\Sigma}$ 
and goes from $\widehat{z}$ to $z$, 
where $\widehat{z}$ lies on the second sheet} and the second sheet is reached by encircling the branch point $\tau$.
This integral is a well defined convergent integral.
Finally, it can be shown  by induction that all the correction terms $S_{2j+1}(z)\d z$, $j\geq 0$ are differentials on the Riemann surface $\overline{\Sigma} $ with poles of increasing order at each of the  ramification points, but always without residue.
For more details see \cite{Iwaki_Nakanishi_2014}.
\end{remark}

It is useful to relate the periods of $S_1$ to the periods of $S_{-1}$. To this end we have the following 
\begin{proposition}
\label{propsublead}
    If $I({s},E)$ denotes the period of $S_{-1} = \sqrt{Q}$ along a closed contour $\gamma$, then the corresponding period of $S_{1}$ is
    \bea
    \oint_{\gamma} S_{1}(z) \dd z =\left( - \frac {\partial^2}{\partial {s} \partial E} - \frac {s} 6 \frac {\partial^2}{\partial E ^2}\right) I({s},E)
    \eea
\end{proposition}
\noindent{\bf Proof.}
Using \eqref{defQ} we see that 
\bea
\partial_E \sqrt{Q(z;{s},E )} = \frac 1{2 \sqrt{Q(z;{s},  E )}}.
\eea
Now, we can write (we drop the indication of the dependence on $x,{s},E$)
\bea
S_{1}  = \frac 1{48} \frac{Q''}{Q ^\frac 3 2} - \frac{5}{24} \left(\frac 1{\sqrt{Q}}\right)''.
\label{520}
\eea
The periods of the second term in \eqref{520} vanish because this gives an exact differential; the first term reads 
\bea
 \frac 1{48} \frac{Q''}{Q^\frac 3 2}  = \frac {6 z^2 + {s}}{24 Q^\frac 32}
 =-(6\partial_{s} + {s} \partial_E)\left(\frac {1}{12 Q^\frac 12}\right)
 =-(6\partial_{s} + {s} \partial_E)\partial_E \left(\frac {1}6 Q^\frac 12\right)
. \label{521}
\eea
Integrating \eqref{520} along $\gamma$ and using the identity \eqref{521} completes the proof.
\QED

\subsection{Stokes graphs and connection formul\ae.}
The WKB series are asymptotic to actual solutions of 
\eqref{eq:schrodinger_wkb_equation} in certain regions of the complex plane that we presently define. 
We start making by fixing a choice of $\sqrt{Q(z)}$ and introducing the notion of Stokes' graph.
\begin{definition}[Square root of $Q$]
\label{defBranch}
Choosing the branch cuts $\mathcal{B}$ of  $\sqrt {Q(z)}$ in the {\it finite part} of the complex plane,  the function $\sqrt {Q(z)}$  becomes single valued in the  complement of the branch cuts and we fix 
it in such a way that  $\sqrt {Q(z)} \sim z^2$ as $|z|\to\infty$.   This choice identifies   the first sheet of the Riemann surface $\overline{\Sigma}$.
The second sheet correspond to the choice  $\sqrt {Q(z)} \sim -z^2$ as $|z|\to\infty$. 
\end{definition}

To minimize confusion when performing integrations along the branch cuts we will make the following explicit definition.
\begin{definition}[Branch cut integration] \label{def:branch-orientation}
Let $\tau$ and $\wt \tau$ be two zeroes of $Q(z;s,E)$ joined by a branch cut.
We denote by 
\begin{equation}
    \int_{\tau}^{\wt \tau} \sqrt{Q(z_{+};s,E)}\dd z
\end{equation}
to be the integral of $\sqrt{Q(z;s,E)}$ along the $+$ side of the branch cut {\it  oriented from $\tau$ to $\wt \tau$}.
We denote with a minus sign $-$ the corresponding integral along the $-$ side of the branch cut. 
As usual, the $+$ and $-$ sides correspond to left side and right side of the oriented contour, respectively.
\end{definition}
Now we introduce the notion of Stokes' curve, which we will use to build Stokes' graphs.

\begin{definition}[Stokes' curve]
\label{DefStokescurve}
    A \textbf{Stokes curve} of the potential $Q(z)$ is a horizontal trajectory of 
    the quadratic differential $Q(z) \dd z^{2}$ where one of the end-points is a turning point.
    In other words, in a local coordinate $z$ of $\overline{\Sigma}$
    it is a curve emanating from a turning point $\tau$ and satisfying
    \begin{equation} \label{eq:stokes_curve}
       \Im \int_{\tau}^z \sqrt{Q (\z)} \dd \z = 0.
    \end{equation}
    The support of the curve is independent of choice of determination of $\sqrt{Q}$.
    Furthermore:
    \begin{enumerate}
    \item The \textbf{orientation of a Stokes curve} is defined by the direction in which $\Re \int_{\tau}^z \sqrt{Q (u)} \dd u $ is {\it increasing}. 
    
    \item The directions near  the point at  $\infty$ are indicated by  $\oplus$ or $\ominus$ if 
    the function $\Re \int_{\tau}^z \sqrt{Q (u)} \dd u$ is increasing or decreasing,
    respectively, along the corresponding Stokes curve.
    We say the Stokes curve is {\it oriented towards} $\oplus$ or {\it oriented away from} $\ominus$, respectively.
    \color{black}
\end{enumerate}
\end{definition}

\begin{figure}
\begin{minipage}{0.32\textwidth}
    \centering
    \includegraphics[width=1\textwidth]{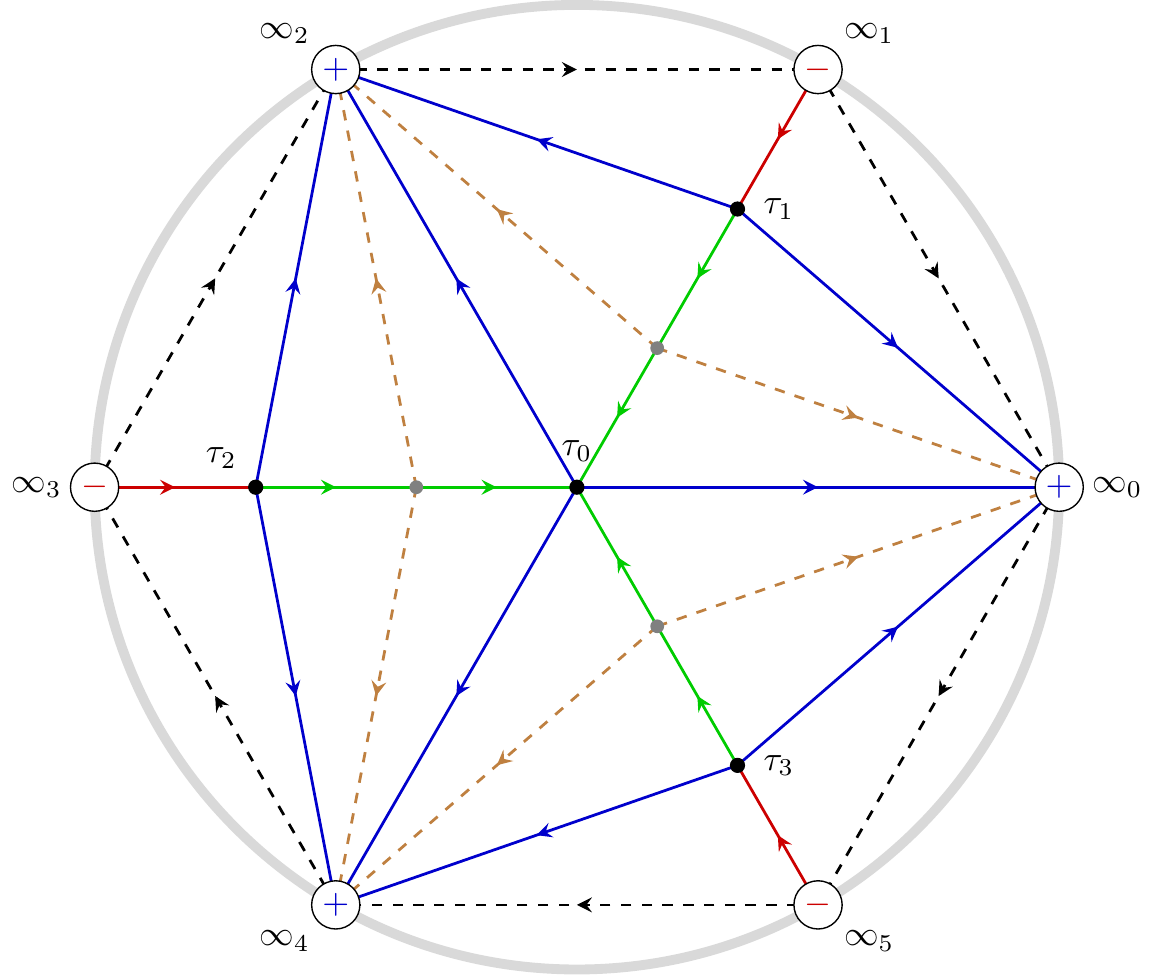}\\
        {(a) RHP for graphs of type $D$.}
\end{minipage}%
\hfill
\begin{minipage}{0.32\textwidth}
    \centering
    \includegraphics[width=1\textwidth]{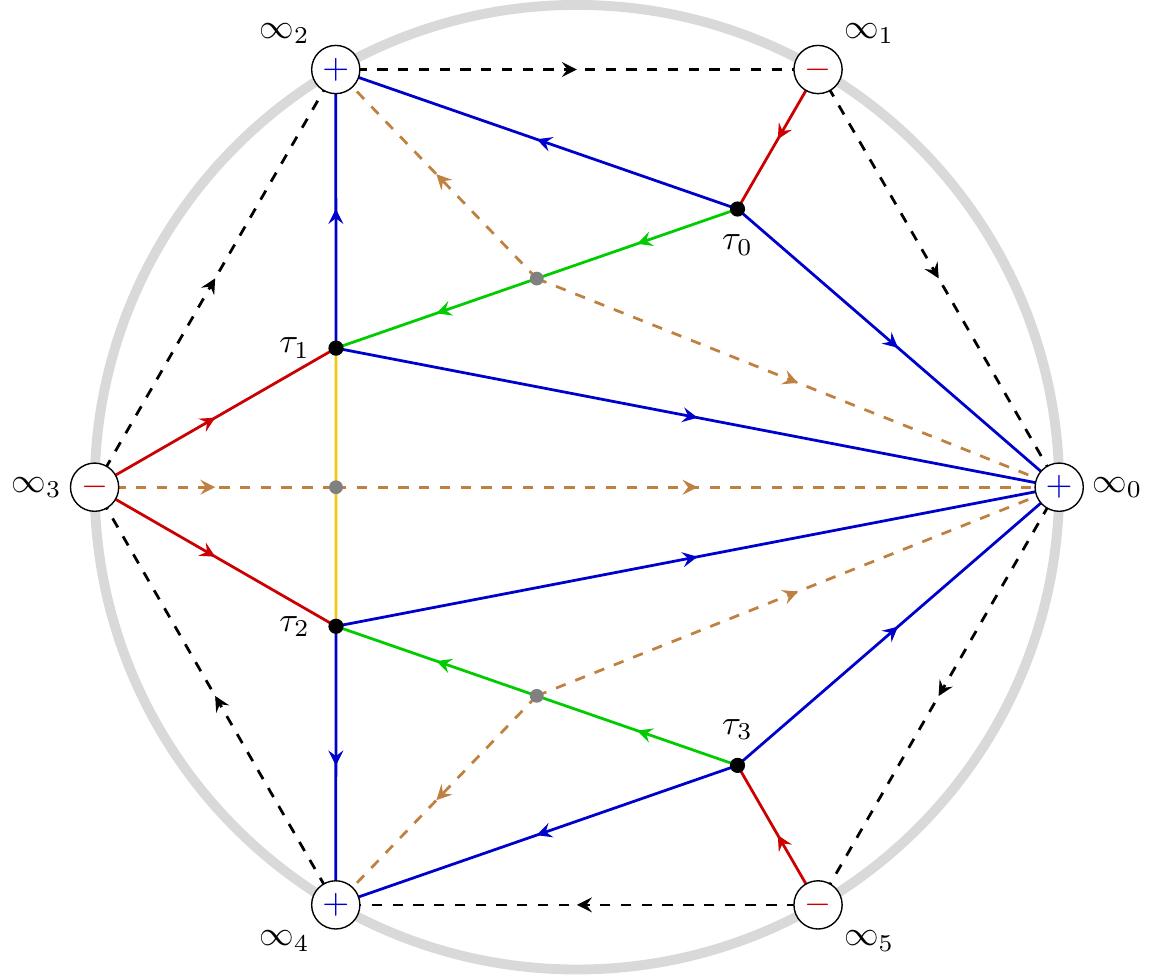}\\
    {(b) RHP for graphs of type $E$.}
\end{minipage}
\hfill
\begin{minipage}{0.32\textwidth}
    \centering
    \includegraphics[width=1\textwidth]{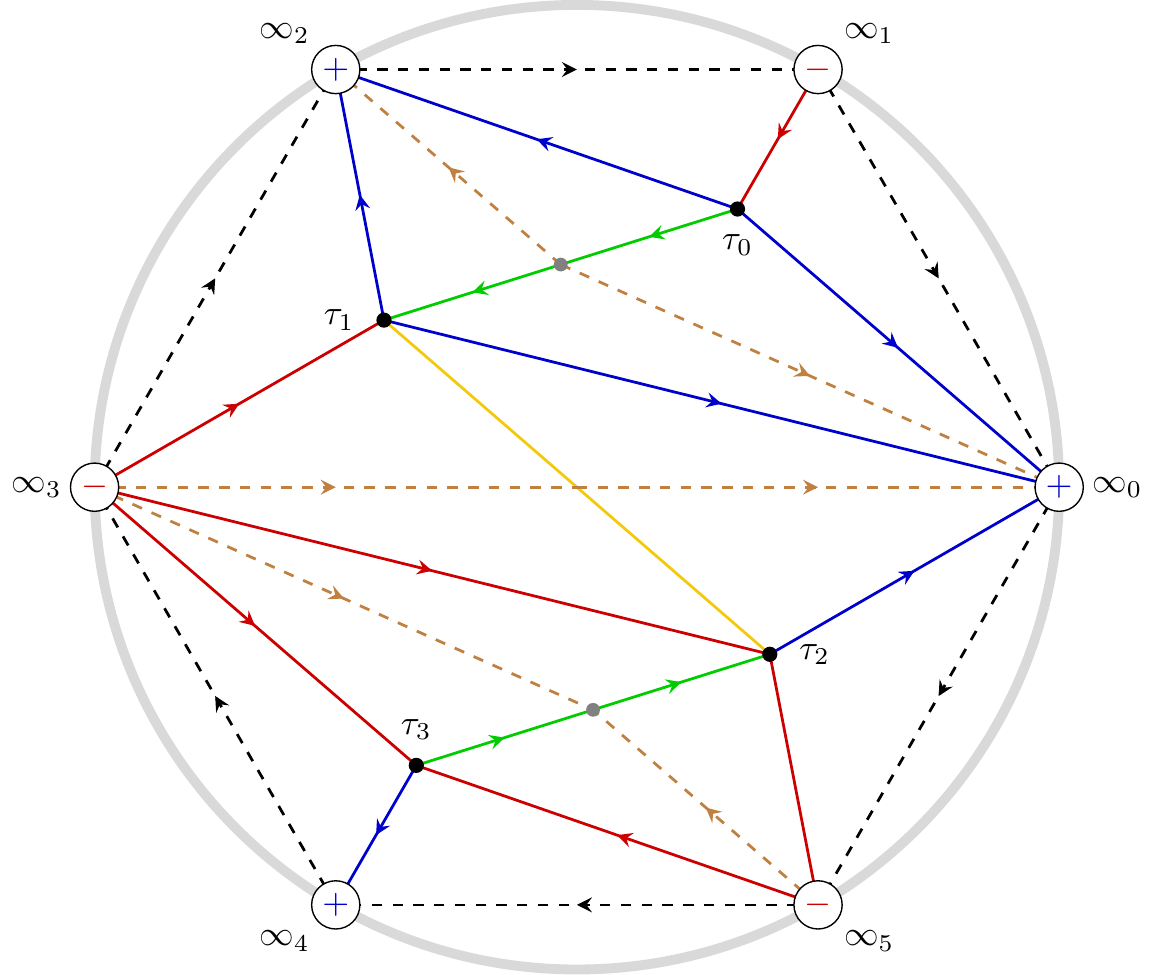}\\
        {(c) RHP for graphs of type $Z$.}
\end{minipage}
\caption{Generic WKB Riemann-Hilbert problem corresponding to each Stokes' graph configuration of the quartic polynomial potential $Q(z)$.}
\label{fig:generic_RHP_configurations}
\end{figure}

\begin{definition}[Stokes graph]
\label{DefStokesGraph}
       The \textbf{Stokes' graph} $\mathcal{G}$
    associated to the potential $Q(z)$ is the   graph embedded in $\overline{\C} $ where the edges are the Stokes curves and
    the vertices are the turning points $\Tau = \{\tau_j\}$ of the potential and the directions near the point at infinity $\infty_j, j=0, \dots, 6$. The graph is oriented according to definition~\ref{DefStokescurve}.  { The Stokes curves are indicated  with  black lines in Fig. \ref{fig:generic_stokes_graph_configurations} and with blue and red lines in Fig.\ref{fig:generic_RHP_configurations}.}
   \end{definition}

\noindent Furthermore, we decorate the Stokes graph with the following additional contours.
\begin{enumerate}
    \item
    {\bf Branch cuts.}
    We draw additional paths called {\it branch-cuts} between all pairs of turning points in such a way that they do not intersect any Stokes curve.
    Their orientation  is fixed in an arbitrary way that we shall  specify in each case.
    These branch cuts are indicated by green lines in Fig.~\ref{fig:generic_stokes_graph_configurations} 
    for each possible Stokes graph configuration.
    
    \item
    {\bf Ideal paths.}
    We draw  arbitrary (smooth) paths connecting the different  $\infty$ in all possible ways that do not intersect any of the Stokes curves.
    These paths are indicated by dashed lines in Fig. \ref{fig:generic_stokes_graph_configurations} and determine an {\it ideal triangulation} of an ideal hexagon.
    We call these paths {\it ideal paths} and the resulting partition of the plane the ideal triangulation.
    We orient  the {\bf outer }ideal paths that form the hexagon in the clockwise way (see Fig.~\ref{fig:generic_RHP_configurations}), while the remaining {\bf inner} ideal paths are oriented 
     towards the $\oplus$ directions.
    Note that the ideal paths separating Stokes regions intersect one and only one branch-cut, and the orientation of an ideal path on the two sides of a branch cut is opposite.
    Refer to  Fig.~\ref{fig:generic_RHP_configurations}.
   
    \item
    {\bf Stokes' regions and external regions.}
    Consider the  connected components of the complement of the Stokes graph,  the ideal paths, and the branch-cuts.
    Amongst them the components  that have at least one Stokes curve on the boundary will be called {\it Stokes' regions}.
    The remaining ones (unbonded regions bounded by a  ideal path) will be called {\it external regions}
    \footnote{These are the regions on the outside of the ideal ``hexagon'' in Fig. \ref{fig:generic_stokes_graph_configurations} and Fig. \ref{fig:generic_RHP_configurations}.}.
\end{enumerate}

\color{black}

The construction in Def. \ref{DefStokesGraph}, under the Assumptions~\ref{assumption-stokes-graph}, is crafted so that each Stokes region $\scr D$ has precisely one and only one turning point on its boundary. 


In each Stokes   region $\scr D$ one can select a fundamental  basis of solutions of the ODE \eqref{eq:schrodinger_wkb_equation} as we now explain.
Consider a Stokes curve $\gamma$ originating at the turning point $\tau$ and
oriented towards $\oplus$  
(with similar considerations for the $\ominus$  curves ): 
since the real part of 
$\int^z_\tau \sqrt{Q(u)}du$ 
is increasing, there is a region  around $\gamma$ near $\infty$ where the real part is positive; then the formal solution $\wkb{\tau}{-}$   is {\it recessive} (i.e. exponentially small as $\hbar \to0_+$).
  Then there is a unique solution $\Psi(z;\hbar)$ which is asymptotic to this  $\wkb{\tau}{-}$ in both Stokes regions on the two sides of $\gamma$. If $\scr D$ is one of these regions, we will denote this first selected solution by 
  $\Psi^{(\scr D)}_-$. To uniquely determine the other solution $\Psi^{(\scr D)}_+$, it is not sufficient to examine its asymptotic behaviour near $\gamma$ because its asymptotics there is {\it dominant} (i.e. exponentially large  as $\hbar \to 0_+$).
  However, the same Stokes region must be bounded also by either a $\ominus$ trajectory or a branch-cut. In the former case, in a neighbourhood of the $\ominus$ trajectory the  formal solution $\wkb{\tau}{+}$ is now recessive and this allows to uniquely determine $\Psi^{(\scr D)}_+$. 

If, instead,  the other boundary is a branch-cut, we need to consider the Stokes region, $\scr D'$ on the other side of the cut: in this region, due to having crossed the branch-cut, the formal solution $\wkb{\tau}{+}$ is now somewhere recessive and this allows to fix $\Psi^{(\scr D)}_+$ uniquely. We refer to \cite{Berto-Kor-WKB}, Section 5 for more details. 
The following theorem of V\"oros \cite{voros_return_quartic_oscillator}, \cite{kawai_takei_algebraic_analysis_sing_perturbation} 
relates the WKB solutions of different Stokes regions near the same turning point.

\begin{theorem}[ \cite{voros_return_quartic_oscillator}, \cite{kawai_takei_algebraic_analysis_sing_perturbation}] \label{thm:voros_connection_formula}
Let $\scr D$ be a Stokes region.
Then there exist   unique solutions $\Psi^{(\scr D)}_\pm$,  of \eqref{eq:schrodinger_wkb_equation},  that we refer to as {\bf normalized solutions}, that  are asymptotic to the WKB solutions in Def.~\ref{def:wkb_sols_turning_pt} uniformly in $z$ in the Stokes region, that is:
\begin{equation}
    \Psi^{(\scr D)}_\pm (z) \sim \wkb{\tau}{\pm}(z;\hbar) \quad \hbar \to 0, \quad x \in \scr D.
\end{equation}

Furthermore, let $\scr D_\ell, \scr D_r$  be two adjacent Stokes regions separated by the  Stokes curve $\gamma$ oriented as in Def. \ref{DefStokesGraph} with $\scr D_\ell$ on the left and $\scr D_r$ on the right of $\gamma$.
Then the corresponding solutions $\Psi_\pm^{\scr D_{\ell}}, \Psi_\pm^{\scr D_{r}}$ are related by:
\begin{equation}
\begin{split}
   & \left[\Psi^{(\scr D_\ell)}_+(z), \Psi^{(\scr D_\ell)}_-(z) \right]  =    
    \left[\Psi^{(\scr D_r)}_+(z), \Psi^{(\scr D_r)}_-(z) \right] 
\begin{cases}
    B 
    \quad \text{ if $\gamma$ is oriented towards $\oplus$} \\[10pt]
    R
    \quad \text{ if  $\gamma$ is oriented away from $\ominus$},
\end{cases}\\
&B := \begin{bmatrix}
    1 & 0 \\
    -i & 1 
    \end{bmatrix},\quad R:= \begin{bmatrix}
    1 & i \\
    0 & 1 
    \end{bmatrix}
\end{split}
\end{equation}
\end{theorem}
The relationship between solutions  in regions separated by an ideal path is simply a different scaling as given by the following proposition.

\begin{proposition} \label{thm:connection_turning_pts}
Let $\scr D_\ell,\scr D_r$ be two Stokes regions separated by an ideal path $\sigma$, with $\scr D_\ell$ on the left and $\scr D_r$ on the right of $\sigma$ with the orientation given in Def. \ref{DefStokesGraph}.
Let $\tau_{\ell}, \tau_r$ be the (unique) turning points on the boundaries of $\scr{D}_{\ell}, \scr{D}_r$, respectively. Then we have the following connection formula
\begin{equation}
[\wkb{\tau_\ell}{+} , \wkb{\tau_\ell}{-}] = [\wkb{\tau_r}{+}, \wkb{\tau_r}{-}] 
\exp(\sigma_3 v_{\ell r})
\end{equation}
where
\begin{equation}
\label{Vorosellr}
v_{\ell r}=v_{\ell r}(\hbar) := \int_{\tau_\ell}^{\tau_r} \So(z_+,\hbar) \dd z 
\end{equation}
with the integration along the branch cut according to Def.~\ref{def:branch-orientation}.

The corresponding actual solutions $\Psi^{(\scr D_{\ell})}_\pm, \Psi^{(\scr D_{r})}_\pm$ given in Thm.~\ref{thm:voros_connection_formula} are similarly related:
\bea
 [\Psi^{(\scr D_\ell)}_+,\Psi^{(\scr D_\ell)}_-] = [\Psi^{(\scr D_r)}_+,\Psi^{(\scr D_r)}_-] 
    \exp(\sigma_3 \hat v_{\ell r})
\eea
where now $\hat v_{\ell r}(\hbar)$ is a function of $\hbar$ that is asymptotic, in the Poincar\'e\ sense, to $v_{\ell r}(\hbar)$ in \eqref{Vorosellr}.
\end{proposition}

A similar proposition can be stated for the relation between a Stokes region $\scr D$ and an external region $\scr B$. Indeed, if $\tau$ is the turning point on the boundary of $\scr D$, we have 
\begin{equation} \label{eq:connection-infinity}
    [\wkb{\infty}{+}, \wkb{\infty}{-}] = [\wkb{\tau}{+}, \wkb{\tau}{-}]
    \exp \left( \sigma_3 w_\tau\right) 
\end{equation}
where $\tau$ is a turning point in the boundary of a Stokes region extending to infinity, and
 \begin{equation} \label{eq:w-Voros symbol}
        w_\tau(\hbar) := R(z;\hbar) - \int_\tau^z\So(\z,\hbar) \dd \z
 \end{equation}
 where $R(z;\hbar)$ is defined in \eqref{defR}.  {Note that $ w_\tau(\hbar)$  is a constant with respect to $z$  since $\dfrac{\dd}{\dd z}w_\tau(\hbar)=0$}
 


In the  case of the ODE \eqref{eq:schrodinger_wkb_equation} we will work under the assumption that the turning points are simple and there are no ``saddle trajectories'' as specified below.

\begin{assumption} \label{assumption-stokes-graph}
The following assumptions shall prevail.
 \begin{itemize}
    \item 
    \textit{Simplicity.} 
    The roots of the potential are all simple.
    In the case of the potential $Q(z;s,E)$, there are no repeated roots if and only if $(s,E)$ satisfy:
    \begin{equation}
    \label{discrimQ}
        E s^4 - 8 E^2 s^2 + 16 E^3 - s^3 + 36 E s - 27 \neq 0.
    \end{equation}    
    \item 
    \textit{Genericity.} 
    There are no saddle trajectories i.e. there are no Stokes' curves connecting two turning points.
    Saddle trajectories can only occur if there is  $\gamma $ in the homology group   of the Riemann surface $\overline{\Sigma}$  for which  
    \begin{equation}
        \Im \oint_\gamma \sqrt{Q(z)} \dd z = 0. 
    \end{equation}
   
\end{itemize}
\end{assumption}
The simplicity assumption means there are exactly three Stokes' curves emanating from each turning point.
The genericity assumption implies that all the Stokes' curves must extend to $\infty$. 
With these assumptions we can classify all the possible Stokes graphs.

In what follows we draw the trajectories on one sheet of the Riemann surface $\Sigma$ relative to the choice of branch as in Def. \ref{defBranch}.
\begin{proposition}[Classification of generic Stokes graph]
Under the Assumptions~\ref{assumption-stokes-graph} and with the choice of
branch cuts in Definition~\ref{defBranch}, the Stokes graphs are in one-to-one 
correspondence with the triangulations of the hexagon, so there are 14 such 
configurations. 
Three of them are topologically distinct (as graphs), they are depicted in 
Fig. \ref{fig:generic_stokes_graph_configurations} and named $E, D$ and  $Z$.
The remaining configurations can be obtained from the graphs of types $E, D$ and $Z$ shown in 
Fig.~\ref{fig:generic_stokes_graph_configurations}
by a $\Z_6$ rotation and by a reflection along the line from $\infty_3$ to $\infty_0 $.

\begin{figure}
\begin{minipage}{0.32\textwidth}
    \centering
    \includegraphics[width=1\textwidth]{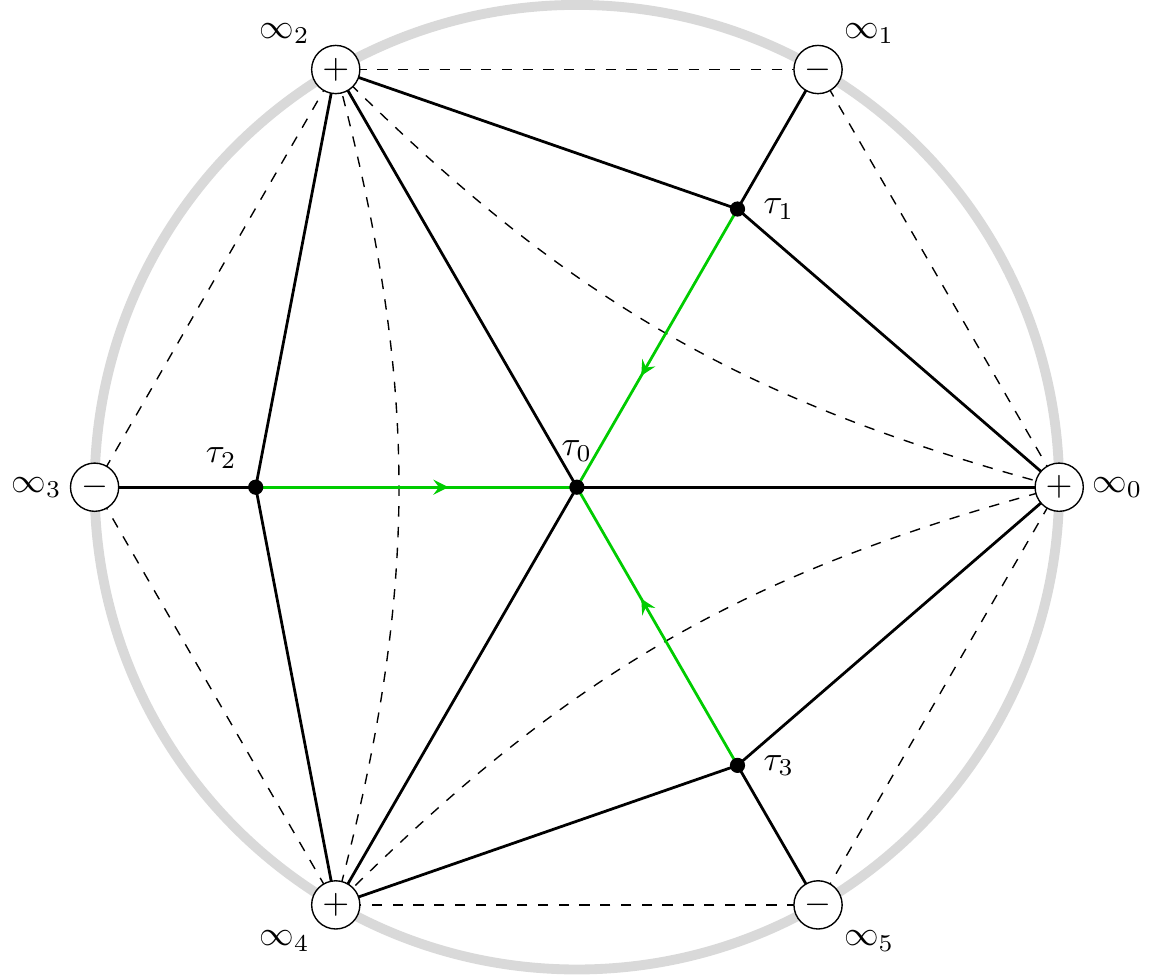}\\ 
    {(a) Graph of type $D$}
\end{minipage}
\begin{minipage}{0.32\textwidth}
    \centering
    \includegraphics[width=1\textwidth]{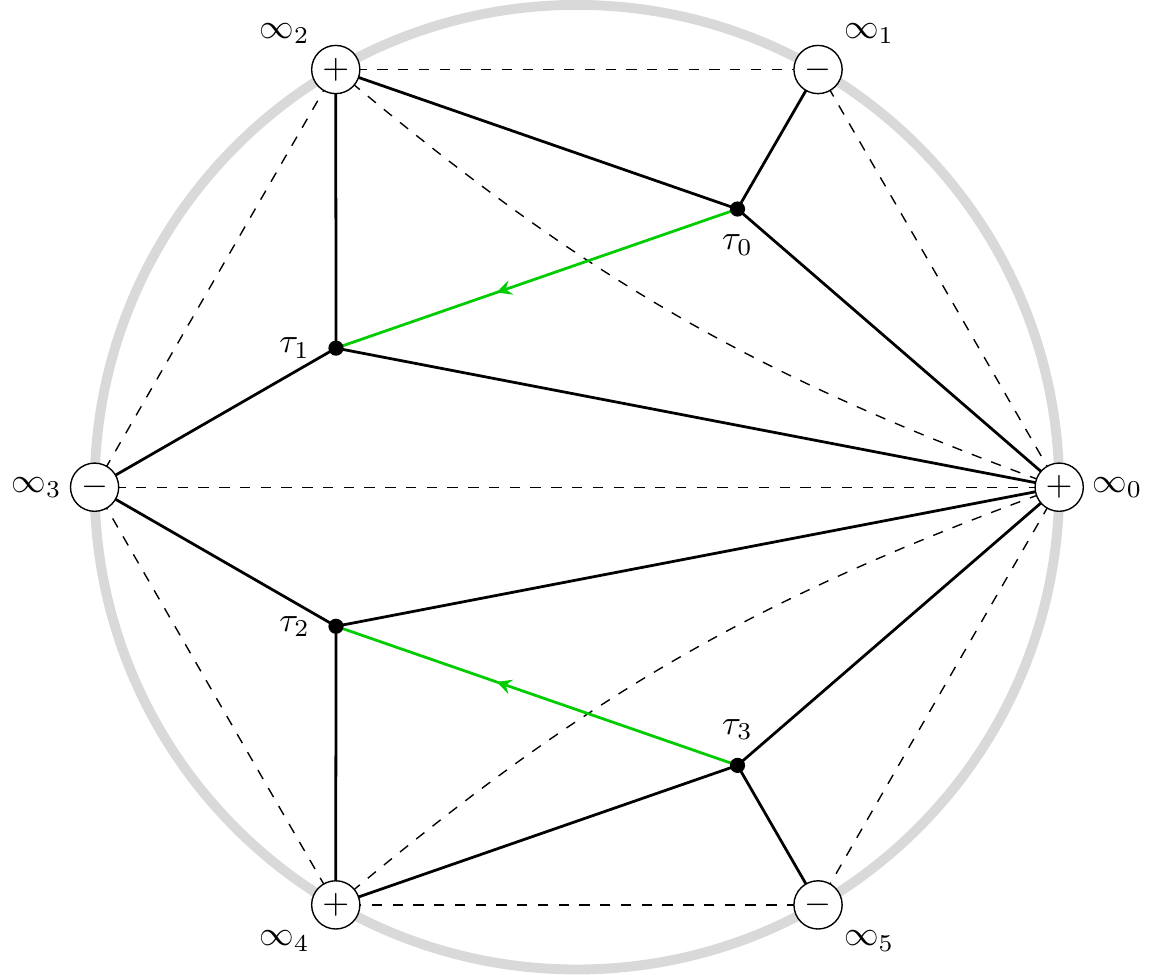}\\
    {(b) Graph of type $E$}
\end{minipage}
\hfill
\begin{minipage}{0.32\textwidth}
    \centering
    \includegraphics[width=1\textwidth]{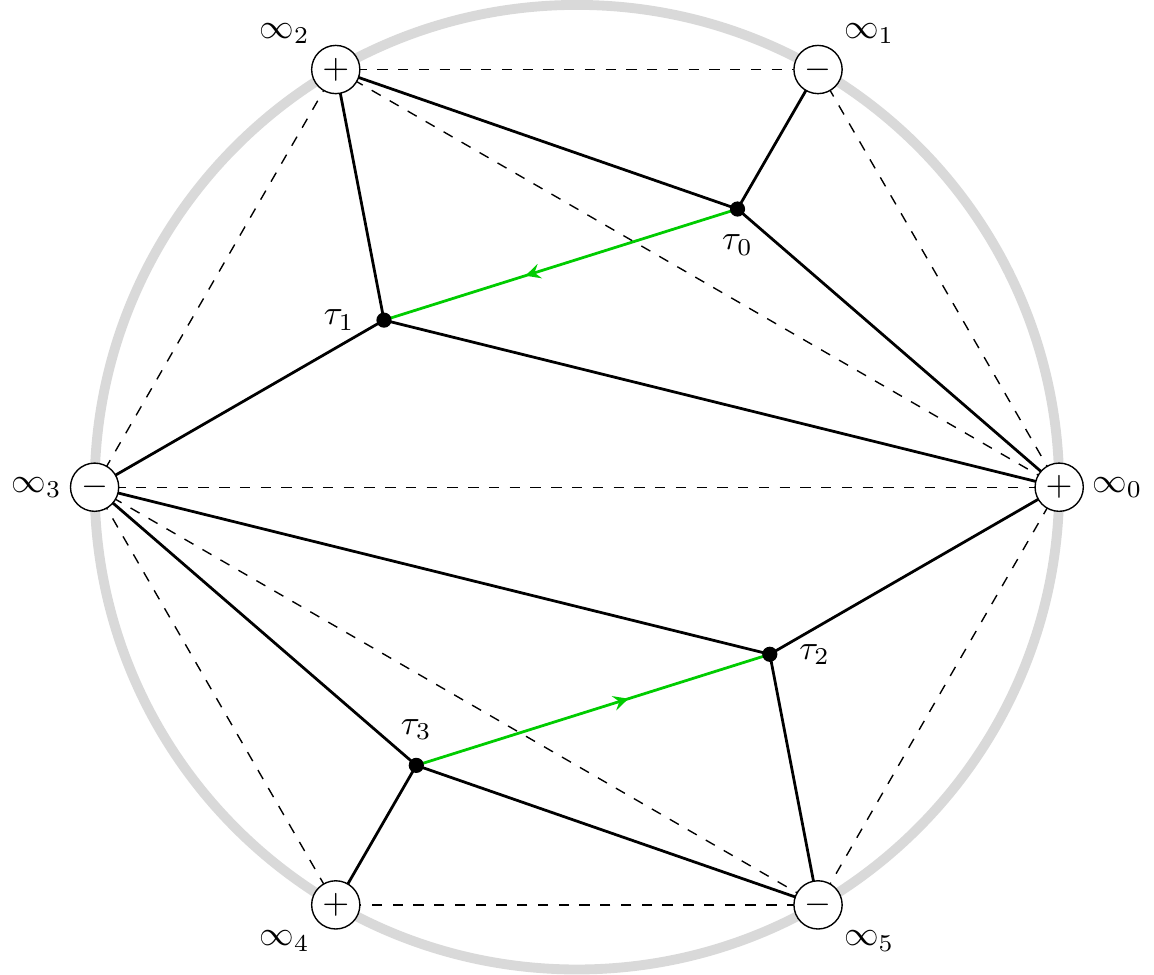}
    {(c) Graph of type $Z$}
\end{minipage}
\caption{Generic Stokes graph configurations for a quartic polynomial $Q(z)$. Solid lines depict the Stokes curves emanating from the turning points $\tau_j$, dashed lines denote the triangulation of the hexagon and green curvy lines correspond to our choice of branch cuts. We remark that for various $(s,E)\in \C^2$ the potential $Q(z;s,E)$ may have a different different Stokes \textit{lines} in $\overline{\C}$ but the underlying Stokes graph will be topologically identical to one of the types depicted.}

\label{fig:generic_stokes_graph_configurations}
\end{figure}

\end{proposition}

\noindent{\bf Proof.}
Assuming simplicity, from each turning point there are exactly three Stokes curves emanating
from it. These Stokes curves end either at infinity or at another turning point.
Assuming genericity the latter cannot happen, therefore the Stokes curves determine an
ideal triangulation of the Riemann sphere  with a small disk around infinity removed and with 6 marked points in the boundary (corresponding to the asymptotic directions at infinity). These triangulations correspond to triangulations of the hexagon, and there  are 14 such triangulations.

Indeed, we can view these 6 marked points as determining a (topological) hexagon. Furthermore, 
each of the turning points $\tau_j$ determines a triangle inside the hexagon 
by connecting the asymptotic directions at infinity with a   Stokes curve originating  from $\tau_j$. Up to rotations and reflections of the hexagon there are 3 distinct such triangulations,  named $E$, $D$ and $Z$  as depicted in Fig.~\ref{fig:generic_stokes_graph_configurations}.
The 14 possible configuration can be obtained as follows.
From configuration $D$ we obtain one other configuration rotating 
by $2 \pi/6$. From configuration $Z$ we obtain 6 distinct configurations, 3 of them corresponding to a $\Z_3$ (i.e. $2\pi/3$) rotation, and another 3 corresponding to a reflection followed by a $\Z_3$ rotation. Finally from configuration $E_+$ we obtain
6 distinct configurations corresponding to a $\Z_6$ (i.e. $2\pi /6$) rotation. 
Finally,  $2 + 3 + 3 + 6 = 14$ as claimed.
\QED

Finally we can coordinate the Theorem \ref{thm:voros_connection_formula} and Proposition \ref{thm:connection_turning_pts}  into a Riemann Hilbert problem as follows.

\begin{rhp}[Quartic WKB jumps]
\label{RHPWKB}
    Fix a quartic potential $Q$ and suppose that it has one of the generic Stokes graph 
    configuration from Fig. \ref{fig:generic_stokes_graph_configurations}.
    The Riemann-Hilbert problem  for the vector valued function $  \Psi$  such that  $  \Psi|_{\scr D}= (\Psi^{(\scr D)}_+, \Psi^{(\scr D)}_-)$    with  $  \Psi|_{\scr D}$ defined in Theorem~\ref{thm:voros_connection_formula},  consists of the following oriented contours in $\overline{\C}$ with their associated jump matrices.
\begin{enumerate}
    \item \textbf{Square root branch cuts}:
    To each square-root branch cut (coloured green in Fig. 
    \ref{fig:generic_RHP_configurations}
), with the orientation indicated in  Fig. \ref{fig:generic_RHP_configurations}, we  associate the jump matrix
    \begin{equation}
            G := \begin{bmatrix} 0 & i \\ i & 0 \end{bmatrix}.
    \end{equation}
  
    \item \textbf{Fourth-root branch cuts}:
    in configurations $E$ and $Z$ there is an extra jump contour corresponding to
    the fact that 
    \begin{equation}
        \sqrt{\So(z,\hbar)}  \sim h^{-1/2}Q (z)^{1/4} + \bigO(\hbar^{1/2}).
    \end{equation}
    To each fourth-root branch cut we associate the jump matrix $Y = -\mathbf 1$. 
    The fourth-root branch cuts are coloured in yellow in Fig. \ref{fig:generic_RHP_configurations}. Note there is no need to specify the orientation of these contours.

    \item \textbf{Stokes curves}: 
    along each Stokes curve oriented  towards $\oplus$  and away from $\ominus$ we assign the jump matrices $B$ and $R$ (respectively)
    \begin{align}
        B :=& \begin{bmatrix} 1 & 0 \\ -i & 1 \end{bmatrix}, \qquad
        R := \begin{bmatrix} 1 & i \\  0 & 1 \end{bmatrix}.
    \end{align}
    The Stokes curves are coloured blue or red, respectively, in Fig.
    \ref{fig:generic_RHP_configurations}.

    \item
    \textbf{Inner ideal paths:} 
    along the inner ideal paths with the orientation in Def. \ref{DefStokesGraph} (and indicated in Fig. \ref{fig:generic_RHP_configurations})  we associate the following jump matrix corresponding to the connection formula in
    Prop.~\ref{thm:connection_turning_pts}: 
    \begin{equation} \label{eq:voros_jumps_turning_pts}  
        V_{jk} := 
        \exp \left( \sigma_3 v_{jk} \right) = 
        \begin{bmatrix} e^{v_{jk}} & 0 \\ 0 & e^{-v_{jk}} \end{bmatrix}, 
        \quad v_{jk}(\hbar) =\int_{\tau_j}^{\tau_k} \So(z_{\color{black}+},\hbar) \dd z    
    \end{equation}
    where $j,k \in \{0,1,2,3\}$  and $\tau_j, \tau_k$ are turning points.
    These contours are denoted by a brown dashed line in
    Fig. \ref{fig:generic_RHP_configurations}.

    \item \textbf{Outer ideal paths:} along the outer ideal paths separating the external regions from the Stokes regions we associate the  jump matrix corresponding to the connection formula \eqref{eq:connection-infinity} between turning points and infinity:    
\begin{equation} \label{eq:voros_jumps_infinity}
        W_{j}:= \exp(\sigma_3 w_j) = 
        \begin{bmatrix} e^{w_j} & 0 \\ 0 & e^{-w_j} \end{bmatrix}, \quad w_j(\hbar) := R(z;\hbar) - \int_{\tau_j}^z \So(\z,\hbar) \dd \z.
    \end{equation}
    where $j \in \{0,1,2,3\}$, $\tau_j$ is a turning point and $R(z;\hbar)$ is the particular antiderivative in \eqref{defR}. These contours are denoted by a black dashed line in
    Fig. \ref{fig:generic_RHP_configurations}.
\end{enumerate}
\end{rhp}
\begin{remark}
We observe that $w_j(\hbar)$ is a constant in $x$. It can be thought as the regularized   integral $\int^{\tau_j}_{\infty} \So(z,\hbar) \dd z.$
\end{remark}

This construction gives us, up to rotation and reflections, three distinct WKB Riemann-Hilbert problems as shown in Fig. \ref{fig:generic_RHP_configurations}.

In order to simplify the upcoming computations we make some notational definitions.
Given turning points $\tau_j$ and $\tau_k$ connected by a branch cut we denote
\begin{equation} \label{eq:xi-symbol-definition}
    \xi_{jk} := \exp(2v_{jk}) =     \exp\left(2\int_{\tau_j}^{\tau_k} \So(z_+,\hbar) \dd z\right),
\end{equation}
with the determination of $\So$  given Def. \ref{defBranch}, and the boundary value $z_+$ is  in accordance to Def.~\ref{def:branch-orientation}.
If the two branchpoints are not connected by a branch-cut we take the integration on the main sheet (i.e. with the determination as above). We call the above parameters the {\it Fock--Goncharov} parameters.

Additionally we will use the same notation to indicate the ``exact'' Fock--Goncharov parameters, namely, the result of the Borel resummation. To phrase it differently we will not distinguish in the notation the Borel resummation by its asymptotic expansion in $\hbar$. 
Now we obtain the following results about the Stokes matrices for each configuration shown in Fig.
\ref{fig:generic_RHP_configurations}.

\begin{theorem} \label{thm:Stokes_matrices_computation}
The Stokes matrices 
\bea
\S_{j} =& \begin{bmatrix} 1 & 0   \\ s_{j} & 1 \end{bmatrix},\;\;j=0,2,4,\hspace{9mm}
 \S_{j} = \begin{bmatrix} 1 & s_{j} \\ 0   & 1 \end{bmatrix},\;\;\;j=1,5,\hspace{9mm}
  \S_{3} = \begin{bmatrix}1 & s_{3} \\ 0   & 1 \end{bmatrix}  {\rm e}^{\frac {2i\pi}\hbar\sigma_3},
\eea
in each of the WKB Riemann-Hilbert problems in Fig. \ref{fig:generic_RHP_configurations}  are expressed  in terms of the contour integrals 
$v_{jk}$, $w_j$ and $\xi_{jk}$ as defined in \eqref{eq:voros_jumps_turning_pts}, \eqref{eq:voros_jumps_infinity} and \eqref{eq:xi-symbol-definition} as follows:
\bea
\begin{array}{c|c|c}
 \hbox{Configuration $D$}  & \hbox{Configuration $E$}  & \hbox{Configuration $Z$} \\[4pt]
\begin{array}{l} 
              s_0 = -ie^{2 w_1} (\xi_{10} \xi_{30} + \xi_{10} +1), \nn\\ [4pt]
             s_1 = -i e^{-2w_1}                                   \nn\\[4pt]
             s_2 = -ie^{2w_2} (\xi_{20} \xi_{10} + \xi_{20} +1), \nn \\[4pt]
             s_3 =-i e^{-2w_2}                             \nn \\[4pt]
            s_4 = -ie^{2w_3} (\xi_{30} \xi_{20} + \xi_{30} +1),\nn\\ [4pt]
            s_5 =-i e^{-2w_3}.
\end{array}
&
\begin{array} {l}
             s_0 = -ie^{2 w_0} (\xi_{01} \xi_{12} \xi_{32} + \xi_{01}\xi_{12} + \xi_{01} + 1),\nn \\[4pt] 
            s_1 = -i e^{-2w_0}          \nn   \\[4pt]
             s_2 =-ie^{2w_1} (\xi_{01} + 1),     \nn    \\[4pt]
             s_3 = -i e^{-2w_2} (\xi_{12} +1)   \nn         \\[4pt]
          s_4 = -ie^{2w_3} (\xi_{32}+1),       \nn       \\[4pt]
            s_5 = -i e^{-2w_3}.
\end{array}
&
\begin{array}{l} 
            s_0 = -ie^{2 w_0} (\xi_{01} \xi_{12} + \xi_{01}+ 1),   \nn  \\[4pt] 
           s_1 = -i e^{-2w_0}                                    \nn   \\[4pt]
            s_2 = -ie^{2w_1} (\xi_{01} + 1),                     \nn    \\[4pt]
          s_3 =-i e^{-2w_3} (\xi_{32} \xi_{12} + \xi_{32} +1)  \nn   \\[4pt]
             s_4 = -i e^{2w_3}                                   \nn     \\[4pt]
             s_5 = -ie^{-2w_2} (\xi_{32}+1).
\end{array}
\end{array}
\eea
In the configuration $D$ the following {identities}  hold 
\bea 
{\rm e}^{2(w_3-w_1)}  = \xi_{10} \xi_{30}, \ \ \ 
{\rm e}^{2(w_1-w_2)} = \xi_{10}\xi_{20},\ \ \ 
{\xi_{10} \xi_{20}\xi_{30}} = {\rm e}^{\frac{2i\pi} \hbar}.
\label{extra}
\eea 
{\color{black}
In configuration $E$ and $Z$ the following identity holds
\begin{equation}
\label{extra1}
\xi_{01} \xi_{23}= {\rm e}^{\frac{2i\pi} \hbar}.
\end{equation}
}
\end{theorem}
\noindent{\bf Proof.}
We compute the Stokes matrices associated to Configuration $D$ of the Riemann-Hilbert problem  in  Fig. \ref{fig:generic_RHP_configurations}.
Consider the vertex $\infty_0$ in Configuration D, there are seven edges incident on it. 
The Stokes matrix $\S_0$ is given by the
clockwise (about the vertex $\infty_0$) product of all the jump matrices  corresponding to the  edges incident to $\infty_0$ as follows
\begin{equation}
    \S_0 = W_{3}^{-1} B V_{30} B V_{10} B W_1 
\end{equation}
Notice that the jump matrix $V_{30} $ follows from the fact that  we have   
$$\exp\left(\int_{\tau_3}^{\tau_0} \So(z_+,\hbar) \dd z\right)=\exp\left(\int_{\tau_0}^{\tau_3} \So(z_-,\hbar) \dd z\right)  ,$$
where $z_{\pm}$ are the boundary values of $ \So(z,\hbar)$ on the left and right side of the segment  $ [\tau_0,\tau_3]$ oriented from $\tau_3$ to $\tau_0$.
Therefore we have 
\[
   \S_0 
    = \exp\left[\sigma_3 (w_1 - w_3 + v_{10} + v_{30})\right]\ \begin{bmatrix} 1 & 0 \\ -ie^{2 w_1} (\xi_{10} \xi_{30} + \xi_{10} +1) & 1\end{bmatrix}.
\]
We observe that 
\begin{equation}
\begin{split}
    w_1 - w_3 &= 
    - \int_{\tau_1}^{z} \So(\z,\hbar) \dd \z + \int_{\tau_3}^{z} \So(\z,\hbar) \dd \z \\
    &=
    { - \int_{\tau_1}^{\tau_0} \So(\z_+,\hbar) \dd \z  - \int_{\tau_0}^{z} \So(\z,\hbar) \dd \z+
      \int_{\tau_3}^{\tau_0} \So(\z_-,\hbar)\dd \z+\int_{\tau_0}^{z} \So(\z,\hbar)}\dd \z\\
    &={-\int_{\tau_3}^{\tau_0} \So(\z_+,\hbar) \dd \z-\int_{\tau_1}^{\tau_0} \So(\z_+,\hbar) \dd \z}
    \end{split}
\end{equation}
and 
\begin{equation}
    v_{10}+v_{30} = \int_{\tau_1}^{\tau_0} \So(z_+,\hbar) \dd z + \int_{\tau_3}^{\tau_0} \So(z_+,\hbar) \dd z
\end{equation}
where  the integral from $\tau_1$ to $\tau_3$ is with the determination  in Def.\ref{defBranch}. 
Therefore we obtain
\begin{equation}
    w_1 - w_3 + v_{10} + v_{30} =0,
\end{equation} 
so that
\begin{equation}
    \exp\left( \sigma_3 [w_1 -w_3 + v_{10} + v_{30}]\right) =\mathbb{I},
\end{equation}
{\color{black} which is equivalent to the first  identity in \eqref{extra}}.
Similar calculations compute the remaining Stokes matrices for each possible configuration.
The extra ${\rm e}^{2i\pi \hbar^{-1} \sigma_3}$ in the form of $\mathbb S_3$ is due to our choice of branch-cut for the logarithm in \eqref{defR}.
The fact that the product of all Stokes matrices is trivial follows by construction.

{ The second  equation in \eqref{extra} for configuration $D$ follows from the definitions of $\xi_{ij}$ in \eqref{eq:xi-symbol-definition} and $w_j$ in \eqref{eq:voros_jumps_infinity}. The last equation  follows from the fact   that $\gamma_{10}+\gamma_{20}+\gamma_{30}$ is homologous to $\Gamma_\infty$, the contour at infinity with winding number equal to one. The residue at infinity of the (formal series in $\hbar$) $\So(z,\hbar)\dd z$  is  $2i\pi/\hbar$   coming from the leading term of the series only (it is shown in \cite{kawai_takei_algebraic_analysis_sing_perturbation}, which can be also verified by induction,  that all the higher  terms in the $\hbar$--series expansion of $\So$ vanish to order $\mathcal O(z^{-2})$ and hence do not contribute to the residue). 
Regarding the relation \eqref{extra1}  since the Stokes graph of $E$ and $Z$ have the same topology of branch points, we consider both cases simultaneously.  One verifies that the contour $\gamma_{01}+\gamma_{23}$ is homologous  to
 $\Gamma_\infty$. Thus  in a similar way as the previous case
 \[
\left( \oint_{\gamma_{01}}+\oint_{\gamma_{23}}\right)\So(z,\hbar)\d z=\Res_{z=\infty}\hbar^{-1}\sqrt{Q(z,s,e)}dz=\dfrac{2\pi i}{\hbar}.
 \]
   }
\QED         
\paragraph{Gauge arbitrariness.}
In terms of the Stokes phenomenon, we must point out that the Stokes parameters $s_0,...,s_5$ are not intrinsically defined because we can conjugate  the fundamental matrix  by an arbitrary diagonal matrix. This freedom translates to the following scaling equivalence for the Stokes parameters
\bea
\label{gauge}
s_{2j+1} \mapsto \lambda s_{2j+1},\ \ \ s_{2j}\mapsto \lambda^{-1}s_{2j},\ \ \ \ \lambda \in \C^\times.
\eea
Using this freedom we can rewrite (in all cases) the Stokes parameters in such a way that only the Vor\"os' symbols $v_{jk}$ enter. For example, in the $D$ configuration (which will be the only one that is relevant later on), we have 
{
        \begin{align} 
              s_0 =& i (\xi_{10} \xi_{30} + \xi_{10} +1), \nn\qquad \qquad
            s_1 = i                                   \\
        s_2 =& ie^{2w_2-2w_1} (\xi_{20} \xi_{10} + \xi_{20} +1) =i\frac{1}{\xi_{20}\xi_{10}} (\xi_{20} \xi_{10} + \xi_{20} +1)  ,  \nn\\\nn
         s_3 =&i e^{-2w_2+2w_1} = i \xi_{10}\xi_{20}       \\\nn
          s_4 =& ie^{2w_3-2w_1} (\xi_{30} \xi_{20} + \xi_{30} +1) = i \xi_{30}\xi_{10} (\xi_{30} \xi_{20} + \xi_{30} +1)   \\
          s_5 =&i e^{-2w_3+2w_1} = i \frac{1}{\xi_{10}\xi_{30}}.
          \label{ConfDstokes}
        \end{align} 
}

\section{Quantization conditions}
In this section we derive ``exact'' quantisation conditions for both the zeroes of Vorob'ev--Yablonskii polynomials (the Jimbo-Miwa case) and for the points of the ES spectrum corresponding to repeated eigenvalues (the Shapiro-Tater case).
To leading order these quantisation conditions yield a system that describes both sets of points in terms of contour integrals of $\sqrt{Q(z;s,E)}$.

In the Jimbo-Miwa case this is achieved by matching the Stokes' phenomenon of the Jimbo-Miwa Lax pair representation of \ref{eq:painleve2} with the Stokes' phenomenon of the quartic WKB Riemann-Hilbert problem \ref{thm:Stokes_matrices_computation}. This result is contained in Theorem~\ref{thm:quantisation-rational-PII}.

In the Shapiro-Tater case it is not enough matching Stokes phenomenon in Theorem~\ref{thm:stokes_quasi_polys} to obtain quantisation conditions, this can seen from Theorem~\ref{thmQESWKB}. For this reason we additionally impose the condition \eqref{458} which yields the correct quantisation equations \eqref{QuantST} from an application of Theorem~\ref{integralA1}.


\subsection{The Jimbo-Miwa case}
\label{quantcondJM}
The parameters $s,E$ in the potential $Q(z;s,E)$ i.e. the parameters $t,\Lambda$ corresponding to the pole with residue $-1$ of the rational solution $u_n$ and ``eigenvalue'' of the \eqref{eq:JM_potential}, are determined by the implicit requirement that the Stokes phenomenon for the ODE matches the one indicated below.  

\begin{figure}
    \centering
    \includegraphics[width=.65\textwidth]{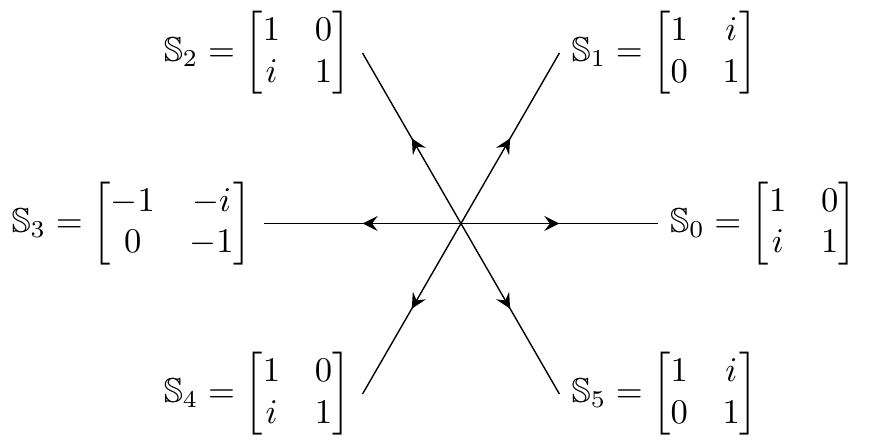}\\
    \caption{Stokes data for the Lax pair \eqref{eq:JM_Lax_pair} corresponding to
    rational solutions of Painlev\'e II}  
    \label{fig:JM_Stokes_data}
\end{figure}

Indeed it was shown in \cite{BM14} that  rational solutions of the Painlev\'e\ II equation correspond to a particular Stokes phenomenon as  shown in Fig. \ref{fig:JM_Stokes_data}. 

We recall that the map to the Stokes' data for general solution of the Painlev\'e\  II transcendent was obtained originally in \cite{its-novok}, see also \cite{FIKN}.

\begin{theorem}[\cite{BM14}]
The matrices $\S_0, \dots, \S_5$ in Fig.~\ref{fig:JM_Stokes_data} form the monodromy data of the Jimbo-Miwa 
Lax pair \eqref{eq:JM_Lax_pair} corresponding to the rational solutions of PII.
\end{theorem}

Consider a rational solution $u_n(t)$ \eqref{ratlsoln} of the \ref{eq:painleve2} equation with  $\alpha = n\in\Z$. Let $a$ be a pole of residue $-1$,  namely a zero of the Vorob'ev-Yablonskii polynomial $Y_n(t)$. 
Let $b$ be the coefficient in the expansion as in \eqref{expu}, and recall $\Lambda = \frac{7a^2}{36} + 10 b$ as in \eqref{eq:JM_potential}. Then, according to Prop. \ref{propMasoero}, the Stokes phenomenon of the ODE \eqref{217} must be as in Fig. \ref{fig:JM_Stokes_data}. Viceversa \cite{Masoero_2010},   if the anharmonic potential \eqref{217} exhibit the Stokes phenomenon as in Fig.  \ref{fig:JM_Stokes_data}, then the pair of values $a,b$ characterize (uniquely) the (rational, in this case) solution $u(t)$ of the Painlev\'e II equation with a pole at $t=a$.

Thus, to find the positions of a pole, we can find for which values of $a,b$ in \eqref{217} (with $\alpha=n$) the Stokes phenomenon matches Fig. \ref{fig:JM_Stokes_data}. Of course the map that associates to the parameters $a,b$ in \eqref{217} the Stokes data is highly transcendental.  

It is the nature of our problem, however, that we are interested in the behaviour when $n$ is large and also in the re-scaled plane. Thus we can apply our (exact) WKB analysis to obtain asymptotic information on the Stokes parameters and set up an implicit equation for $a,b$, or rather  their rescaled counterparts $s,E$ as in \eqref{eqsJM} with the identifications \eqref{scalingJMU}. 

In order to apply the exact WKB method, we set our large parameter to be $\hbar^{-1} = (n+1/2)$.
By the general theory in Section \ref{section:exact-wkb-analysis}, 
we construct the WKB solutions associated to (\ref{eq:schrodinger_wkb_equation}) as formal
power series in $\hbar$:
\begin{equation}
    \psi_{\pm}^{(\tau)}(z,\hbar)= \frac{1}{\sqrt{\So(z;\hbar)}}\exp\left( \pm \int_\tau^z \So(u;\hbar) \dd u \right)
\end{equation}
normalized near a turning point $\tau$.

\begin{theorem}[Vorob'ev-Yablonskii quantisation]\label{thm:quantisation-rational-PII}
Suppose that $(a,b) \in \C^2$ determine a rational solution of \ref{eq:painleve2} with $\alpha = n$ and Laurent expansion \eqref{eq:P2-laurent}.
Let $s = \hbar^{2/3}a$ and $E = 7s^2/36 + \hbar^{4/3}b$ with $\hbar^{-1} = n+1/2$ as in the scaling \eqref{scalingJMU}.
Then the Stokes' graph for the potential $Q(z;s,E)$ must be of type $D$ and the corresponding Fock--Goncharov parameters \eqref{eq:xi-symbol-definition} must satisfy
\begin{equation}
\xi_{10} = \xi_{20} = \xi_{30} = -1.    
\end{equation}
In terms of the Vor\"os symbols we obtain the following leading order estimates:
\begin{align}
    \pi i (2k_1 +1) = 2\int_{\tau_1}^{\tau_0} \So(z_+,\hbar) \dd z  \simeq (2n+1) \int_{\tau_1}^{\tau_0} \sqrt{Q(z_+;s,E)} \dd z + \bigO(\hbar)\\
    \pi i (2k_2 +1) = 2\int_{\tau_2}^{\tau_0} \So(z_+,\hbar) \dd z  \simeq (2n+1) \int_{\tau_2}^{\tau_0} \sqrt{Q(z_+;s,E)} \dd z + \bigO(\hbar)\\
    \pi i (2k_3 +1) = 2\int_{\tau_3}^{\tau_0} \So(z_+,\hbar) \dd z  \simeq (2n+1) \int_{\tau_3}^{\tau_0} \sqrt{Q(z_+;s,E)} \dd z + \bigO(\hbar) 
\end{align}
where $k_1, k_2, k_3$ are integers and $\hbar = (n+1/2)^{-1}$
\end{theorem}
\noindent{\bf Proof.}
We take the Stokes data for the Jimbo-Miwa Lax pair corresponding to the rational solutions of PII in Fig. \ref{fig:JM_Stokes_data}
and we equate it to the Stokes matrices from the WKB Riemann-Hilbert 
problem in each configuration  of Theorem \ref{thm:Stokes_matrices_computation} i.e.:
\begin{equation}
\label{55}
    \S_{2j} = \begin{bmatrix} 1 & 0 \\ i & 1 \end{bmatrix},
    \quad
    \S_{2j+1} =(-1)^{\delta_{j1}} \begin{bmatrix} 1 & i \\ 0 & 1 \end{bmatrix}
    \quad 
    (j = 0,1,2)
\end{equation}
From each configuration we obtain a system of 6 equations involving the exponentials of the periods $v_j$, for example in 
\textbf{Configuration $D$} we have from  \eqref{ConfDstokes} the conditions: 
\bea
 s_0 =&i= i (\xi_{10} \xi_{30} + \xi_{10} +1), \nn\qquad 
            s_1 = i                                   \\
        s_2 =&i=i\frac{1}{\xi_{20}\xi_{10}} (\xi_{20} \xi_{10} + \xi_{20} +1),\quad           
        s_3 =i = i \xi_{10}\xi_{20} \nn      \\
        \nn
          s_4 =& i =i \xi_{30}\xi_{10} (\xi_{30} \xi_{20} + \xi_{30} +1),  \quad
          s_5 =i = i \frac{1}{\xi_{10}\xi_{30}}\,.
\eea

The only solution of the above system is $\xi_{20} = \xi_{10} = \xi_{30} =-1$ which  is consistent with the fact that the product is $-1$. 
Direct inspection of the formulas in Theorem \ref{thm:Stokes_matrices_computation} shows that it is impossible to satisfy the constraints \eqref{55} in all of the other configurations. 
\QED

\subsection{The Shapiro-Tater case}
\label{quantcondST}
\begin{theorem}
\label{thmQESWKB}
Suppose that the quasi-polynomial $y=p_n(x) e^{\theta(x;t)}$ is a solution to the boundary problem \eqref{eq:ST_eval_problem}-\eqref{boundary1} with $J=n+1$.
Then the ``exact''  Fock--Goncharov parameters $\xi_{jk}$  in \eqref{eq:xi-symbol-definition}  satisfy one of the following systems, depending on the indicated  Stokes graph configuration.

\begin{itemize}
    \item Configuration $D$ (Fig. \ref{fig:generic_stokes_graph_configurations}){  or its $\Z_3$ rotations}:
        \begin{equation}
        \label{FGD}
        \begin{cases}
        \xi_{10} \xi_{30} + \xi_{10} + 1 = 0 \\
        \xi_{20} \xi_{10} + \xi_{20} + 1 = 0 \\
        \xi_{30} \xi_{20} + \xi_{30} + 1 = 0
        \end{cases} 
        \end{equation}

      This systems cuts an affine rational curve in $\C^3$  given by:
        \begin{equation}
            \xi_{10} = \rho , \quad \xi_{20} =-\frac{1}{\rho+1}, \quad \xi_{30} = -\frac{\rho + 1}{\rho}\,.
            \label{eq:p-parametrisation}
        \end{equation}
    
    \item Configuration $E$ (Fig. \ref{fig:generic_stokes_graph_configurations}) or its $\Z_3$ rotations:
        \begin{equation}
        \label{FGE}
            \begin{cases}
           \xi_{01} \xi_{12} (\xi_{23} + 1) + \xi_{01} + 1 = 0\\ 
            \xi_{01}+1 = 0 \\
            \xi_{23}+1 = 0\,.
            \end{cases}
        \end{equation}
\end{itemize}
Furthermore all other configurations cannot occur. 
\end{theorem}
\noindent{\bf Proof.}
In Proposition \ref{propQES} it was shown that if $J=n+1$ and $(t,\Lambda)$ belong to the ES spectrum (i.e. there is a solution of the boundary problem \eqref{eq:ST_eval_problem} and \eqref{boundary1}) then the Stokes parameters $s_0,s_2,s_4$ all vanish simultaneously. 

In Theorem \ref{thm:Stokes_matrices_computation}  and using the gauge transformation \eqref{ConfDstokes} we have expressed the Stokes parameters in terms of the exact Fock-Goncharov parameters $\xi_{jk}$: thus we have to see which configurations are compatible with the three equations $0=s_0=s_2=s_4$.  {\color{black} Note that equations \eqref{FGD} and \eqref{FGE} corresponds exactly to the condition $0=s_0=s_2=s_4$ in configuration $D$ and $E$  as in Fig.\eqref{fig:generic_RHP_configurations} or  its $Z_3$ reflections. }  Note that $\hbar= 1/(n+1)$ and hence ${\rm e}^{\frac {2i\pi}\hbar}=1$ in the Theorem. 
Direct inspection and simple algebra then allows us to rule out configuration $Z$ as well as all the other configurations obtained from 
{it}
 by a  $\Z_3$ rotation or reflection, { and  those obtained from $D$ or $E$ by a reflection around the imaginary axis followed by any $\Z_3$ rotation.  \QED}

\subsubsection{Repeated eigenvalue condition}
Theorem \ref{thmQESWKB} establishes the conditions for the V\"oros symbols to yield a  point in the ES spectrum; together with Theorem \ref{thmQESWKB} the conditions are equivalent to the statement that the Stokes' graph is either of $D$ or $E$ type and the Fock-Goncharov parameters $\xi_{jk}$ satisfy the corresponding conditions specified in Theorem \ref{thmQESWKB}.

\begin{figure}[t]
    \begin{minipage}{0.49\textwidth}
        \centering
        \includegraphics[width=1\textwidth]{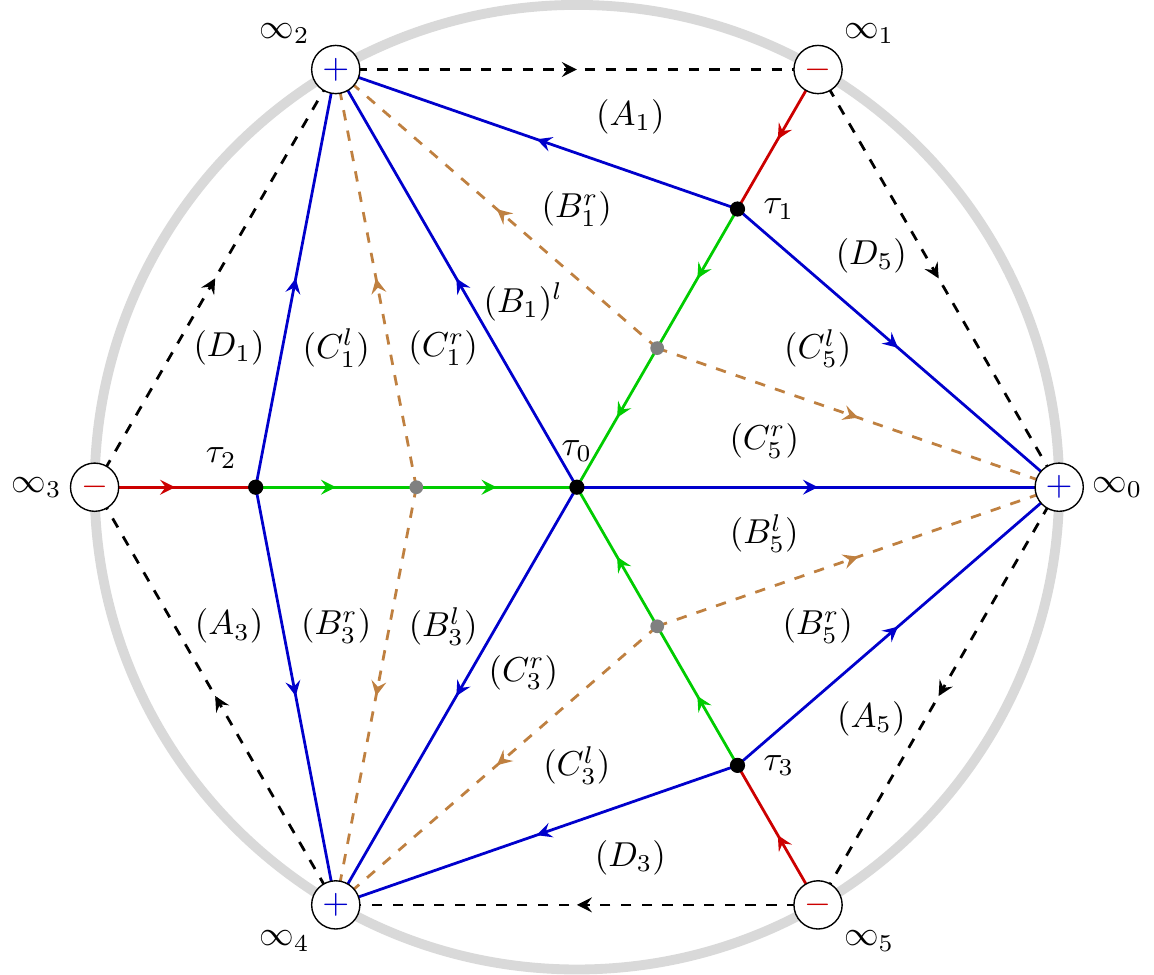}\\
        {(a) Configuration $D$}
    \end{minipage}  
    \begin{minipage}{0.49\textwidth}
        \centering
        \includegraphics[width=1\textwidth]{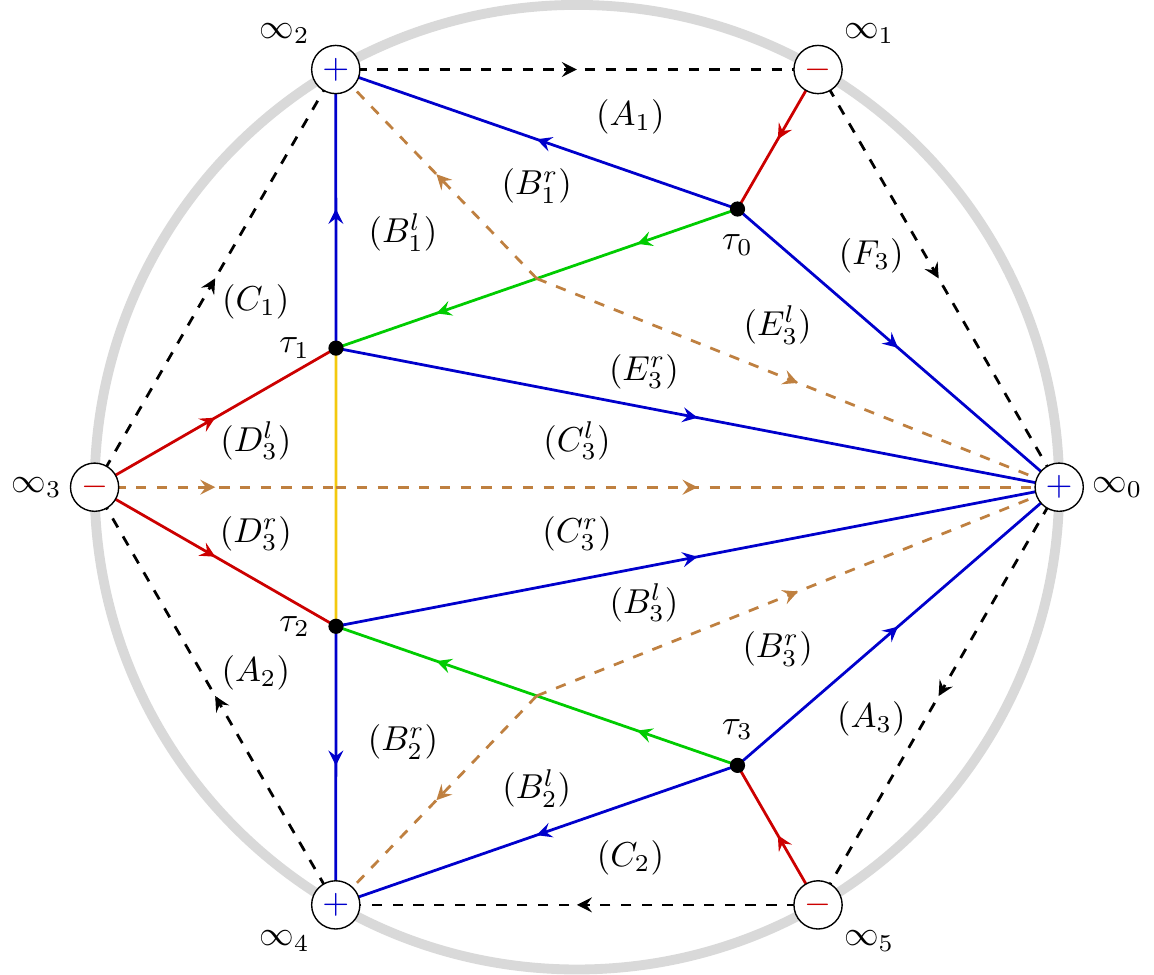}\\
        {(b) Configuration $E$}
    \end{minipage}
    \caption{Labelled regions in the WKB Riemann-Hilbert problem.}
    \label{WKB-regions}
\end{figure}

In addition we must now impose the  condition that the eigenvalue is a repeated one:
as proved in Theorem \ref{thm:repeated_eigenval_condition} this requires that all the integrals of $p_n^2{\rm e}^{2\theta}$ between $\infty_{2k+1}$ vanish. It was also explained in the  theorem that it suffices to impose one of the two  vanishing conditions and the other one will follow.
Thus the strategy now is to compute 
\bea
I_{13} = \int_{\infty_1}^{\infty_3} p_n(x)^2{\rm e}^{2\theta(x;t)} \d x\,,
\eea
using the asymptotic expansion in terms of formal WKB solutions obtained so far. 
{\color{black}In order to simplify the upcoming computations  we will label the regions in the WKB RH problem for configuration $D$ and $E$ according to
 Fig.~\ref{WKB-regions}. This will help us distinguish between  the entire solutions of the differential equation
\eqref{eq:schrodinger_wkb_equation} 
 that are asymptotics to  WKB solutions $\psi_\pm^{\tau_j}$ in Def. \ref{def:wkb_sols_turning_pt}
 in different regions in accordance with Thm.~\ref{thm:voros_connection_formula}.}  Namely
 \[
 \psi_{\pm}^\mathcal{B}(z)\simeq\psi_\pm^{\tau_j}(z;\hbar)= \frac{1}{\sqrt{\So(z,\hbar)}}
        \exp \left( \pm \int^z_\tau \So(u,\hbar) \dd u \right),\quad u\in\mathcal{B},\;\hbar\to0
        \]
        where $\mathcal{B}$ is one of the labelled regions in Fig.~\ref{WKB-regions} ad $\tau_j$ is one of the turning 
        points in its boundary.
We observe that to compute $I_{13}$  we can equivalently compute the integral of $\Psi_+^{(A_1)}$ because this function  is proportional to $ p_n(x)^2{\rm e}^{2\theta(x;t)}$ since they are both recessive in the direction $\infty_1$.
Thus, the main aim  of this section is to prove the following Theorem.
\begin{theorem}
\label{integralA1}
Let $(t,\Lambda)$ belong to the ES spectrum. Let $s = t\hbar^\frac 23$ and $E= \Lambda \hbar^\frac 4 3$, with $\hbar = (n+1)^{-1}$. 
Then 
\begin{enumerate}
\item In configuration $D$  as in Fig. \eqref{WKB-regions} we have:
\begin{equation}
\label{leadingestimate}
\int_{\infty_1}^{\infty_3} \Psi_+^{(A_1)} (z;\hbar) ^2 \d z \simeq 2i\hbar \le(
{\rm e}^{2v_{12} } \int_{\tau_2}^{\tau_0}\frac {\d z}{\sqrt{Q(z_+;s,E)} }
-
\int_{\tau_1}^{\tau_0}\frac {\d z}{\sqrt{Q(z_+;s,E)} }
 \ri) + \mathcal O(\hbar^2)\,
\end{equation}
where $v_{12}$ is defined in \eqref{Vorosellr}.
The rescaled parameters $(s,E)$   of the ES spectrum {   are asymptotic to a }  repeated eigenvalue  provided that 
\be
\exp\le[{\frac 2{\hbar}\int_{\tau_1}^{\tau_2} \sqrt{Q(z;s,E)}\d z } \ri]
=
\boldsymbol \tau(s,E) + \mathcal O(\hbar)
\label{2eigs}
\ee
where 
\be
\label{tau}
\boldsymbol \tau(s,E) = \frac
{\ds \int_{\tau_1}^{\tau_0}\frac {\d z}{\sqrt{Q(z_+;s,E)} }}{\ds \int_{\tau_2}^{\tau_0}\frac {\d z}{\sqrt{Q(z_+;s,E)} }},\quad \Im(\boldsymbol \tau(s,E))>0.
\ee
\item
In configuration  $E$  as in Fig. \eqref{fig:generic_RHP_configurations} we have 
\be
\int_{\infty_1}^{\infty_3} \Psi_+^{(A_1)} (z;\hbar) ^2 \d z \simeq 2i
 \int_{\tau_2}^{\tau_0}\frac {\hbar\d z}{\sqrt{Q(z_+;s,E)} }
 + \mathcal O(\hbar^2).
\ee
Then the rescaled parameters $(s,E)$   of the ES spectrum cannot be a double eigenvalue for large $n$. 
\end{enumerate}
\end{theorem}

The proof is relatively straightforward but significantly delicate and technical and it is postponed to  the Appendix \ref{section_proof}.
We can take the  naive approach in which we replace $\Psi_+^{(A_1)}$ with the appropriate  combinations of the WKB formal solutions along the pieces of the contour of integration that traverse each Stokes region.
In fact this approach yields  the correct result, but over-estimates the error term. It is however appealing for its simplicity  to give here the heuristic derivation and defer technicalities to later. 
Consider the case of configuration $D$: we split the integration at the turning points $\tau_1,\tau_0,\tau_2$.
The unbounded integrals are then along steepest descent paths for the integrand and can be neglected. In the integrations along  $[\tau_1,\tau_0]$ and $[\tau_0,\tau_2]$ we observe that, as a consequence of the Riemann-Hilbert problem \ref{RHPWKB} we can express $\Psi^{(A_1)}_+$ as suitable linear combinations of $\wkb{\tau_j}{\pm}$; indeed 
\be
\Psi_+^{(A_1)}(z;\hbar) &\simeq \left\{
\begin{array}{ll}
\wkb{\tau_1}{+}(z;\hbar) + i\wkb{\tau_1}{-}(z;\hbar) &  z\in (B_1^r)\cup (B_1^\ell)
\\[4pt]
 {\rm e}^{v_{12}} \le(\wkb{\tau_2}{+}(z;\hbar) - i\wkb{\tau_2}{-}(z;\hbar)\ri)& z\in (C_1^\ell)\cup (C_1^r)
 \end{array}
 \ri.
 \ee
 where we have used already the fact that we are on the ES spectrum so that Theorem \ref{thmQESWKB} applies  and the Stokes matrices $\mathbb S_{j}, \,j\in\{0,2,4\}$ are trivial. 
Computing the square of $\Psi_+^{(A_1)}$, we have that  the cross--products yield non-oscillatory functions that contribute to the leading order while the squares of the ``pure'' WKB solutions give oscillatory integrals which can be neglected to leading order. 
 
 Thus one is lead to the rough estimate
 \be
 \int_{\infty_1}^{\infty_3} \le(\Psi_+^{(A_1)}\ri)^2\d z \simeq 2i \int_{\tau_1}^{\tau_0}  \wkb{\tau_1}{+}\wkb{\tau_1}{- }\d z
- 2i{\rm e}^{2v_{12}} \int_{\tau_0}^{\tau_2}  \wkb{\tau_2}{+}\wkb{\tau_2}{- }\d z
\simeq\nn
\\
\simeq
2i \int_{\tau_1}^{\tau_0}  \frac {\hbar \d z}{\sqrt{Q(z_-;s,E)}}
- 2i{\rm e}^{2v_{12}} \int_{\tau_0}^{\tau_2}  \frac {\hbar \d z}{\sqrt{Q(z_+;s,E)}}
 \ee
where the boundary values of $\sqrt{Q}$ are due to our choice of orientations for the branch-cuts in Fig. \ref{WKB-regions} { (i.e. not according to Def. \ref{def:branch-orientation})}.
Rearranging the endpoints and boundary values yields \eqref{leadingestimate}.

The reason why the above reasoning is defective is that it replaces the formal WKB expansions also in the neighbourhoods of the turning points, where the formal WKB solutions have a singularity.
One may still make sense of the resulting integrals because they involve a singularity of type $(z-\tau)^{-\frac 12}$ which is integrable.
However, approaching the integrals in this way and using a (formal) application of the Laplace method would suggest that the subleading order is $\mathcal O(\hbar^\frac 4 3)$.
The careful analysis, instead, of the contribution near the turning points reveals that the subleading correction is of order is $\mathcal O(\hbar^2)$. 

Unfortunately we could not find a reference in the vast literature on exact WKB analysis that helps us on this issue. For this  reason  we have postponed this part of the proof to the appendix.

\subsection{Comparison of quantization conditions}
\label{comparing-quantization}
\paragraph{The ST case.}
In view of Theorem \ref{integralA1} and in particular \eqref{2eigs}, we can now express, to within the leading order, the quantization conditions that characterize those points $(s,E)$ in the ES spectrum with a double eigenvalue. Indeed from \eqref{eq:p-parametrisation} it follows that 
\bea
-\frac {\xi_{10} }{1+\xi_{10}} = \frac 1 {\boldsymbol \tau}
\qquad \Rightarrow\qquad
\le\{
\begin{array}{l}
\ds 
\xi_{10} = \frac {-1}{1+\boldsymbol \tau}
\\[8pt]
\ds
\xi_{20} = -1 -\frac 1{\boldsymbol \tau}
\\[8pt]
\ds
\xi_{30} = {\boldsymbol \tau}
\end{array}
\ri.
\eea
In terms of the Voros' symbol we then have, to leading order
\begin{align}
&2(n+1)\int_{\tau_1}^{\tau_0} \sqrt{Q(z_+;s,E)} \d z = \ln \le( \frac {-1}{1+\boldsymbol \tau(s,E)}
\ri) - 2i\pi (m_1+1)
\nn
\\
&2(n+1)\int_{\tau_2}^{\tau_0} \sqrt{Q(z_+;s,E)} \d z = \ln \le( -1 -\frac 1{\boldsymbol \tau(s,E)}
\ri) - 2i\pi (m_2+1)
\nn
\\
&2(n+1)\int_{\tau_3}^{\tau_0} \sqrt{Q(z_+;s,E)} \d z = \ln \le( {\boldsymbol \tau(s,E)}
\ri)- 2i\pi (m_3+1)
\label{QuantST}
\end{align}
where the three integers satisfy  $m_1+m_2+m_3 = n-1$ due to the fact that the sum of the three integrals on the left is $-2i\pi(n+1)$ while the sum of the three logarithms is $2i\pi$ (principal determination) due to the definition of $\boldsymbol \tau(s,E) $
as in \eqref{tau}.
\paragraph{The JM case.}
On the other hand, the quantization conditions for the Vorob'ev-Yablonskii zeroes, to the same order of approximation, read
\bea
&\le(2n+1\ri)\int_{\tau_j}^{\tau_0} \sqrt{Q(z_+;s,E)} \d z =-{i\pi} - 2i\pi k_j
\nn\\
&k_1+k_2+k_3=n-1.
\label{QuantJM}
\eea
Both conditions \eqref{QuantST}, \eqref{QuantJM} involve a triple of positive  integers adding to $n-1$ but they differ notably in the multiplicative factor $2(n+1)$ vs. $(2n+1)$  on the left side, and on the values on the right side.  We now analyze the two lattices to explain their similarity which is apparent from the numerical experiments.

{
\subsection{The elliptic region}
\label{ellreg}
\begin{figure}
\begin{center}
\includegraphics[width=0.99\textwidth]{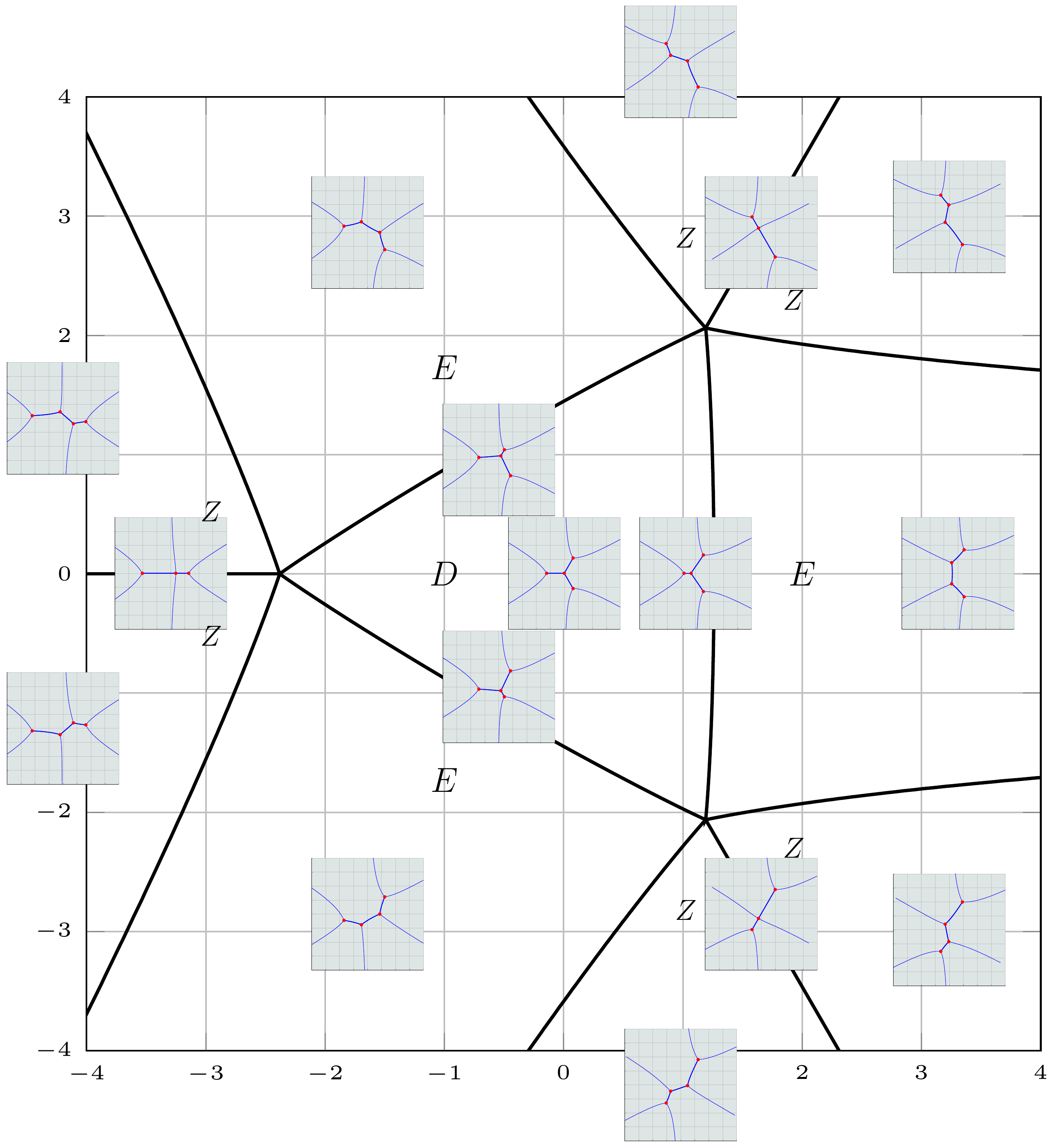}
\end{center}
\caption{The phase space in the $s$--plane; depicted are the curves defined by \eqref{regions}. The inserted vignettes depict the critical graph $\mathfrak C$ corresponding to a sample value of $s$ in their approximate location (which is at the centre of each vignette). The types of topologies in each region is marked by the corresponding letter in the figure. The central triangular region corresponds to  the $D$ topology;  the three ``corridors'', adjacent to the sides, in the directions $\arg(s) = 0,\tfrac {\pm 2\pi}3$ correspond to  the three $\Z_3$ rotation of a $E$, while in the six remaining regions we find the various $\Z_2\times \Z_3$ versions of the $Z$ topologies. Also indicated some of the topologies with coalesced turning points, which are found when $s$ is on any of the boundaries between regions.  }
\label{EllFig}
\end{figure}

By the term ``elliptic region'' we refer, with a nod to the terminology in \cite{BM14, BM15, BB15}, to the triangular region containing the (scaled) zeros. The boundary of this region is determined in loc. cit. for the zeroes of the Vorobev--Yablonski polynomials. We want to show that (at least asymptoticallyl) also the lattice (described in detail in the next section) of rescaled zeroes of $D_n(t)$ are contained in the same region.\footnote{We thank the anonymous referee for pointing out this aspect of the conjecture.}
In order to argue that the (rescaled) zeros of the discriminant $D_n(t)$ and of the Vorob'ev-Yablonski polynomials $Y_n(t)$ fill the same region of the $s$--plane, we should show that the $D$--configuration occurs only within the elliptic region.
For the Vorob'ev-Yablonski polynomials this issue has been addressed in the above literature. For the Shapiro-Tater problem we give only a semi-rigorous  argument in this section (with numerical support).

We have established that only Stokes' graphs of topology $D$ can give rise to a solution of the repeated eigenvalue condition.
In fact the  equations \eqref{QuantST}  imply that, to leading order as $n\to\infty$, the three periods $\int_{\tau_j}^{\tau_0} \sqrt{Q(z)} \dd z$  are all {\it purely imaginary}:
\be
\int_{\tau_j}^{\tau_0} \sqrt{Q(z_+;s,E)} \d z = -i\pi \frac {m_j+1}{n+1} + \mathcal O(n^{-1}),\quad j=1,2,3.
\ee
We  take these equations to leading order; 
\be
\int_{\tau_j}^{\tau_0} \sqrt{Q(z_+;s,E)} \d z = -i\pi \mu_j,\ \ \ \mu_1 + \mu_2 + \mu_3 = 1, \ \ \mu_j\in \R. 
\label{periods}
\ee
This means that the differential $\sqrt{Q} \dd z$ is (to leading order) an imaginary normalized differential and the algebraic elliptic curve $ w^2= Q(z;s,E)$ is termed, in the literature \cite{Bertola:Boutroux}, a {\it Boutroux curve}, with the condition that all the periods \eqref{periods} are imaginary being the {\it Boutroux condition}. Under this condition we know  that the critical graph, $\mathfrak C$, is connected. This is, by definition, the graph  determined by the ``vertical critical trajectories\footnote{The "vertical" here refers to the fact that in the plane of the locally defined conformal map $\xi(z):= \int^z \sqrt{Q} \dd w$ these are vertical straight lines.}'' \cite{Strebel}, namely, the maximal integral curves of the direction field 
\be
\label{ReQ0}
\Re \sqrt{Q(z;s,E)} \dd z=0
\ee
that issue from each of the turning points. 
Observe that the integral curves of \eqref{ReQ0} form a foliation that is  orthogonal to the Stokes' graph \eqref{eq:stokes_curve}, in the sense of the Riemannian metric $\dd \sigma^2:= |Q(z;s,E)| |\dd z|^2$  (with conical singularities at the turning points).  Briefly  this can be seen by considering the real--valued function $\Phi(z):=\Re\le( \int_{\tau_0}^z w \dd z\ri)$, which is easily shown to be 
\begin{enumerate}
\item [--] harmonic on the Riemann surface $\Sigma$ \eqref{Rsurf} minus the two points above $z=\infty$;
\item [--]   odd under the holomorphic involution $(w,z)\to (-w,z)$ (i.e. under the flip of sign of the square-root). 
\end{enumerate}
The connectedness of $\Phi^{-1}(\{0\})$ on $\Sigma$ (and hence also of its $z$--projection) follows from the fact that $\Phi$ is a Morse function whose critical points (the turning points) belong all to the zero level-set (due to the Boutroux conditions). 
These two properties imply that the zero level set $\Phi^{-1} (\{0\})$ is a well--defined subset of the complex $z$--plane (i.e. of the projection from $\Sigma$ to the $z$--plane).
It is then not hard, following similar arguments to those used in  analyzing the topology of the Stokes' graph, that $\frak C$ can only have three different topologies  corresponding to the three configurations $D,E,Z$ (and ``dual'' to the Stokes' graph). 
The graph $\frak C$ is however also a metric space with the distance induced by the Riemannian metric $\dd \sigma^2$; in fact the geodesic length between two turning points (along the graph) is precisely the (absolute value) of the purely imaginary integral of $\int_{\tau_i}^{\tau_j} \sqrt{Q} \dd z$ along any path homotopic to the edge of $\frak C$ that connects the two turning points. In other words the  numbers $\pi |\mu_j|$ are distances, and the vanishing of any of them indicates that the two turning points have coalesced; therefore this happens when the topology of the critical graph undergoes a change. 
In particular this can only happen if at least two turning points coincide, namely on the critical locus where the discriminant of $Q$  \eqref{discrimQ} vanishes. Summarizing the above discussion, the boundary separating between different configurations is given by system of equations (of mixed algebraic/transcendental nature)
\be
\label{sysEQ}
\le\{\begin{array}{cc}
E s^4 - 8 E^2 s^2 + 16 E^3 - s^3 + 36 E s - 27=0 \\
\Re \int_{\tau_0}^\mu \sqrt Q \dd z=0
\end{array}\ri.
\ee
where $\mu$ denotes the multiple root of $Q(z;s,E)$ and $\tau_0$ a remaining simple root.

It is possible to simplify the system \eqref{sysEQ}: the first equation can be rationally parametrized
\be
E = \frac {8a^3+1}{16 a^4} ,\ \  s= \frac {4a^3-1}{2a^2}; \ \ \ 
Q(z;s({a}),E({a})) = \big(z+\tfrac 1{2a}\big)^2 \le(z^2 -\frac z a +\frac {1+8a^3}{4a^2}   \ri)
\ee
A further degeneration is when three turning points coalesce, and from the above we see that this happens if and only if $2a^3+1=0$. The three corresponding values of $s$  are 
\be
s_0 = -\frac 3{2^\frac 1 3} \simeq -2.381101,\ \ \  s_k = s_0{\rm e}^{\frac {2i\pi k}3}, \ \ k=1,2.
\ee
and the boundary of the regions are determined by the image in the $s$--plane (under the map $s =\frac {4a^3-1}{2a^2} $) of the curves defined implicitly in the $a$--plane by 
\be
\Re \int_{-\frac 1{2a}}^{\tau_0} \sqrt{Q(z;s(a), E(a))} \dd z=
\Re \le[
\frac{4a^3-1}{6a^3} \sqrt{2a^3+1} + \ln \le(
\frac{1-\sqrt{2a^3+1}}{a^\frac 32 \sqrt{2} }
\ri)
\ri]=0.
\label{regions}
\ee
{We observe that the above condition does not depend on the choice  of $\tau_0$, namely  one of the two remaining roots of the quartic polynomial, $Q(z;s(a)),E(a))$.}
We remark that these are  the same equations that appeared in \cite{BB15} (Definition 1.3 ibidem) defining the boundary of the elliptic region.
The curves defined by \eqref{regions} partition the $s$--plane into 10 regions (see Fig. \ref{EllFig}); in each region the topology of the critical graph of the Boutroux curve is the same, and the boundary between regions corresponds to two turning points coalescing. The ``elliptic region'' is the triangular shaped region containing the origin and it is the only region where the topology of the critical graph is of type $D$. This was shown in \cite{BM14} and we will not repeat the arguments here because it is a rather lengthy affair. We rather offer a numerical investigation of the ``phase space'' summarized in Fig. \ref{EllFig}.  A complete proof can be obtained following the ideas in \cite{Berto-Tovbis-SIGMA, Bertola:Boutroux} of performing a ``continuation in parameter space'' and carefully tracking the change in topology of the critical graph $\mathfrak C$.

The important issue for us is that since only the elliptic region $D$ is compatible with the solution of the Shapiro-Tater problem, we deduce that  in any compact sets of the complement of the elliptic region there are (at least for $n$ large enough) none of the rescaled zeros of $D_n(t)$.

This also addresses another aspect of the Shapiro-Tater conjecture.
}

\subsection{Analysis of the two lattices}
\begin{figure}
\centering
\includegraphics[scale=1]{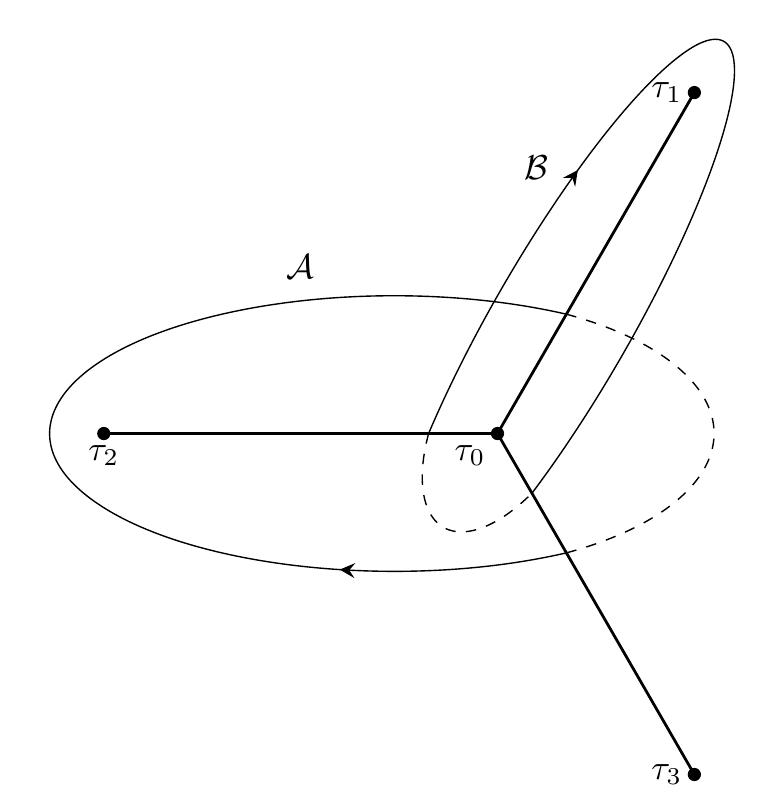}
\caption{Canonical basis of cycles for the homology of the elliptic Riemann surface $\overline{\Sigma}$ with Stokes graph configuration of type $D$.
The points $\tau_j$ are the branch points of $\sqrt{Q(z;s,E)}$ and the lines connecting them are the branch cuts.}
\label{homology}
\end{figure}

Both lattices involve implicit equations for the parameters $(s,E)$ via the periods  of the differential $\sqrt{Q(z;s,E)} \d z$.
We introduce a  canonical basis of cycles $\mathcal A,\mathcal B$   of the elliptic  Riemann surface  $\overline{\Sigma}$   as in the Fig.\ref{homology}  so that 

\be
\boldsymbol \tau(s,E)= \frac
{\ds \int_{\tau_1}^{\tau_0}\frac {\d z}{\sqrt{Q(z_+;s,E)} }}{\ds \int_{\tau_2}^{\tau_0}\frac {\d z}{\sqrt{Q(z_+;s,E)} }}=\frac
{\ds \int_{\mathcal B}\frac {\d z}{\sqrt{Q(z;s,E)} }}{\ds \int_{\mathcal A}\frac {\d z}{\sqrt{Q(z;s,E)} }}.
\ee
Notice that the quantities $v_{j0}$ defined in \eqref{eq:voros_jumps_turning_pts} can be written in the form

\be
 v_{10}=\frac{1}{2} \int_{\mathcal B} \sqrt{Q(z;s,E)}\d z,\;\; v_{20}= \frac{1}{2}\int_{\mathcal A} \sqrt{Q(z;s,E)}\d z,\;\; v_{30}=-\frac{1}{2} \int_{\mathcal A+\mathcal B} \sqrt{Q(z;s,E)}\d z
 \ee
The Jacobian determinant is a constant as we prove in the next lemma.
\begin{lemma}
\label{lemmaJac}
 Let $\mathcal A$ and $\mathcal B$ be the canonical homology basis as in  Fig.~\ref{homology}   and consider the periods 
 \begin{equation}
 I_{\mathcal A} = \oint_{\mathcal A} \sqrt {Q(z;s,E)}\d z,
 \quad 
I_{\mathcal B} = \oint_{\mathcal B} \sqrt {Q(z;s,E)}\d z.
 \end{equation}
Then
\be
\det \begin{bmatrix}
\ds \frac {\pa I_{\mathcal A}}{\pa s}  & \ds \frac {\pa I_{\mathcal A}}{\pa E}   \\[9pt]
\ds \frac {\pa I_{\mathcal B}}{\pa s} &\ds \frac {\pa I_{\mathcal B}}{\pa E}  
\end{bmatrix}=i\pi.
\ee
\end{lemma}
\noindent{\bf Proof.}
The determinant gives 
\be
\det \mathbb J= \frac {\pa I_{\mathcal A}}{\pa s}\frac {\pa I_{\mathcal B}}{\pa E}-
  \frac {\pa I_{\mathcal B}}{\pa s}\ds \frac {\pa I_{\mathcal A}}{\pa E}.     
\ee
Since  $Q(z;{s},E) = z^4 + {s} z^2 +2z +E,$ the derivative w.r.t. $E$ gives the holomorphic periods and the derivative in $s$ gives second-kind periods. Thus we can compute the above expressions with the Riemann bilinear identity to give 
\begin{align*}
\det \mathbb J &=\frac{1}{4}\int_{\mathcal A} \frac {z^2\d z}{\sqrt{Q(z;s,E)}} \int_{\mathcal B} \frac {\d z}{\sqrt{Q(z;s,E)}} -\frac{1}{4}\int_{\mathcal A} \frac {z^2\d z}{\sqrt{Q(z;s,E)}} \int_{\mathcal B} \frac {\d z}{\sqrt{Q(z;s,E)}}\\
&=2i\pi  \frac 12 \mathop{\rm res}_{z=\infty^+}\frac {z^2\d z}{\sqrt{Q(z;s,E)}} \int_{\tau_1}^z \frac {\d \xi}{\sqrt{Q(\xi;s,E)}} = i\pi
\end{align*}
where we have used that 
\be
\frac {z^2 \d z}{\sqrt{Q(z;s,E)}} = \le(1+  \mathcal O(z^{-2})\ri) \d z\qquad
\int_{\tau_1}^z \frac {\d \xi}{\sqrt{Q(\xi;s,E)}} = \int_{\tau_1}^\infty \frac {\d \xi}{\sqrt{Q(\xi;s,E)}} - \frac 1 z + \mathcal O(z^{-2}).
\ee
and the contribution from the point $\infty^-$ is the same as the point $\infty^+$.
\QED
We observe that 
\begin{equation}
\label{omega}
\omega:=\frac{\pa I_{\mathcal A}}{\pa E}= \int_{\tau_2}^{\tau_0}\hspace{-8pt}\frac{\d z}{ \sqrt{Q(z_+;s,E)} },\quad
\omega':=\frac{\pa I_{\mathcal B}}{\pa E}=\int_{\tau_1}^{\tau_0}\hspace{-8pt}\frac{\d z}{ \sqrt{Q(z_+;s,E)} } 
\end{equation}
 are the half periods of the holomorphic differential $\frac {\d z}{\sqrt{Q(z;s,E)}}$.
The lemma is useful in that it allows us to explore the geometry of the quantization conditions;

\begin{proposition}
\label{locallattice}
Let $(s_0,E_0)$ correspond to the first-order quantization conditions \eqref{QuantST} or \eqref{QuantJM} in the bulk, namely, $m_j/n \simeq c_j \neq 0$. Then the neighbour points in the $s$--plane form a slowly modulated hexagonal lattice in the sense that the six closest neighbours of $s_0$ are
\begin{equation}
\label{Deltas}
s_0 + 2 \hbar\le(\omega \Delta m_1 -\omega'\Delta m_2\ri)
\end{equation}
where $\omega$ and $\omega'$ are the half periods of the holomorphic differentials in \eqref{omega} and 
\be
\Delta m_j \in \{-1,0,1\}, \qquad \ |\Delta m_1 + \Delta m_2 |\leq 1, \qquad |\Delta m_1|+|\Delta m_2|\geq 1.
\ee
\end{proposition}
\noindent
{\bf Proof.}
Let $(m_1,m_2,m_3)$ be a triple of quantization numbers for either \eqref{QuantST} or \eqref{QuantJM}. The neighbour points correspond to adding/subtracting $1$ from each, subject to the constraints  
\be
\Delta m_1 + \Delta m_2 + \Delta m_3 =0, \ \ \ \ \Delta m_j \in \{-1,0,1\}.
\ee
There are six elementary possibilities $$
(\Delta m_1 , \Delta m_2 , \Delta m_3) \in \bigg\{(1,-1,0), (1,0,-1), (0,1,-1),(-1,1,0), (-1,0,1), (0,-1,1)\bigg \}.
$$
The values of the periods $  \int_{\tau_\ell}^{\tau_0} \sqrt{Q(z_+;s,E)}\d z$, $ \ell=1,2,3$,  change by $\hbar  {\Delta m_\ell}$, where $\hbar  = (n+1)^{-1}$ in the ST case and $\hbar ( n  + 1/2)^{-1}$ in the VY case.

Let $(s,E) = (s_0,E_0) + (\Delta s, \Delta E)$
be a neighbour point in the lattice. We want to estimate $( \Delta s, \Delta E)$: 
we observe that 
\[
 \int_{\tau_2}^{\tau_0} \sqrt{Q(z_+;s,E)}\d z=\frac{1}{2} I_{\mathcal A},\quad  \int_{\tau_1}^{\tau_0} \sqrt{Q(z_+;s,E)}\d z=\frac{1}{2} I_{\mathcal B},\quad  \int_{\tau_3}^{\tau_0} \sqrt{Q(z_+;s,E)}\d z=-\frac{1}{2} I_{\mathcal A}  -\frac{1}{2} I_{\mathcal B}  
\]
where $ I_{\mathcal A}$ and $ I_{\mathcal B} $   are defined in Lemma~\ref{lemmaJac}.
  If we take  for example   the first two periods    and expand   \eqref{QuantST} or \eqref{QuantJM} to linear order,  we obtain 
\bea
\begin{bmatrix}
\Delta s\\ \Delta E
\end{bmatrix}
\simeq  2\hbar
\begin{bmatrix}
\ds \frac {\pa I_{\mathcal A}}{\pa E}  & \ds \frac {-\pa I_{\mathcal B}}{\pa E}   \\[9pt]
\ds \frac {-\pa I_{\mathcal A}}{\pa s} &\ds \frac {\pa  I_{\mathcal B}}{\pa s}  
\end{bmatrix}
\begin{bmatrix}
\Delta m_1\\ \Delta m_2
\end{bmatrix}
\eea
so that, from the definition \eqref{omega}  one recovers \eqref{Deltas}. 
This local lattice generators $\omega$ and $\omega'$  are slowly modulated across the elliptic region.
\QED
\paragraph{Near the origin.}
If $(s,E)=\mathcal O(\hbar)$ then the elliptic surface is $w^2 = z^4+\mathcal O(\hbar) z^2 + 2z + \mathcal O(\hbar)$ and then  a direct computation shows that $\boldsymbol \tau = {\rm e}^{
\frac{2i\pi} 3} +\mathcal O(\hbar)$. Note that the $\Z_3$ symmetry of the limiting elliptic curve gives 
\be
{\rm e}^{2i\pi/3} = \boldsymbol \tau = \frac {-1}{1+\boldsymbol \tau} = -1-\frac 1 {\boldsymbol \tau}.
\ee
We are now going to show that  two quantization conditions yield the same lattices to order $\hbar^2$.

\begin{theorem}
\label{thmorigin}
The rescaled lattices of the zeroes of the VY Polynomials, and the of ST problem coincide to within order $\mathcal O(\hbar^2)= \mathcal O(n^{-2})$ in a $\mathcal O(\hbar)$ neighbourhood of the origin in the $s$--plane. More precisely the quantization conditions \eqref{QuantST}, \eqref{QuantJM}  corresponding to the triples $(m_1,m_2,m_3), \ \ m_1+m_2+m_3= n-1$ and $(k_1,k_2,k_3),\ \ k_1+k_2+k_3=n-1$ with $m_j=k_j$ single out  values of $s, E$ that differ by a discrepancy of order $\mathcal O(\hbar^2)$,  provided that $m_j- \frac{n-1}3$ remain bounded as $n\to\infty$. 
\end{theorem}

\noindent{\bf Proof.}
Let $(s,E)=\mathcal O(\hbar)$. Then 
  the two quantization  conditions \eqref{QuantST}, \eqref{QuantJM} read, to order $\hbar$, 
\be
\begin{split}
&2(n+1)\int_{\tau_j}^{\tau_0} \sqrt{Q(z_+;s,E)} \d z = - 2i\pi \le(m_j + \frac 2 3\ri)\\
&(2n+1)\int_{\tau_j}^{\tau_0} \sqrt{Q(z_+;s,E)} \d z = - 2i\pi \le(k_j + \frac 1 2\ri)
\label{ST0}
\end{split}
\ee
{ Let us now set $s = \hbar \delta s$, $E = \hbar \delta E$ for some fixed values $\delta s, \delta E$. Then we can use the following  linear approximation }
\be
\int_{\tau_j}^{\tau_0} \sqrt{Q(z_+; s, E)}\d z &\simeq \int_{\tau_j}^{\tau_0} \sqrt{Q(z_+;0,0)}\d z  +  \hbar\delta s\int_{\tau_j}^{\tau_0} \frac {z^2\d z}{2\sqrt{Q(z_+;0,0)}} +\hbar \delta E\int_{\tau_j}^{\tau_0} \frac {\d z}{2\sqrt{Q(z_+;0,0)}} =\\
&=-\frac {i\pi} 3 +\hbar \le(c_2{\rm e}^{\frac{2i\pi }3 (j-1) }\delta s  +c_0 {\rm e}^{-\frac{2i\pi }3 (j-1)}\delta E \ri) {\rm e}^{\frac{ 7i\pi} 6} 
\\
c_\ell &= \int_{-2^\frac 1 3}^0 \frac {z^\ell\d z}{2\sqrt{|z^4+2z|}}, \ \ \ c_0\simeq 1.9276, \ \ c_2 \simeq 0.9409.
\ee
where $\hbar$ means either $(n+1)^{-1}$ or $(n+1/2)^{-1}$ depending on the case we are considering. 
Replacing  the above expansions  in \eqref{ST0} we obtain (all to within $\mathcal O(\hbar)$);
\be
&2(n+1) \le(-\frac {i\pi} 3 +\frac 1{n+1}\le(c_2{\rm e}^{\frac{2i\pi }3 (j-1) }\delta s  +c_0 {\rm e}^{-\frac{2i\pi }3 (j-1)}\delta E \ri) {\rm e}^{\frac{ 7i\pi} 6} \ri) =  -2i\pi \le(m_j+ \frac 2 3\ri)
\\
&(2n+1)\le( -\frac {i\pi} 3 +\frac{1}{n+\frac 1 2}\le(c_2{\rm e}^{\frac{2i\pi }3 (j-1) }\delta s  +c_0 {\rm e}^{-\frac{2i\pi }3 (j-1)}\delta E \ri) {\rm e}^{\frac{ 7i\pi} 6} \ri) = - 2i\pi \le(k_j+ \frac 12\ri)
\ee
Simplifying we get the quantization rules for $\delta s, \delta E$ in identical  form provided we identify $m_j = k_j$: 
\be
&2  \le(c_2{\rm e}^{\frac{2i\pi }3 (j-1) }\delta s  +c_0 {\rm e}^{-\frac{2i\pi }3 (j-1)}\delta E \ri) {\rm e}^{\frac{ 7i\pi} 6}  =  -2i\pi \le(m_j-\frac {n-1}3\ri) + \mathcal O(\hbar).
\ee
{ Since the uncertainty on $\delta s ,\delta E$ is of order $\mathcal O(\hbar)$, the uncertainty on $s = \hbar \delta s, E= \hbar \delta E $ is of order $\mathcal O(\hbar^2)$, namely, } the two quantization conditions give two approximate lattices that  differ by $\mathcal O(\hbar^2)$ as long as $m_j-\frac {n-1}3$ remain bounded as $n\to\infty$.
\QED

\begin{remark}
The   differential equation \eqref{eq:ST_eval_problem} for $  t= 0 = \Lambda$ can be solved in terms of Whittaker ${\rm W}_{\mu,\nu}, {\rm M}_{\mu,\nu}$ functions (i.e. confluent hypergeometric functions)  [\href{https://dlmf.nist.gov/13.14}{DLMF 13.14}] as follows:
\bea
y'' - (z^4+2Jz) y =0\ \ , \ 
y_1 = \frac 1 z {\rm M}_{-\frac J 3, \frac 1 6 } \le(\frac{2 z^3} 3 \ri),\ \ 
y_2 = \frac 1 z  {\rm W}_{-\frac J 3, \frac 1 6 } \le(\frac{2 z^3} 3 \ri).
\eea
Writing $J= n+1$, $n = 0,1,2,\dots$ we have that for $n \equiv 1 \mod 3$ ($J\equiv 2\mod 3$) the solution $y_1$ is our quasi-polynomial solution. For example
\be
J=2   && y_1(z) = \le(\frac 2 3\ri)^\frac 2 3 z\, {\rm e}^{z^3/3}\nn \\
J=5  && y_1(z) = \le(\frac 2 3\ri)^\frac 2 3 \frac z 2 (z^3+2)\, {\rm e}^{z^3/3} \nn \\
J=8  && y_1(z) = \le(\frac 2 3\ri)^\frac 2 3 \frac z 7 (z^6+ 7z^3 + 7)\, {\rm e}^{z^3/3} \nn \\
J=11  && y_1(z) = \le(\frac 2 3\ri)^\frac 2 3 \frac z {70} (2z^9 + 30 z^6 + 105 z^3 + 70)\, {\rm e}^{z^3/3}\nn 
\ee
et cetera.
 This corresponds to the quantization conditions $m_j = \frac {n-1}3$, $ j=1,2,3$. 
\end{remark}

\appendix
\section{Contribution of the integrals near the turning points}
In order to estimate integrals $\int \Psi^2\d z$ near turning points we need, as a model, to estimate the corresponding integrals for the solutions of the normal form of the ODE near a simple turning point. Namely, we need to estimate integrals involving squares of solutions of the Airy equation. 
This simple lemma will be expedient.
\begin{lemma}
\label{lemmaNoice}
Let  $f=f(x), g=g(x)$ be two arbitrary solutions of $ y(x)'' - x y(x)=0$.
There are polynomials $A_n, B_n, C_n\in \C[x]$ of degrees $n+1, n,n-1$, respectively,   such that 
\be
\label{A1}
x^n f^2(x) = \frac {\d }{\d x} \le( A_n(x) f(x)^2 -B_n(x) (f'(x))^2 + C_n(x) f(x)f'(x)\ri).
\\
x^n f\,g  = \frac {\d }{\d x} \le( A_n\, f\,g -B_n\, f'g' + C_n\frac { fg'+g f'}2\ri).
\ee
They are related to each other by the following equations:
\be
A_n(x) = x B_n(x)- \frac {1} 2 B_n''(x),\qquad C_n(x) = B_n'(x), \qquad
B_n(x)+ 2x B_n'(x) - \frac {1} 2 B_n'''(x)=x^n.
\ee
The first few are:
\bea
\begin{array}{cc|c|c}
n& A_n & B_n \\
\hline
0 & x & 1 \\
\hline 
1 & \frac {x^2}3 & \frac x 3\\
\hline
2 & \frac {x^3}5-\frac {1}5 & \frac {x^2}5\\
\hline
3 & \frac {x^4}7 & \frac {x^3}7- \frac {3}7\\
\hline
4 & \frac {x^2( x^3-2)}9 & \frac {x(x^3+4)}9
\end{array}.
\eea

The leading coefficients of $A_n, B_n$ are 
\be
\label{lcoeff}
A_n(x) =\frac {x^{n+1}}{2n+1} + \mathcal O(x^n),\ \ \ B_n = \frac {x^n}{2n+1} + \mathcal O(x^{n-1}).
\ee
\end{lemma}
\noindent {\bf Proof.} 
It suffices to prove the first formula since the second is obtained by polarization.
Let $A,B,C$ be polynomials of degrees at most $n+1,n,n-1$ respectively and consider the Anzatz
\be
\label{Ans}
J = A f^2 - B (f')^2 + C f f'.
\ee
Then, using the ODE, we get 
\be
J' = \big( A' + x C\big) f^2  + \big(2A -2x B +  C'\big) ff' +\big(C-B' \big) (f')^2.    
\ee
Thus we seek polynomials such that 
\be
A' + xC = x^n,\ \ \ \ C = B', \ \ \ 2A-2xB+C'=0
\\
A' + xB' = x^n,\ \ \ \ C = B', \ \ \ 2A-2xB+B''=0
\\
A = xB- \frac {1} 2 B'', \ \ C = B',\ \ \ 
B+ 2xB' - \frac {1} 2 B'''  = x^n.
\ee
The latter equation for $B$ has a unique polynomial solution obtained by inverting the (finite-dimensional) linear operator on the space of polynomials of degree $\leq n$.  The relationship of the leading coefficients follows by plugging into \eqref{A1}.
\QED
\begin{proposition}
\label{propasympAiry}
Let $f(x)$ be an analytic function in a neighbourhood of the origin. 
\begin{enumerate}
\item[{\bf [1]}] Let $a, b>0$ . Then 
\be
\int_{-a}^b \Ai^2(\hbar^{-\frac 2 3}x) f(x) \d x 
= i \hbar^\frac 13\int_{-a}^0 \frac {f(x)\d x}{2\pi\sqrt {x}_+}  + \mathcal O(\hbar^\frac 43),
\ee
where $\sqrt {x}$ is analytic in $\mathbb{C}\backslash (-\infty,0]$  and   positive for $x>0$ and $\sqrt{x}_+$  denotes the boundary value  from the upper half plane.

\item[{\bf [2]}] Let $a, b>0$. Then 
\be
\label{A12}
&\int_{-a}^{{\rm e}^{i\pi/3}b}\hspace{-15pt} \Ai\le(\hbar^{-\frac 2 3}\omega^k x\ri) \Ai\le(\hbar^{-\frac 2 3}\omega^jx\ri) f(x) \d x = 
q_{jk}\int_{-a}^0 \frac {  i\hbar^\frac 1 3f(x)\d x}{2\pi\sqrt{x}_+} 
+r_{jk} \int_0^{{\rm e}^{i\pi/3}b} \frac {  i\hbar^\frac 1 3f(x)\d x} {2\pi\sqrt x} + \mathcal O(\hbar^\frac 43).
\ee
Here  $\omega=e^{\frac{2\pi i}{3}}$ and  $\sqrt{x}$ and $\sqrt{x}_+$ are defined as above and 
\be
q_{jk}=\le\{
\begin{array}{cc}
1  
& j=k=0
\\[4pt]
\frac {{\rm e}^{\frac{i\pi}3}}2 & j=0, \ k=1
\\[4pt]
\frac {{\rm e}^{-\frac{i\pi}3}}2 & j=0, \ k=2
\\[4pt]
0 & j=1=k
\\[4pt]
\frac 12 &j=1,\ k=2
\\[4pt]
0 & j=k=2
\end{array}
\ri.
\ \ \ \ \ 
r_{jk} =\le\{
\begin{array}{cc}
0  
& j=k=0
\\[4pt]
-\frac {{\rm e}^{\frac{i\pi}3}}2 & j=0, \ k=1
\\[4pt]
\frac {{\rm e}^{-\frac{i\pi}3}}2 & j=0, \ k=2
\\[4pt]
{\rm e}^{-\frac{i\pi}3} & j=1=k
\\[4pt]
\frac {1}2 &j=1,\ k=2
\\[4pt]
0 & j=k=2
\end{array}
\ri. .
\label{qr}
\ee
\end{enumerate}
\end{proposition}
\noindent{\bf Proof.} 
We will consider  $f(x) = x^n$, with the proof being completed simply by summing the series. 
\\
{\bf [1]} 
Then we need to compute 
$
\int_{-a}^b \Ai^2(\hbar^{-\frac 2 3}x) x^n \d x.
$
We can use Lemma \ref{lemmaNoice} and setting $\alpha = a\hbar^{-\frac 23}, \beta = b\hbar^{-\frac 2 3}$ we obtain
\be
\int_{-a}^b \Ai^2(\hbar^{-\frac 2 3}x) x^n \d x=
\hbar^{\frac {2(n+1)}3} \int_{-\alpha}^\beta \Ai^2(\xi) \xi^n \d \xi=
\nn
\\
=\hbar^{\frac {2(n+1)}3} \bigg( A_n(\xi)\Ai^2(\xi)- B_n(\xi)\, (\Ai'(\xi))^2\ + C_n(\xi) \Ai(\xi) \Ai(\xi)'\bigg) \bigg|_{-\alpha}^\beta.
\label{A24}
\ee
Recall now the asymptotic expansion of the Airy function 
\be
\label{Airy+}\Ai(s) &= \frac { {\rm e}^{-\frac 2 3 s^\frac 32}}{2\sqrt \pi  s^\frac 1 4}\big(1+ \mathcal O(s^{-\frac 32})\big), \ \ \ |s|\to +\infty,  \ \ |\arg (s)|< \pi-\epsilon\\
\label{Airy-}
\Ai(-s) &= 
\frac { \sin\le(\frac 23 s^\frac 32 + \frac \pi 4\ri)}{\sqrt \pi  s^\frac 1 4}\big(1+ \mathcal O(s^{-3})\big)
-
\frac { \cos\le(\frac 23 s^\frac 32 + \frac \pi 4\ri)}{\sqrt \pi  s^\frac 1 4}\mathcal O(s^{-\frac 32})
, \ \ \ |s|\to\infty, |\arg s|< \frac {2\pi} 3.
\ee
It is then clear that the evaluation at $\beta= \hbar^{-\frac 2 3} b$ yields exponentially small terms thanks to \eqref{Airy+}.

For the evaluation at $-\alpha  =-\hbar^{-\frac 2 3} a$ we use instead \eqref{Airy-}; thanks to \eqref{lcoeff} we see that 
\be
\hbar^{\frac {2(n+1)}3} \int_{-\alpha}^\beta \Ai^2(\xi) \xi^n \d \xi
=
\hbar^{\frac {2(n+1)}3}\frac {(-\alpha)^{n+1}} {2n+1} \frac {1}{\pi \sqrt{\alpha}}\le(1 + \mathcal O(\alpha^{-\frac 32})\ri)
=\nn
\\=
\hbar^\frac 1 3\frac {(-a)^{n+1}} {2n+1} \frac {1}{\pi \sqrt{a}}\le(1 + \mathcal O(\alpha^{-\frac 32})\ri)
=
\hbar^\frac 1 3 \int_{-a}^0 \frac{x^{n}\d x}{2\pi\sqrt{-x}} + \mathcal O(\hbar^\frac 43).
\ee
Finally, we observe that $\sqrt{-x} = -i\sqrt{x}_+$ for $x\in \R_-$. \\
{\bf [2]} All these integrals involve oscillatory direction for each of the arguments. The computation is of entirely similar nature, where it is particularly important to pay great care at the relative phases.

Using Lemma \ref{lemmaNoice} we need to estimate integrals of the form ($\Ai_j(\xi) := \Ai(\omega^j\xi)$)
\be
\int_{-a}^{{\rm e}^{\frac {i\pi}3}b} x^n \Ai_{j}(x \hbar^{-\frac 23})\Ai_{k}(x \hbar^{-\frac 23})\d x = 
\hbar^{\frac {2(n+1)}3} \int_{-\alpha}^{{\rm e}^{\frac {i\pi}3}\beta} \xi^n  \Ai_{j}(\xi)\Ai_{k}(\xi)\d \xi
=
\hbar^{\frac {2(n+1)}3}H_{jk}(\xi)\Bigg|_{-\alpha}^{{\rm e}^{\frac {i\pi}3}\beta }\nn\\
H_n^{(j,k)} := \le[
A_n \Ai_{j}\Ai_{k} - B_n \Ai_j'\Ai_k' -C_n \frac {\Ai_j'\Ai_k + \Ai_j \Ai_k'}{2}  
\ri]
\ee
At the point $\alpha = \hbar^{-\frac 2 3} a$ a direct computation using the asymptotic properties \eqref{Airy+}, \eqref{Airy-} yields:
\be
\label{HAsy}
-\hbar^{\frac {2(n+1)}3}H_n^{(j,k)}(-\hbar^{-\frac 2 3}a)
\simeq
 i \int_{-a}^0 \frac{\hbar^{\frac 1 3}x^n\d x}{2\pi \sqrt{x_+}} \times
q_{jk} + \mathcal O(\hbar^\frac 4 3)
\ee
Similarly one finds
\be
\label{HAsy2}
\hbar^{\frac {2(n+1)}3}H_n^{(j,k)}(\hbar^{-\frac 2 3}b{\rm e}^{i\pi/3})
\simeq
 i \int_{0}^{b{\rm e}^{i\pi/3}} \frac{\hbar^{\frac 1 3}x^n\d x}{2\pi \sqrt{x}} \times r_{jk} +  \mathcal O(\hbar^\frac 4 3)
\ee
with $q_{jk}, r_{jk}$ as in \eqref{qr}.
Now  the integral \eqref{A12}  with $f(x)=x^n$ becomes 
\be
\int_{-a}^{{\rm e}^{i\pi/3}b} \Ai(\hbar^{-\frac 2 3}\omega^k x) \Ai(\hbar^{-\frac 2 3}\omega^jx) x^n \d x  
=
\hbar^{\frac {2(n+1)}3} 
H_n^{(j,k)}(\xi)\bigg|_{-\hbar^{-\frac 2 3 }a}^{\hbar^{-\frac 2 3 } {\rm e}^{\frac {i\pi}3}b}.
\ee
Putting together the two terms we obtain the proof. \QED

%
As a consequence of the Lemma \ref{lemmaNoice} we can estimate the contribution of integrals involving $\Psi^2$ in a neighbourhood of a turning point.

\begin{theorem}
 \label{mainestimate}
 Let $\tau$ be a simple turning point for the potential $Q(x)$, $\gamma_d$  a steepest  descent  $\ominus$ path and $ \gamma_o$ the oscillatory path on the opposite side of $\tau$: we choose two points $\tau_+\in \gamma_d, \tau_-\in \gamma_o$, in the neighbourhood of $\tau$ at finite positive distance. Let $\Psi$ be a solution of the ODE 
 \be
\hbar^2 \Psi''(x)  - Q(x)\Psi(x)=0.
 \ee
 Suppose that $\Psi$ is asymptotic to the recessive formal solution $\wkb{\tau}{+}$.  Then
 \be
 \int_{\tau_-}^{\tau_+} \Psi^2\d z= \int_{\tau_-}^\tau \frac {2i\hbar\, \dd z}{\sqrt {Q(z_+;s,E)}} + \mathcal O(\hbar^2).
 \ee
 with the branch-cut of $\sqrt {Q}$ running along the oscillatory path and the determination the one that has regative real part on the descent path $\gamma_d$.
 \end{theorem}
\begin{figure}
\centering
\includegraphics{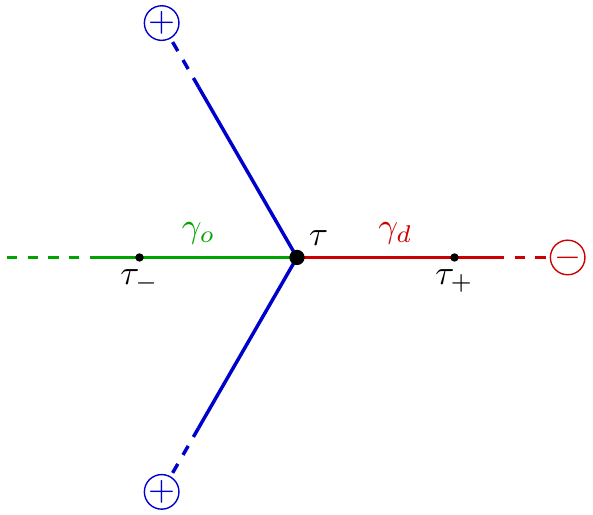}
 \caption{The integration for Theorem \ref{mainestimate}.}
\end{figure}

\noindent {\bf Proof.}
In \cite{voros_return_quartic_oscillator} it is shown that there is a conformal mapping $\xi(z;\hbar^2)\in \mathcal O(z)\otimes \C[[\hbar^2]]$ that maps $z=\tau(=0)$ to $\xi(\tau;\hbar^2) \equiv 0$ (identically in $\hbar$) that 
maps the equation to the Airy equation:
\bea
\hbar^2 \pa_z^2 \Psi(z)- Q(z)\Psi(z)=0\ \ \ \mapsto\ \ \ \ 
\hbar^2 \pa_\xi^2 f(\xi) - \xi f(\xi) =0.
\eea
The relationship between the two equations is as follows 
\be
\label{Sch}
Q(z) =\le(\frac {\d \xi}{\d x} \ri)^2\ \xi -\frac {\hbar^2} 2 \{\xi, z\}, \ \ \ \ \Psi(z) = \frac {f(\xi)}{ \sqrt{\d \xi /\d z}}
\ee
where $\{,\}$ is the Schwarzian derivative 
\be
 \{\xi, z\} =\frac {\xi'''}{\xi'} - \frac 3 2 \le(\frac{\xi''}{\xi'}\ri)^2.
\ee
The formal series $\xi(z;\hbar^2)$ is  Borel--resummable to an analytic conformal mapping whose asymptotic expansion as $\hbar\to 0_+$ coincides with the formal series \cite{kawai_takei_algebraic_analysis_sing_perturbation}.

The arc $[\tau,\tau_+]$ on the contour $\gamma_d$ is mapped to a segment $[0,\xi_+]\subset \R_+$ in the $\xi$--plane, and similarly the arc $[\tau_-,\tau]\subset \gamma_o$ maps to $[\xi_-,0]\subset \R_-$.

The function $\Psi$ is the recessive solution on $\gamma_+$ and hence it must  map to a multiple of the $\Ai$ function within a whole neighbourhood of $\tau$. 
 Thus the integral to be estimated translates to 
\be
\int_{\tau_-}^{\tau_+} \Psi(z;\hbar )^2 \d z = C(\hbar) \int_{\xi_-}^{\xi_+}  \frac {\Ai\le(\hbar^{-\frac 23 }\xi\ri)^2}{ \xi'(z^{-1}(\xi);\hbar^2)^2} \d \xi,
\label{616}
\ee
where $C(\hbar)$ is an appropriate proportionality constant that depends on the chosen normalization of $\Psi$. Our choice is to normalize $\Psi$ so that it is asymptotic to the recessive formal solution $\wkb{\tau}{+}$. To fix $C(\hbar)$ we can consider the asymptotics of the Airy function at $\tau_+$  \eqref{Airy+} and compare it to the WKB solution.
Indeed we find that 
\be
\frac {\Ai\le(\hbar^{-\frac 43 }\xi_+\ri)^2}{ \xi'(z^{-1}(\xi_+);\hbar^2)} \simeq
\frac {\hbar^\frac 1 3 {\rm e}^{-\frac 4{3\hbar} \xi_+^\frac 32} }{4\pi \xi'(z^{-1}(\xi_+;\hbar^2) \sqrt{\xi_+}} \simeq 
\frac {\hbar^\frac 1 3 {\rm e}^{\frac 2{\hbar} \int_\tau^{\tau_+} \sqrt{Q}\d z }} {4\pi \sqrt{Q(\tau_+;s,E)} } \simeq \frac {1}{4\hbar^{\frac 23}\pi} \le(\wkb{\tau}{-}\ri)^2,
\ee
so that $C(\hbar) = 4 \hbar^\frac 23 \pi (1+ \mathcal O(\hbar^2))$.
The  integral involving the Airy function in \eqref{616}  is  of the general form   of Proposition \ref{propasympAiry} and hence, to within $\mathcal O(\hbar^\frac 43)$ we have
\be
\int_{\xi_-}^{\xi_+}  \frac {\Ai\le(\hbar^{-\frac 23 }\xi\ri)^2}{ \xi'(z^{-1}(\xi);\hbar^2)^2} \d \xi=\frac {i \hbar^\frac 13}{2\pi}
\int_{\xi_-}^{0}  \frac {\d \xi}{ \xi'(z^{-1}(\xi);\hbar^2)^2 \sqrt{\xi(z^{-1}(\xi)_+;\hbar^2)}}.
\ee
Now from \eqref{Sch} we get 
\be
\xi' \sqrt {\xi} = \sqrt{Q(z)} + \mathcal O(\hbar^2), 
\ee
and hence, to within $\mathcal O(\hbar^2)$, we have 
\bea
\int_{\xi_-}^{0}  \frac {\d \xi}{ \xi'(z^{-1}(\xi);\hbar^2)^2 \sqrt{\xi(z^{-1}(\xi);\hbar^2)^2}}
=\int_{\xi_-}^{0}  \frac {\d \xi}{ \xi'(z^{-1}(\xi);\hbar^2) \sqrt{Q(z^{-1}(\xi))}} 
=\int_{\tau_-}^{\tau}  \frac {\d z}{ \sqrt{Q(z)}}.
\eea
Using the estimate of  $C(\hbar)$ and substituting into \eqref{616} we obtain the statement of the theorem.
\QED

\subsection{Proof of Theorem \ref{integralA1}}
\label{section_proof}
We will consider in detail only the case $D$ because it contains all the intricacies.

Since we want to impose that the integral $I_{13}$ vanishes, it suffices to determine the integral of a function that is proportional to $y_n= p_n {\rm e}^{2\theta}$. Keeping this in mind we then observe that in the region $(A_1)$ along the Stokes curve $(\tau_1,\infty_1)$ the function is {\it recessive}  (as $n\to\infty$ and as $x\to\infty_1$) and hence it must be proportional to $\Psi^{(A_1)}_+(z;\hbar)$ and thus asymptotic to $\wkb{\tau_1}{+}$. We will then forget about $y_n$ and integrate instead the function $\Psi^{(A_1)}_+(z;\hbar)^{2}$ along $\gamma$: recall that this function is entire since it is the solution of the ODE \eqref{eq:schrodinger_wkb_equation}. 

Let us refer to Fig. \ref{WKB-regions}. We choose the contour $\gamma$ to follow the steepest descent  from $\infty_1$ to $\tau_1$, then the branch-cut to $\tau_0$, then to $\tau_2$ and the descent path to $\infty_3$. The branch-cut will be chosen to follow the  anti--Stokes curve in a neighbourhood of each turning point.  
\begin{figure}
\resizebox{0.6\textwidth}{!}{
    \includegraphics{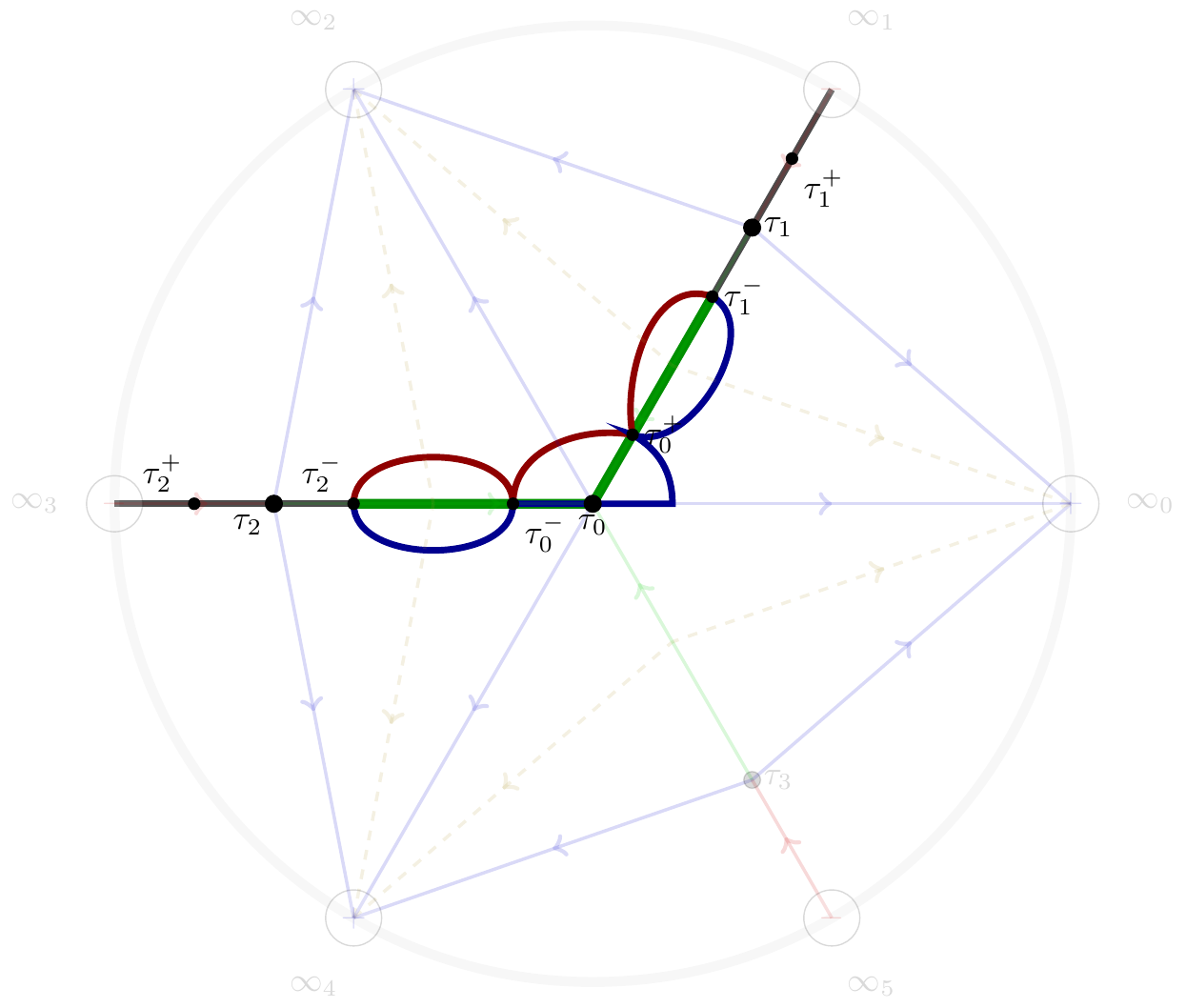}
}
\resizebox{0.4\textwidth}{!}{
    \includegraphics{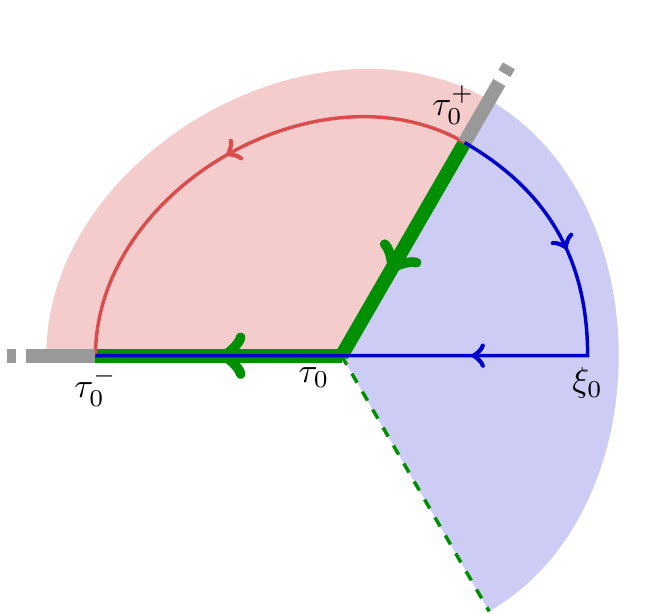}
}
\caption{The splitting of the integration of \eqref{abc} along the three contours for each of the three integrands near $\tau_0$.}
\label{figurebest}
\end{figure}

We then  split  our contour $\gamma = \bigcup_{j=1}^7 \gamma_j$ in several pieces
\be
\infty_1\mathop{ \longrightarrow}^{\gamma_1}  \tau_1^+ 
\mathop{ \longrightarrow}^{\gamma_2} 
\tau_1^- 
\mathop{ \longrightarrow}^{\gamma_3}
 \tau_0^+
 \mathop{ \longrightarrow}^{\gamma_4}
    \tau_0^-
 \mathop{ \longrightarrow}^{\gamma_5}
       \tau_2^+ 
 \mathop{ \longrightarrow}^{\gamma_6}
     \tau _2^-
 \mathop{ \longrightarrow}^{\gamma_7}
      \infty_3.
\ee

Observe next that the solution $\Psi:= \Psi^{(A_1)}_+$ is recessive both on the Stokes line $\infty_1\to \tau_1$ as well as on the Stokes line $\tau_2\to \infty_3$ because  we are already on a point of the ES spectrum, where the Stokes matrices satisfy $\mathbb S_0=\mathbb S_2= \mathbb S_4=\mathbf 1$. More precisely  it follows that $\Psi_+^{(A_1)} = \Psi_{+}^{(D_3)} = {\rm e}^{v_{12}} \Psi_+^{(D_1)} =
 {\rm e}^{v_{12}} \Psi_+^{(A_2)}$.

\paragraph{Contributions of $\gamma_1,\gamma_7$.} On both these contours the integral is exponentially small as $\hbar\to 0_+$ since $\Psi=\Psi_+^{(A_1)}$ is recessive on $\gamma_1$ and $\Psi = {\rm e}^{v_12} \Psi_{+}^{(D_1)}$ is recessive on $\gamma_7$. 

\paragraph{Contributions of $\gamma_2,\gamma_6$.} 
These  are estimated by the use of Theorem \ref{mainestimate} 
\be
\label{gamma2}
 \int_{\tau_1^+}^{\tau_1^-} \Psi^2\d z&=  \int_{\tau_1^+}^{\tau_1^-} \frac {2i\hbar\, \dd z}{\sqrt {Q(z_-;s,E)}} + \mathcal O(\hbar^2),
 \\
 \int_{\tau_2^+}^{\tau_2^-} \Psi^2\d z&=  {\rm e}^{2v_{12}} \int_{\tau_2^+}^{\tau_2^-} \le(\Psi_+^{(D_1)} \ri)^2\d z= 
 {\rm e}^{2v_{12}} \le(  \int_{\tau_2^+}^{\tau_2} \frac {2i\hbar\, \dd z}{\sqrt {Q(z_-;s,E)}} + \mathcal O(\hbar^2)\ri).
 \label{gamma6}
\ee
The boundary value is the $-$ because the orientation of the branch-cut that we have chosen is the one egressing from the turning point, while in Theorem \ref{mainestimate} it was the one towards the turning point (and the boundary value was the $+$ one). 
\paragraph{Contributions of $\gamma_3,\gamma_5$.} 
Along either  paths we are away from the turning points and can substitute the asymptotic expressions in terms of the WKB solutions.  
We can use the Riemann--Hilbert problem \ref{RHPWKB} to see that  the function $\Psi = \Psi_+^{(A_1)}$ has the asymptotic behaviour 
\be
\Psi_+^{(A_1)}(z;\hbar) &\simeq \left\{
\begin{array}{ll}
\wkb{\tau_1}{+}(z;\hbar) + i\wkb{\tau_1}{-}(z;\hbar) &  z\in (B_1^r)
\\[4pt]
{\rm e}^{-v_{10}}\wkb{\tau_0}{+}(z;\hbar) + i{\rm e}^{v_{10}}\wkb{\tau_0}{-}(z;\hbar) &  z\in (B_1^\ell)
\\[4pt]
 {\rm e}^{v_{12}} \le({\rm e}^{v_{20}}\wkb{\tau_0}{+}(z;\hbar) - i{\rm e}^{-v_{20}}\wkb{\tau_0}{-}(z;\hbar)\ri)& z\in (C_1^r)
\\[4pt]
 {\rm e}^{v_{12}} \le(\wkb{\tau_2}{+}(z;\hbar) - i\wkb{\tau_2}{-}(z;\hbar)\ri)& z\in (C_1^\ell)
 \end{array}
 \ri.
\ee
The different expressions in $(B_1^{\ell, r}) $ are due simply to the different choice of normalization point for the WKB formal solutions, but we can use either of the two expressions also as asymptotic expansion in the other. Ditto for the pair $(C_1^{\ell, r})$.
Consequently we can estimate the contribution of the whole  $\gamma_3$ using the asymptotic expression for $(B_1^r)$ as follows:
\be
\label{bulk}
\int_{\tau_1^-}^{\tau_0^+} \Psi^2\d z &\simeq \int_{\tau_1^-}^{\tau_0^+} \le((\wkb{\tau_1}{+})^2 -(\wkb{\tau_1}{-})^2 \ri)\d z + 2i 
 \int_{\tau_1^-}^{\tau_0^+}
 \wkb{\tau_1}{+}\wkb{\tau_1}{-} \, \d z ,
\ee
with similar expression for the integral along $\gamma_5$.
The contribution of the first integral in the right side of \eqref{bulk} is of order $\mathcal O(\hbar^2)$ as we show a few lines below. 
The main contribution comes instead  from the last integral and it yields
\be
  2i 
 \int_{\tau_1^-}^{\tau_0^+}
 \wkb{\tau_1}{+}\wkb{\tau_1}{-} \, \d z  =  2i 
 \int_{\tau_1^-}^{\tau_0^+}
\frac {\hbar \d z}{\sqrt{Q(z_-, s,E)}} +\mathcal O(\hbar^2),
\label{gamma3}
\ee
where the $-$ boundary value in the last integral is due to the fact that the regions $(B_1^{\ell, r})$ lie on the $-$ side of the branch-cut, with the orientation that we have chosen. In particular this contribution together with the contribution \eqref{gamma2} yield a single integration from $\tau_1$ to $\tau_0^+$. Similar considerations apply to the integral along $\gamma_5$ which yield 
\be
\int_{\tau_0^-}^{\tau_2^+} \Psi^2 \d z \simeq
\xi_{12} \le( \int_{\tau_0^-}^{\tau_2^+}\frac { 2i  \hbar \d z}{\sqrt{Q(z_+;s,E)}} + \mathcal O(\hbar^2)\ri)
\label{gamma5}
\ee
To see that the first integral in \eqref{bulk} is sub-dominant, for example consider the integration of $(\wkb{\tau_1}{+})^2$. We can deform the path of integration (at fixed endpoints) into the region $(C_3)$ where the formal solution is recessive. Then a simple estimate shows that the only contribution come from the neighbourhood of the endpoints of integration (at which points  the real part of the exponential is  zero). Then we can use Laplace's method to easily estimate their contribution of order\footnote{Note that the function being integrated has already an $\hbar$ in front, and the integral contributes another order $\mathcal O(\hbar)$.} $\mathcal O(\hbar^2)$. A similar reasoning applies to the integration of $(\wkb{\tau_1}{-})^2$ where instead we deform the contour within the region $(B_1)$. 

\paragraph{Contribution of $\gamma_4$.}
We are going to show that this integral is estimated as follows:
\be
\label{gamma4}
-\int_{\tau_0^+}^{\tau_0}  \frac {2i\hbar( 1 + \mathcal O(\hbar)) \dd z}{\sqrt{Q(z_+;s.E)}} -\le(1+ {\rm e}^{2v_{12}}\ri) \int_{\tau_0}^{\tau_0^-}  \frac {2i\hbar( 1 + \mathcal O(\hbar)) \dd z}{\sqrt{Q(z_+;s.E)}}
\ee

Our function $\Psi = \Psi^{(A_1)}_+ = \Psi_+^{(B_1^l)}  + i\Psi_-^{(B_1^l)}$. 
Near $\tau_0$ the function $\Psi$ is, along the contours of integration, an oscillatory solution. 
We express it in terms of the normalized solutions in $(B_1^r)$:
\be
\Psi^{(A_1)}_+ = {\rm e}^{v_{10}}\Psi_+^{(B_1^r)} + i{\rm e}^{-v_{10}}\Psi_-^{(B_1^r)}
 \ \ \Rightarrow\ \ \ 
 \nn
 \\
(\Psi^{(A_1)}_+)^2  = {\rm e}^{-2v_{10}}\underbrace{\le(\Psi_+^{(B_1^r)}\ri)^2}_{{\bf (a)} } -{\rm e}^{2v_{10}}\underbrace{\le(\Psi_-^{(B_1^r)}\ri)^2}_{{\bf (b)}} +2i \underbrace{\Psi_+^{(B_1^r)} \Psi_-^{(B_1^r)}}_{\bf (c)}.
\label{abc}
\ee
Refer now to Fig. \ref{figurebest}: the term marked ${\bf(a)}$ is recessive in the blue-shaded region, the term {\bf (b)} in the pink-shaded region while {\bf (c)} is oscillatory throughout. 
Consequently we will split the integration of $\Psi^2$ into three paths from $\tau_0^{+}$ to $\tau_0^-$ and integrate {\bf(a)} along the blue path, {\bf (b)} along the red path and {\bf(c)} along the green path. 

The whole contribution of ${\bf(b)}$ is then sub-leading and of order $\mathcal O(\hbar^2)$ where the main contributions come from the neighourhoods of $\tau_0^\pm$. 

The contribution of ${\bf (a)}$ near $\tau_0^+$ and $\xi_0$ similarly is of order $\mathcal O(\hbar)^2$ (with the main contribution coming solely from the neighbourhood of $\tau_0^-$). The integration from $\xi_0$ to $\tau_0^{-}$ is achieved by Theorem \ref{mainestimate} and  we have 
\be
{\bf (a)} \mapsto
{\rm e}^{-2v_{10}}
\int_{\xi_0}^{\tau_0^-}\le(\Psi_+^{(B_1^r)}\ri)^2\d z
 = {\rm e}^{-2v_{10}}\int_{\tau_0}^{\tau_0^-} \frac {2i\hbar(1+\mathcal O(\hbar))  \d z}{\sqrt{Q(z_+;s,E)}}.
 \label{A41}
\ee
We then have the contribution of {\bf(c)} along the green path; for this we need to use a similar reasoning as in the proof of Theorem \ref{mainestimate}.
The function $ \Psi_+^{(B_1^r)}$ is recessive in the blue-shaded region of Fig. \ref{figurebest}; thus in the coordinate $\xi(z;\hbar)$ along the direction $\arg \xi=0$ and is therefore proportional to $\Ai(\hbar^{-\frac 2 3} \xi)/\sqrt{\xi'} $ after the change of coordinates. Similarly the function $ \Psi_-^{(B_1^r)}$ is recessive in the pink-shaded region and hence proportional to  $\Ai(\hbar^{-\frac 2 3}\omega^2 \xi)/\sqrt{\xi'} $ , with $\omega={\rm e}^{2i\pi/3}$
\be
\int_{\tau_0^+}^{\tau_0^-} \Psi_+^{(B_1^r)} \Psi_-^{(B_1^r)}\d z =
C(\hbar) \int_{\xi_+}^{\xi_-} \frac {\Ai(\hbar^{-\frac 2 3}\xi )\Ai (\hbar^{-\frac 2 3}\omega^2 \xi)\d \xi}{(\xi')^2} \simeq C(\hbar) \le(\int_{\tau_0^+}^{\tau_0^-} \frac {i\hbar^\frac 13 \d z}{4\pi\sqrt{Q(z_+;s,E)}}  + \mathcal O(\hbar^\frac 4 3)\ri)
\ee
The latter integral was estimated as in the first part of this proof using  Lemma \ref{propasympAiry} by using part {\bf [2]} of Proposition \ref{propasympAiry}: it requires the case $j=0,k=2$ in \eqref{qr}. 
The proportionality constant $C(\hbar)$ is now estimated by matching the asymptotic behaviour; 
\be
 \frac {\hbar}{\sqrt{Q(z;s,E)_+}}   \simeq\Psi_+^{(B_1^r)} \Psi_-^{(B_1^r)}\simeq C(\hbar) \frac {\hbar^\frac 1 3{\rm e}^{\frac{i\pi} 3}}{4\pi \xi^\frac 12\ \xi'} 
\ee
So,  the coefficient $(1+ {\rm e}^{2v_{12}})$ in \eqref{gamma4} comes from the contribution of {\bf (c)} and {\bf (a)} while the first term in \eqref{gamma4} is coming only from the integration of {\bf (a)} from \eqref{abc}.
We have thus established \eqref{gamma4}. To complete the proof we now recall that the parameters $\xi_{jk} = {\rm e}^{2v_{jk}}$ satisfy the relation \eqref{eq:p-parametrisation} so that $1  + \xi_{10} = -\xi_{20}^{-1}$ so that all the integrals combine nicely. 
Indeed putting together all contributions we obtain 
\bea
\int_\gamma\Psi^2 \d z  =& 
  \overset{(\gamma_2)}{\int_{\tau_1}^{\tau_1^-} \frac {2i\hbar\, \dd z}{\sqrt {Q(z_-;s,E)}}} 
 + 
 \overset{(\gamma_3)}{\int_{\tau_1^-}^{\tau_0^+}
\frac {2i\hbar \d z}{\sqrt{Q(z_-, s,E)}}}  
\overset{(\gamma_4)}{-\int_{\tau_0^+}^{\tau_0}  \frac {2i\hbar( 1 + \mathcal O(\hbar)) \dd z}{\sqrt{Q(z_+;s.E)}}}
 +\\
 &-\overset{(\gamma_4)'}{ \le(1 + {\rm e}^{-2v_{10}}\ri) \int_{\tau_0}^{\tau_0^-}  \frac {2i\hbar( 1 + \mathcal O(\hbar)) \dd z}{\sqrt{Q(z_+;s.E)}} }
 +\overset{(\gamma_5)}{\xi_{12}  \int_{\tau_0^-}^{\tau_2^+}\frac { 2i  \hbar (1+\mathcal O(\hbar)) \d z}{\sqrt{Q(z_+;s,E)}}} 
+ 
\label{A45}
 \\
&+
 \overset{(\gamma_6)}{\xi_{12}   \int_{\tau_2^+}^{\tau_2} \frac {2i\hbar(1+\mathcal O(\hbar))\, \dd z}{\sqrt {Q(z_-;s,E)}} } 
\eea
where above each term we have recalled with contribution it comes from. We now see that the contribution indicated with $(\gamma_4)'$ and $(\gamma_5)$ combine to give a single integral: this is due to the fact that ${\rm e}^{-2v_{10}} = \xi_{10}^{-1}$ and from \eqref{eq:p-parametrisation} we see 
$$
1+\xi_{10}^{-1} = -\xi_{30}.
$$
Then, using the identity
\be
\int_{\tau_3}^{\tau_0} \So(z_+;\hbar)\d z  =2i\pi (n+1) - \int_{\tau_1}^{\tau_0} \So(z_+;\hbar)\d z- \int_{\tau_1}^{\tau_0} \So(z_+;\hbar)\d z  = 2i\pi (n+1) + \int_{\tau_1}^{\tau_2} \So(z_+;\hbar)\d z
\ee
we deduce  that $\xi_{30} = (\xi_{10}\xi_{20})^{-1} = \xi_{12} = {\rm e}^{2v_{12}}$, which allows us to write the two terms on line \eqref{A45}  as a single integral.
Adding all the contributions yields the final statement.
\QED

\section{Numerical verifications}
\label{Numerics}
We performed numerical verifications of the formulas in view of the adage, valid in many walks of life,  of ``trust but verify".
For the JM case we proceed as indicated below.
\begin{enumerate}
\item Compute numerically in arbitrary precision the list  of zeros, $\scr Z_n$,  of the VY polynomial $Y_n(t)$, i.e., the poles of the rational solution $u_n$ \eqref{ratlsoln}. 

For each of $a \in \scr Z_n$, compute the corresponding value of the parameter $b$ appearing in \eqref{expu}. Thus we can define, for each $a\in \scr Z_n$ a corresponding value of $\Lambda=\frac {7a^2}{36} + 10 b$ as in \eqref{eq:JM_potential}. Let us call $\scr L_n$ the list of the corresponding values of $\Lambda$. 
\item Let $\scr S_n:= \scr Z_n/(n+\frac 1 2)^{\frac 2 3}$ and $\scr E_n := \scr L_n/(n+\frac 1 2)^{\frac 4 3}$ as per scaling \eqref{scalingJMU}. 
For each pair $(s_\ell,E_\ell)$ in $\scr S_n, \scr E_n$ we construct the corresponding potential $Q(z;s_\ell,E_\ell)= z^4 + s_\ell z^2+2z +E_\ell $ and evaluate numerically the Voros symbols {\it including the subleading term}, using Prop. \ref{propsublead}, along the relevant cycles of the Riemann surface of $\sqrt Q$. 
The numerical verification consists in checking 
\be
(2n+1) \oint \sqrt {Q(z;s_\ell,E_\ell)}\d z + \frac 1{2n+1} \oint S_1(z)\d z  \simeq i\pi + 2i\pi m_\gamma, 
\ee
i.e. an odd multiple of $i\pi$. We tested up to $n = 26$ and the numerics indeed supports the formula. It is interesting to observe that the leading order computation yields (unsurprisingly) less accurate approximations of odd multiples of $i\pi$, in particular in that it has a small but still non-negligible real part.
\end{enumerate}
For the ST case we proceed similarly as follows.
\begin{enumerate}
\item Compute numerically in arbitrary precision the list  of zeros, $\scr Z_n$,  of the ST discriminant $D_n$ \eqref{STDn}.  

For each of $t \in \scr Z_n$, compute the characteristic polynomial of the matrix $M_n(t)$ \eqref{STMn} and find the double eigenvalue $\lambda$: due to the numerical error in the evaluation of the zero of the discriminant, one has to find the two roots that are very close to each other.  
Once we find $\lambda$ we define $\Lambda = \lambda + \frac {t^2}4$ as per \eqref{eq:polynomial_ODE}. 
We thus construct corresponding lists $\scr Z_n, \scr L_n$ as in the previous case, for the values in the ES spectrum. 
\item We perform the appropriate scaling $\scr S_n:= \scr Z_n/(n+1)^{\frac 2 3}$ and $\scr E_n := \scr L_n/(n+1)^{\frac 4 3}$ as per scaling \eqref{eq:wkb-scaling}. 
For each pair $(s_\ell,E_\ell)$ in $\scr S_n, \scr E_n$ we construct the corresponding potential $Q(z;s_\ell,E_\ell)= z^4 + s_\ell z^2+2z +E_\ell $ and we compute the Voros symbols $v_{jk}$ to subleading order as for the previous case. Let us call  $\xi_{jk} = {\rm e}^{2v_{jk}}$ as in the main text. Then the numerical verification consists in two checks:
\begin{enumerate}
\item We verify that the three Fock--Goncharov parameters $\xi_{10}, \xi_{20}, \xi_{30}$ satisfy the rational parametrization \eqref{eq:p-parametrisation}. 
\item We verify the equation \eqref{2eigs}, which returns numerically correct within less than $1\%$ for $n=80$ with a higher accuracy of up to $0.01\%$, unsurprisingly, when we check zeros that lie away from the boundary of the triangular region (elliptic region) with boundaries described in Section  \eqref{ellreg}. 
\end{enumerate}

\end{enumerate}
\subsection{Overlapping patterns}

\begin{figure}
\includegraphics[width=0.4\textwidth]{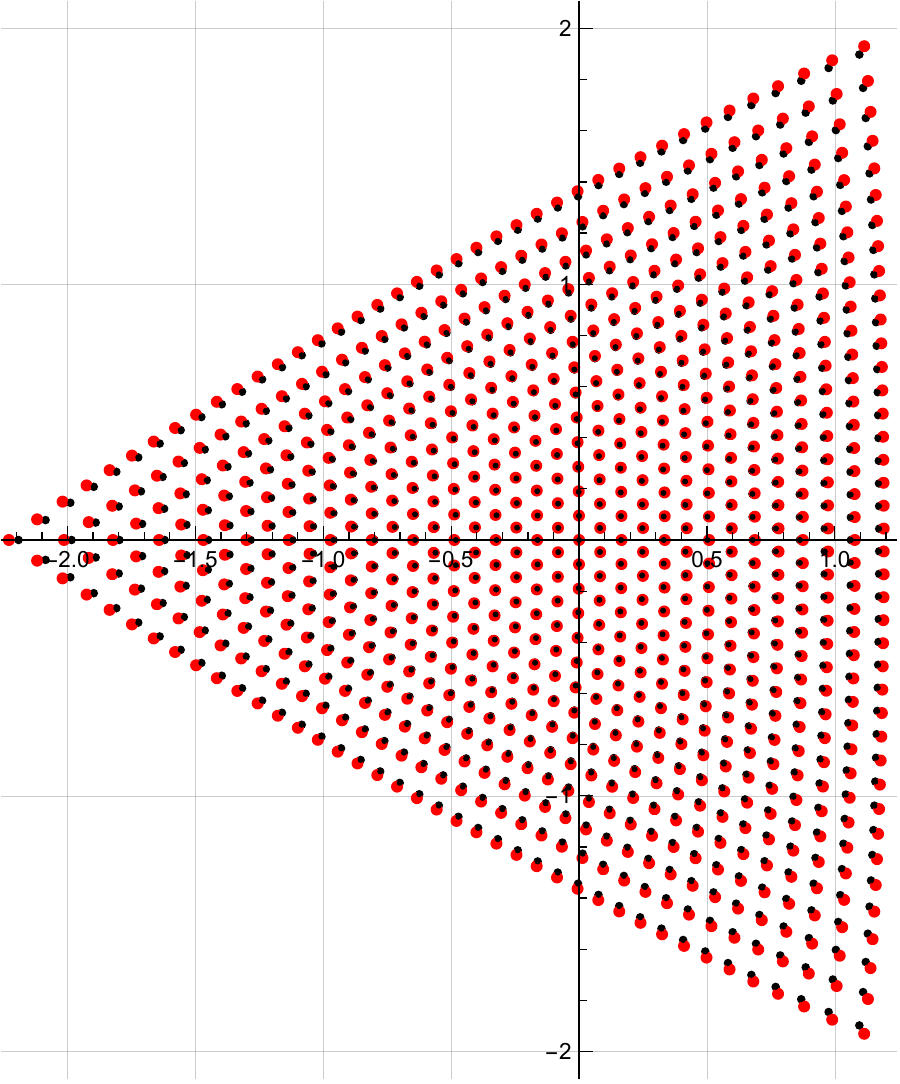}
\hfill
\includegraphics[width=0.4\textwidth]{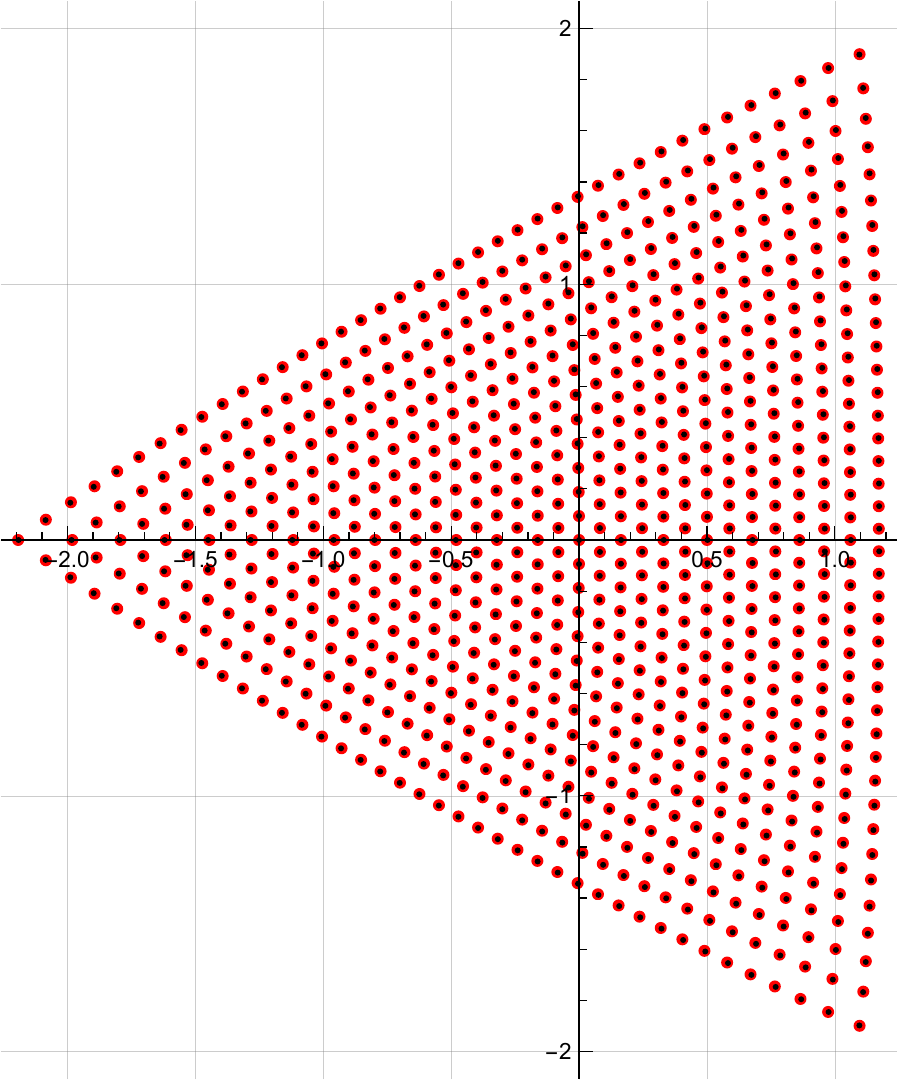}
\caption{The ES spectrum (black) and the VY zeros (red) for $n=40$. On the left the zeros of the ST discriminant are scaled as $(n+1)^{-\frac 2 3}$, which is the ``natural'' scaling. On the right they are scaled by $n^{-\frac 2 3}$. }
\label{comparison}
\end{figure}

According to our analysis, the natural scaling for the points in the ES spectrum is the one indicated in \eqref{eq:wkb-scaling} for the ST case, and \eqref{scalingJMU} for the VY polynomial case.  In particular for the ST case $\hbar = (n+1)^{-1}$ and for the VY case $\hbar = (n+\frac 1 2)^{-1}$. 

If we plot the zeros with these exact scalings, we obtain the two lattices shown in Fig. \ref{comparison}, on the left. 

However, if we use the scaling $\hbar = n^{-\frac 23}$ for the ES spectrum, we obtain an almost perfect match as shown on the right pane of Fig. \ref{comparison}.

We cannot find a mathematical justification of this coincidence, beyond what we already have proved; namely we can justify the coincidence of the two lattices in a $\mathcal O(\hbar)$ neighbourhood of the origin, as well as the matching, slowly modulated,  local geometry of either lattices (Prop. \ref{locallattice}). However we cannot quantify the reason why the ``wrong'' scaling seems to yield a much better match.

We have, however, verified that the discrepancy between the two (scaled) lattices decreases indeed like $\mathcal O(n^{-1})$ in the regions at finite distance from the origin, and as $\mathcal O(n^{-2})$ in the $\mathcal O(\hbar)$--region around the origin. 

\begin{figure}
\phantom{.}\hfill \includegraphics[width=0.3\textwidth]{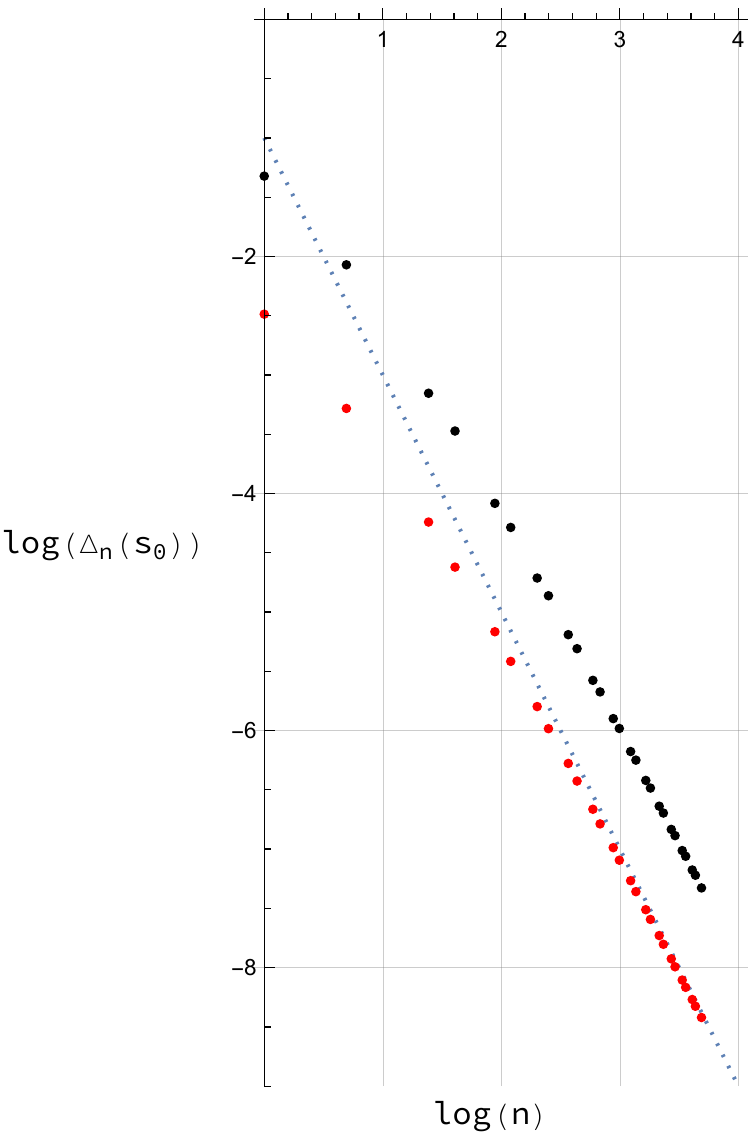}\hfill
\includegraphics[width=0.3\textwidth]{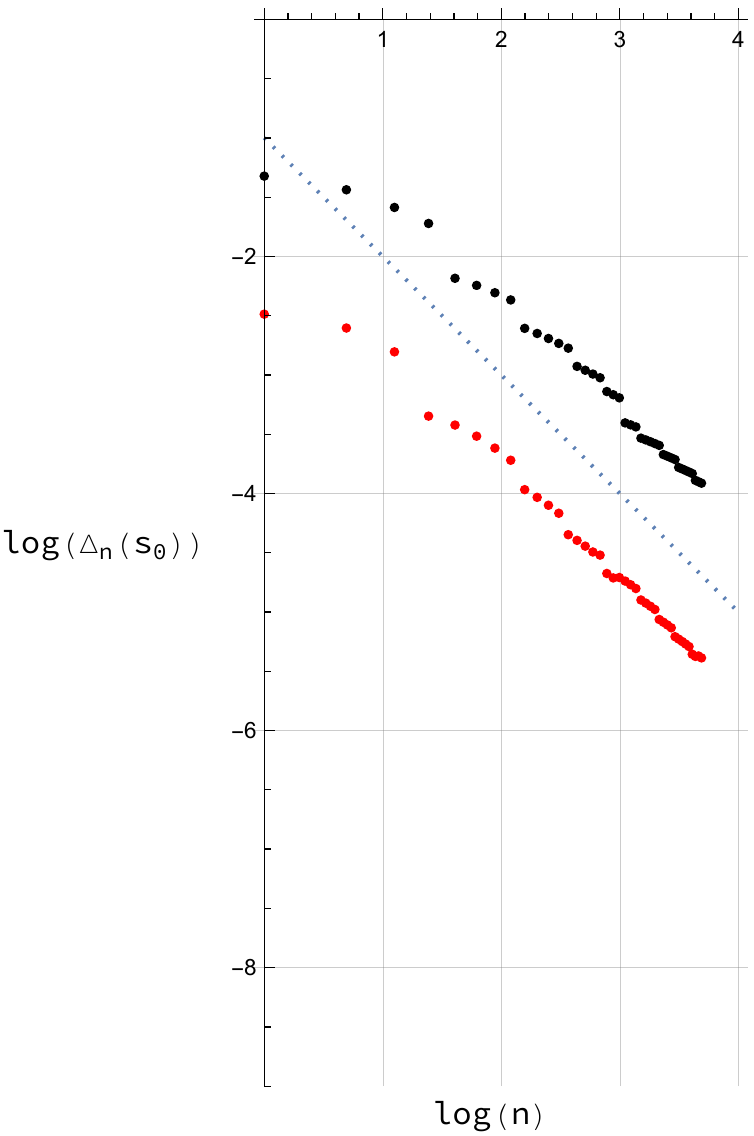}\hfill\phantom{.}
\caption{The local discrepancies $\Delta_n(s_0)$ against $n$ in a log-log plot for two values of $s_0$. The red dots correspond to the discrepancy with the ``wrong'' scaling $n^{-\frac 2 3}$. Although the ``wrong'' scaling is better, the rate of convergence is the same in  the local lattices. In the left picture $s_0=0$ and the slope is $-2$ (the dotted line is plotted for reference). In the right picture $s_0=1+i$ and the slope is $-1$ as expected.}
\label{discrepancy}
\end{figure}

To effect this test, we choose a point, $s_0$ in the elliptic region and find the closest points of either lattices to $s_0$. Let $\Delta_n(s_0)$ { denote the distance between these two points belonging to the different lattices}; we call it the ``local discrepancy''.  Then we plot $\ln \Delta_n(s_0)$  against $\ln n$.  
If we choose $s_0$ at finite positive distance from the origin, the slope of this graph is $-1$ while for $s_0\simeq 0$ the slope is $-2$.  This plot of  local discrepancy $\Delta_n(s_0)$ is reported in Fig. \ref{discrepancy} in the two regimes of $s_0\simeq 0$ and $s_0\neq 0$, verifying that the decrease of the discrepancy is $\mathcal O(n^{-2})$ in the first case and $\mathcal O(n^{-1})$ in the second.

\vskip 1cm
\noindent
{\bf Acknowledgments}
This project has received funding from the European Union’s H2020 research and innovation programme under the Marie Skłodowska–Curie grant No. 778010 IPaDEGAN. TG  and MB acknowledge the support of GNFM-INDAM group and the  research project Mathematical Methods in Non Linear Physics (MMNLP), Gruppo 4-Fisica Teorica of INFN. 
The work of  MB was
supported in part by the Natural Sciences and Engineering Research Council of Canada (NSERC)  grant  RGPIN-2016-06660.

\newpage 
\bibliographystyle{alpha}

\end{document}